\documentclass[onecolumn,a4paper,fleqn,usenatbib]{mnras}
\usepackage{newtxtext,newtxmath}
\usepackage[T1]{fontenc}
\usepackage{ae,aecompl}

\usepackage{graphicx}	% Including figure files
\usepackage{amsmath}	% Advanced maths commands
\newcommand\numberthis{\addtocounter{equation}{1}\tag{\theequation}}
\usepackage{cancel}
\usepackage{pdflscape}	% Landscape pages
\usepackage[T1]{fontenc}
\usepackage{ae,aecompl}
\usepackage{glossaries}
\usepackage{amsmath, bm}
\usepackage{svg}
\usepackage{float}
\usepackage{xcolor}

\newcommand{\summa}[3]{\sum^{+\infty}_{#1 = -\infty} \; \sum^{+\infty}_{#2 = 1} \; \epsilon^{#3} \; e^{il(k'x' - \omega't' )}}
\newcommand{\dd}[3]{\frac{\partial f^{(#1)'}_{\alpha,#2}}{\partial #3}}
\newcommand{\integ}[1]{\int^{+\infty}_{-\infty} d#1 }
\newcommand{\ddo}{\frac{\partial f^{(0)'}_{\alpha}}{\partial p'}}

\title[relativistic Langmuir solitons]{Pulsar radio emission mechanism II. On the origin of relativistic Langmuir solitons in pulsar plasma}

\author[Rahaman et al.]{
Sk. Minhajur Rahaman,$^{1}$\thanks{E-mail: rahaman.minhajur93@gmail.com}
Dipanjan Mitra,$^{1,2}$
George I. Melikidze,$^{2,3}$
Taras Lakoba$^{4}$
\\
$^{1}$,National Centre for Radio Astrophysics,Tata Institute of Fundamental Research, Post Bag 3, Ganeshkind,Pune-411007,INDIA\\
$^{2}$Janusz Gil Insitute of Astronomy, University of Zielona G\'ora, ul Szafrana 2, 65-516 Zielana G\'ora, Poland \\
$^{3}$ Abastumani Astrophysical Observatory, Ilia State University, 3-5 Cholokashvili Ave., Tbilisi, 0160, Georgia \\
$^{4}$ Department of Mathematics and Statistics, University of Vermont, Burlington VT 05401, USA}

\date{Accepted XXX. Received YYY; in original form ZZZ}

\pubyear{2022}

\begin{document}
\label{firstpage}
\pagerange{\pageref{firstpage}--\pageref{lastpage}}
\maketitle

\begin{abstract}
Observations suggest that coherent radio emission from pulsars is
excited in  a dense  pulsar plasma by curvature radiation from
charge bunches. Numerous studies propose that these charge bunches are relativistic charge solitions
  which are solutions of the non-linear Schr\"{o}dinger equation
  (NLSE) with a group velocity dispersion ($G$), cubic-nonlinearity
  ($q$) and non-linear Landau damping ($s$). The formation of stable solitons crucially
  depends on the parameters $G, q$ and $s$ and the particle distribution
  function. In this work, we use realistic pulsar plasma
  parameters obtained from observational constraints to explore the
  parameter space of NLSE for two representative distribution
  functions (DF) of particles' momenta: Lorentzian (long-tailed) and
  Gaussian (short-tailed). The choice of DF critically affects the
  value of $|s/q|$, which, in turn, determines whether solitons can
  form. Numerical simulations show that well-formed solitons are
obtained only for small values of $|s/q| \lesssim 0.1$ while for
moderate and higher values of $|s/q| \gtrsim 0.5$ soliton formation is
suppressed. Small values for $|s/q| \sim 0.1$ are readily obtained for
long-tailed DF for a wide range of plasma temperatures. On the other
hand, short-tailed DF provides these values only for some 
narrow range of plasma parameters. Thus, the presence of a
prominent high-energy tail in the particle DF favours soliton
formation for a wide range of plasma parameters. Besides pair plasma,
we also include an iron ion component and find that they make a
negligible contribution in either modifying the NLSE coefficients or
contributing to charge separation.
\end{abstract}


%%%%%%%%%%%%%%%%%%%%%%%%%%%%%%%%%%%%%%%%%%%%%%%%%%

%\baselineskip=22pt plus1pt

%%%%%%%%%%%%%%%%% BODY OF PAPER %%%%%%%%%%%%%%%%%%
\section{Introduction}
 Understanding the mechanisms of coherent radio emission from pulsars
  has been a challenging astrophysical problem since the discovery of
  pulsars. Most models of coherent radio emission involve growth of
  instability in strongly magnetized relativistically streaming pair
  plasma and are broadly classified into maser or antenna
  mechanisms (see e.g. \citealt{1969Ap&SS...4..464G,1991MNRAS.253..377K,
  1995JApA...16..137M}). Recent single pulse polarization observations,
  however, strongly favour the antenna mechanism, where the radio 
  emission is excited in pair plasma by coherent curvature radiation  
  (hereafter CCR) due to motion of charge bunches along curved magnetic
  field lines (\citealt{2009ApJ...696L.141M}). 
  
  Observations have further
  established that the radio emission detaches the pulsar magnetosphere
  from around 500 km above the neutron star surface
(\citealt{1997MNRAS.288..631K}; \citealt{1998MNRAS.299..855K};
\citealt{2017JApA...38...52M}), where the magnetic field topology 
is purely dipolar (\citealt{2004A&A...421..215M}). At the radio emission region, due to enormously strong magnetic field, the motion of plasma particles can be approximated to be one-dimensional. The primary source of pair plasma in pulsars is due to 
magnetic pair production by high energy photons at the polar cap. In our study we consider the scenario for $\vec{\Omega}_\mathrm{Rot} \cdot \vec{B} < 0$ in which a charge-starved inner accelerating region (IAR) region develops above the polar cap where unscreened electric field exists and the primary pairs are formed and accelerated to extremely 
high Lorentz factors $\gamma_\mathrm{p}$ (see \citealt{1971ApJ...164..529S}, 
RS75). One kind of charges is accelerated away from the polar cap, and these charges
can radiate high energy photons, which in turn produces a cascade of secondary 
pair plasma moving with Lorentz factor $\gamma_\mathrm{s}$.
Several lines of evidence suggest a
strongly non-dipolar magnetic field topology at the surface
(\citealt{2017JApA...38...46G}; \citealt{2019MNRAS.489.4589A};
\citealt{2020MNRAS.492.2468M}) and in such strong fields copious pair creation 
can occur. As a 
result, dense and hot pair plasma is produced. The density of the pair plasma exceeds the
co-rotation Goldreich-Julian charge density (\citealt{1969ApJ...157..869G}) by 
a factor $\kappa \sim 10^{4} - 10^{5}$
(\citealt{2002ApJ...581..451A}), streaming with a bulk Lorentz
factor $\gamma_\mathrm{s} \approx 10^{2} - 10^{3}$ in the observer's
frame of reference. Observations of Pulsar Wind nebulae  has
also confirmed the presence of a dense pair plasma
(\citealt{2011ASSP...21..624B}). In the IAR, the charge which accelerates towards
the polar cap can heat the polar cap to high temperatures, and
X-ray observation have revealed the presence of such hot polar cap in several pulsars. However, extremely high temperatures could be expected if the polar cap discharges were to occur under pure vacuum conditions, 
which is not observed. Hence to properly account for the polar
cap temperature, \cite{2003A&A...407..315G} suggested that the
IAR is a Partially Screened Gap (PSG). The 
PSG model is a variant of the pure vacuum models and takes into consideration the binding energy of iron ions on the surface. The heating due to backflowing charges unpins iron ions from the surface and contributes close to 90 $\%$ of the
co-rotational charge density. The flow of ions is thermostatically regulated as follows: if the surface is heated beyond some critical temperature $T_\mathrm{ion}$, the gap closes completely while for surface temperature below $T_\mathrm{ion}$  the gap is partially screened. Under equilibrium conditions, the surface temperature is only slightly offset from the critical temperature and any greater offset is corrected on timescales of few hundred nanoseconds. Owing to a heavier mass, the iron ions are accelerated to Lorentz factors $\gamma_\mathrm{ion}$ close to the Lorentz factor of the secondary
plasma $\gamma_\mathrm{s}$. The PSG model is a very successful
phenomenological model for explaining the subpulse drift rates, mode
changing and thermal X-ray luminosity \citep[see
  e.g.][]{ 2016ApJ...833...29B, 2021MNRAS.500.4139R, 2015MNRAS.447.2295S}. The
presence of an additional iron component in the pulsar plasma is hence an
important ingredient. To summarize, magnetically induced pair cascades and outflow of ions above the polar cap gives rise to an ultra-relativistic, collisionless and multi-component plasma outflow strictly 
along the open magnetic field lines of 
the pulsar (\citealt{1969ApJ...157..869G};
\citealt{1971ApJ...164..529S}; \citealt{1975ApJ...196...51R} hereafter
RS75).

On the theoretical front, the formation of stable charge bunches capable of explaining coherent radio
emission from pulsars has been a long standing puzzle (\citealt{1969Ap&SS...4..464G};
\citealt{1999ApJ...521..351M}). 
Earlier studies suggested that in the radio emission zone linear Langmuir waves can
be unstable due to plasma two-stream instability, and as a result, 
linear charge bunches can radiate coherently
(RS75; \citealt{1977ApJ...212..800C}). However,
it was soon realised that the very high-frequency linear Langmuir
waves disperse the linear bunch well before it can emit coherently
(\citealt{1986FizPl..12.1233L}; \citealt{1999ApJ...521..351M};
\citealt{2000ApJ...544.1081M} hereafter MGP00; \citealt{2018MNRAS.480.4526L}). In order to circumvent
this problem, studies like that of \citealt{1975PhyS...11..271K};
\citealt{1980Afz....16..161M,1980Ap&SS..68...49P,1984Ap.....20..100M}; MGP00 explored the non-linear regime
of Langmuir waves to provide a time-stable charge distribution. A
necessary condition for exploring the non-linear regime is the
presence of strong plasma turbulence in the linear
regime, and  \cite{1998MNRAS.301...59A} 
and more 
recently \cite{2020MNRAS.497.3953R} (hereafter Paper I) showed that
very effective two-stream instabilities can provide this condition
within 1000 km from the neutron star surface. Recent particle-in-cell simulations
by \cite{2021A&A...649A.145M} also established the presence of
strong Langmuir turbulence in pulsar plasma. In the non-linear regime
the linear Langmuir waves with frequency $\omega_1$ interact to produce low-frequency beats ($ \Delta \omega \ll \omega_1 $) that modulates the
envelope $E$ of the high-frequency linear Langmuir waves. Since the linear waves do not maintain a definite phase relationship with each other over the spatial scale. As a result, the envelope electric field $E$ itself  has a white-noise character and the initial envelope electric field is assumed to be completely disordered. 
The envelope $E$ is governed by the
non-linear Schr\"{o}dinger equation (hereafter NLSE) with a non-local
term (see e.g. \citealt{ 1980Afz....16..161M}; \citealt{1980Ap&SS..68...49P}; \citealt{1984Ap.....20..100M};  \citealt{2000ApJ...544.1081M}):
\begin{equation}
    i \partial_{t} E + G \partial^{2}_{xx} E + q |E|^2 E + \mathrm{s} \; \mathcal{P} \int dx' \; V(x,x') = 0. \label{general NLSE}
\end{equation}
The term $G \partial^{2}_{xx} E$ represents the group velocity
dispersion (hereafter GVD) of the linear Langmuir waves. The term $q |E|^2
E$ represents cubic non-linearity (hereafter CNL). The non-local term $
s \; \mathcal{P} \int dx' \; V(x,x') $ represents the non-linear
Landau damping (hereafter NLD). NLD represents a resonant interaction at the group velocity of the Langmuir waves with plasma particles. The interaction at group velocity not only gives rise to NLD but also modifies CNL. The coefficient $q$ represents the strength of local (in space) non-linear interactions. The coefficient $s$ represents a non-local interaction via a cascade of energy from higher length scales (lower wave numbers) to shorter length scales (higher wave numbers) (see subsection \ref{sign_sq}). The time-stable solution of Eq. (\ref{general NLSE}) are referred to as solitons, which are considered as candidates for charge bunches giving rise to CCR at radio wavelengths.

In the absence of NLD, Eq. (\ref{general NLSE}) represents a
purely local NLSE. \cite{1967RSPSA.299...28L} showed that this equation
admits analytical solutions as solitons, provided that the so-called  Lighthill condition represented as
\begin{equation}
    G q > 0,  \label{lighthill condn}
\end{equation}
is satisfied and the initial electric wave field is a phase-coherent plane wave. Previous studies by \citealt{1978SoGru..90...49M}; \citealt{1980Afz....16..161M}; \citealt{1980Ap&SS..68...49P}; \citealt{1984Ap.....20..100M}; MGP00; neglected NLD to get analytical solutions and conjectured that the Lighthill condition can be satisfied in pulsar pair plasma.  \cite{2018MNRAS.480.4526L} (hereafter LMM18) pursued numerical solution of Eq. (\ref{general NLSE}) and confirmed a previously known fact that purely cubic NLSE cannot give rise to long-living solitons from either an initially disordered electric field $E$ (the most natural state for the Langmuir envelope) or even from a phase-coherent plane wave-like initial electric field. More importantly, LMM18 found that for finite but sufficiently weak non-locality
of the nonlinear interactions, i.e., for finite but small values of $|s/q|$, 
formation of long-living solitons did occur. LMM18 estimated a range of $|s/q|$
values where such formation takes place, but did not address the question 
whether that range values of $|s/q|$ could actually exist under generic hot plasma
conditions in pulsar magnetosphere. Answering it
requires modelling of group velocity interaction of Langmuir waves with plasma particles, which depends on the choice of particle momentum distribution function (hereafter DF). Therefore, for such modelling, one needs to consider physically motivated and
representative forms of DF in pulsar plasma. To our knowledge, this has not been
done in any previous studies and thus has been an open issue. 

 In order to obtain solitons that can have properties of a charge bunch, the electron-positron DFs 
of the pair plasma must separate to
create charge-separated structures in the configuration space. These
charge-separated structures have been proposed as candidates for CCR
charge bunches. MGP00 also suggested that the
presence of heavier ion species, that had been proposed by the PSG model, can also aid in charge separation. However, the relative contribution of
the two effects has not been studied before. Thus, the presence of ions is an important ingredient that has not been considered in earlier studies and hence also needs to be explored.

The present study is focused on addressing the two open issues stated above. Namely, it has the following objectives. Firstly, we want to estimate the ratio of $s/q$ and explore the parameter space for pulsar plasma DFs and then simulate numerically the soliton profiles for the $s/q$ range obtained. Secondly, we want to estimate the relative contribution of the separation of the electron-positron DF and the presence of ions in determining the charge separation in soliton profiles.

The paper is organized as follows. We introduce the NLSE in section
\ref{NlSE_intro}. The parameter space and soliton solutions of NLSE is
explored in section \ref{para space}. Typical estimates of the charge
separation of the Langmuir solitons are presented in section
\ref{charge_sep_section}. Our conclusions are summarized in section
\ref{conclusions}.

\section{Introduction to NLSE with NLD}\label{NlSE_intro}

We identify three frames of reference. We have a plasma frame of
reference (hereafter PFR) where the average velocity of the pair
plasma particles is zero. The PFR moves with a Lorentz factor
$\gamma_\mathrm{s}$ with respect to the observer frame of reference
(OFR). The moving frame of reference (MFR) moves with respect to PFR at the
group velocity of the linear Langmuir waves $v_\mathrm{gr}$ in
PFR. Quantities in MFR are primed while the quantities in PFR are
unprimed. The envelope ($E$) of the Langmuir waves is governed by the
NLSE with the NLD,
\begin{equation}
    i \;  \frac{\partial E}{\partial \tau'} +   G \; \frac{\partial^2 E }{\partial \xi'^2} + q \; E \;  |E|^2  +  s \; \frac{1}{\pi} \mathcal{P} \int^{+\infty}_{-\infty} d\xi'' \; \frac{|E(\xi'',\tau')|^2}{\xi' - \xi''}  \; E  = 0 , \label{NLSE}
\end{equation}
where the quantities $\tau'$ and $\xi'$ represents the slow time and space variables in MFR respectively (see Eq. A1 and Eq. A2 in Appendix A). It must be noted that while the equation itself is written in MFR, the coefficients $(G, q,s)$ are computed in PFR. Here the symbol $\mathcal{P}$ stands for the
Principal value Cauchy integral.

A complete formal derivation of Eq. (\ref{NLSE}) is found in
Appendix A. Eq. (\ref{NLSE}) was derived by MGP00 (see also 
\citealt{1980Afz....16..161M}; \citealt{1980Ap&SS..68...49P}; \citealt{1984Ap.....20..100M}), however our 
derivation differs from MGP00 in one crucial aspect. It allows contributions for arbitrary species
of mass $m_{\alpha}$ and charge $e_{\alpha}$ (in particular, ions)
to be taken into
account, while the original derivation of MGP00 was for an electron-positron plasma. 
Inclusion of ions as an additional plasma
component and evaluation of their contribution to the coefficients of the NLSE \eqref{NLSE}
is one of the stated goals of this study. 
It must also be noted that in our derivation, certain integrals where MGP00 missed the charge dependencies, have been updated. The setup for the introduction of ions and tracking the
charge dependence of multiple species is described in Appendix B.

\subsection{The NLSE coefficients}

The coefficients of Eq. (\ref{NLSE}) can be represented in their
dimensionless form as (see MGP00, also Eq. B6, B8 and B11 of Appendix B)
\begin{eqnarray}
&\ G =  \frac{c^2}{\omega_\mathrm{p}} (\gamma^3_\mathrm{gr} \mathrm{g}_\mathrm{d}) =  \frac{c^2}{\omega_\mathrm{p}} \mathrm{G}_\mathrm{d}, \label{GVD} \\
&\ q = \frac{1}{\omega_\mathrm{p}} \left( \frac{e}{m_{e} c}\right)^2   \mathrm{q}_\mathrm{d}, \label{CNL} \\
&\ s = \frac{1}{\omega_\mathrm{p}} \left( \frac{e}{m_{e} c}\right)^2   \mathrm{s}_\mathrm{d}, \label{NLD} 
\end{eqnarray}
where the 
coefficients ($G_\mathrm{d},
q_\mathrm{d},s_\mathrm{d}$) are dimensionless.
We will first present an estimate for the plasma frequency
$\omega_\mathrm{p}$ in \eqref{GVD}--\eqref{NLD} and then discuss factors
that affect ($G_\mathrm{d}, q_\mathrm{d},s_\mathrm{d}$). Values
of these coefficients themselves are discussed in the next Section.

The typical plasma frequency at a distance $r$ from the neutron star surface in OFR is
\begin{equation}
    \omega_\mathrm{p,OFR} =
\sqrt{\frac{4 \pi n_\mathrm{s} e^2}{ m_\mathrm{e}} }
\end{equation}
 where $m_e$ is the mass of electron, $e$ is charge of electron, $n_\mathrm{s} =
\kappa B/ (P c e)$ is the number density of the pair plasma, $\kappa$ is the ratio of the number density of the pair plasma to the Goldreich-Julian number density $n_\mathrm{GJ} = {B}/{(Pce)}$,
$B = B_\mathrm{d} (r/R_\mathrm{NS})^3$ is the magnetic field strength,
$P$ is the period of the pulsar, and $c$ is the speed of light. For typical pulsar
parameters with period $P = 1$ second, dipolar magnetic field
$B_\mathrm{d} = 10^{12}$ gauss and radius $R_\mathrm{NS} = 10 $ km,
the corresponding plasma frequency $\omega_\mathrm{p}$ in PFR can be obtained by the Lorentz transformation to be
\begin{equation}
    \omega_\mathrm{p} = \frac{\omega_\mathrm{p,OFR}}{\gamma_\mathrm{s}}  \approx   10^{8} \left( \frac{200}{\gamma_\mathrm{s}}  \right) \sqrt{ \left( \frac{\kappa}{10^{4}} \right) \; \left( \frac{1 \; \text{s}}{P} \right) \left( \frac{500 \; \text{km} }{ r}\right)^3  } \; \text{rad s$^{-1}$}, \label{plasma_freq}
\end{equation}
where the Lorentz factor $\gamma_\mathrm{s}$ was discussed in  Introduction.

The coefficients ($G_\mathrm{d}, q_\mathrm{d},s_\mathrm{d}$)
depend only on the plasma particles' 
momentum distribution function (DF):
see Eqs. B7, B9, B12 in Appendix B.
Therefore, we now review various relevant models of DF 
so as to justify its representative forms that we will use
in this study. 
As stated in Introduction, it is well established that 
normal-period radio pulsars have a strong non-dipolar surface component ( see \citealt{2019MNRAS.489.4589A} and the references therein) along with an thermionic ion flow from the surface (\citealt{1980ApJ...235..576C}; \citealt{2003A&A...407..315G}).  While some semi-analytical estimates of the pair cascade in strong non-dipolar fields  have been made (\citealt{2015MNRAS.447.2295S}), the generic shape of the pair plasma DF is not known. However, numerical simulations like those by 
\cite{2002ApJ...581..451A} show that the shape of the DF is strongly affected by the opening angle between the ambient magnetic field and the initial seed photon, the strength of the magnetic field and the seed photon energy.
Namely, for low-opening angles, the DF is well described by the J\''{u}ttner-Synge distribution, so that the number of particles with high dimensionless momenta $p$ (defined in (\ref{defmomentum})) falls off as $\exp(- K (\ln{p/p_\mathrm{o}})^2)$, where $K$ is inverse width of the DF and and $p_\mathrm{o}$ is the dimensionless momentum corresponding to the peak of the DF . In this paper we refer to this behaviour of the DF as ``short-tailed''. On the other hand, at large opening angles, the DF of the number of particles was found to fall off as $\exp(-p^{0.2})$ at high momenta. 
In general, 
\cite{2002ApJ...581..451A} 
found these latter DF to be significantly broader than those at small opening angles. Therefore, we refer to this type of DFs as ``long-tailed''. 
  It must be kept in mind that the simulations by \citet{2002ApJ...581..451A}  assume the initial seed photons to be mono-energetic, and relaxing this condition may lead to significant changes in the resulting DFs. 
  Among other pair cascade models, like those
by \citet{2001ApJ...560..871H} 
and by \cite{1973Ap&SS..23..189S} 
exhibit the presence of a 
 power-law ``long-tail'' which falls off inversely as the third power of the particle momentum.
On the other hand, Monte Carlo models used by 
 \cite{1982ApJ...252..337D} show a ``short-tail'' in the particle DF
(see Fig. 5 of \citealt {2001ApJ...560..871H} for comparison). 
Thus, earlier studies demonstrate the possibility of having both types: short- and long-tailed, DF in pulsar plasma.

For the soliton formation based on the NLSE model \eqref{NLSE},
we will show below that
the presence/absence of an extended tail in the DF is of paramount importance. Namely, it eventually determines 
the number of plasma particles contributing to the non-linear Landau damping and cubic non-linearity terms in the NLSE. In order to explore this aspect, we choose two representative forms of particle DF, viz., a Gaussian with an exponentially decaying tail (``short-tail'') and a Lorentzian with a power-law tail (``long-tail'').

The particle
DF $f^{(0)}_{\alpha}$ is taken to be a
function of the dimensionless momentum $p$, which is defined as
\begin{equation}
    p = \frac{P_{\alpha}}{m_{\alpha} c} = \frac{\gamma m_{\alpha} v}{m_{\alpha} c} = \gamma \beta \equiv \frac{\beta}{\sqrt{1 - \beta^2}}.  
    \label{defmomentum}
\end{equation}
where $P_{\alpha}$ is the the relativistic momentum and $m_{\alpha}$
is the mass of the plasma particles of the $\alpha$-th species. For
the rest of the analysis the term `momentum' would be used to refer to
dimensionless momentum of the plasma particles. 
For both Gaussian and Lorentzian DFs,
the term ``temperature'' will be used to refer to their
widths in the momentum space.
We will also sometimes refer to the tail of the Lorentzian DF 
as ``high-energy" tail, since particle energy scales approximately as momentum in the ultra-relativistic regime.

 As seen from the Table B2 and Table B3 of Appendix B, the integrals in the the dimensionless coefficients $G_\mathrm{d}$, $q_\mathrm{d}$ and $s_\mathrm{d}$ require the estimation of the group velocity of the particles. For a given DF the wave group velocity (normalized to speed of light $c$) is estimated from the expression (see Eq. B3 of Appendix B)
\begin{equation}
    \beta_\mathrm{gr} = \frac{1}{c} \frac{d \omega}{d k} = \frac{1 + \sum_{\alpha} \left( \frac{\omega_\mathrm{p,\alpha}}{kc} \right)^2 \int^{+\infty}_{-\infty} dp \; \frac{\partial f^{(0)}_{\alpha}}{\partial p} \frac{\beta}{ (\beta_\mathrm{ph} - \beta)^2}  }{ \sum_{\alpha} \left( \frac{\omega_\mathrm{p,\alpha}}{k c}\right)^2 \int^{+\infty}_{-\infty} dp \; \frac{\partial f^{(0)}_{\alpha}}{\partial p} \frac{1}{(\beta_\mathrm{ph} - \beta)^2} } . 
\end{equation}
where  $\beta_\mathrm{ph}$ corresponds to the non-dimensional phase velocity of the linear Langmuir waves (normalized to the speed of light $c$) and the dimensional wave number $k$  is given by the expression
\begin{equation}
    k = \frac{1}{c} \;\left[ \sum_{\alpha} \omega^2_\mathrm{p,\alpha} \int^{+\infty}_{-\infty} dp \;  f^{(0)}_{\alpha} \frac{1}{\gamma^3 \; (\beta_\mathrm{ph} - \beta)^2} \right]^{1/2}.  \; \label{wave number}
\end{equation}
where  $\omega_\mathrm{p,\alpha}$ is the plasma frequency associated with $\alpha$-th species in the plasma and is defined as
\begin{equation}
    \omega_\mathrm{p,\alpha} = \sqrt{\frac{4 \pi  n_{\alpha} e^2_{\alpha}}{m_{\alpha}}}
\end{equation}

Note that the dependence of $\omega_\mathrm{p,\alpha}$ on the particle species
comes from its dependence on mass $m_{\alpha}$, number density $n_{\alpha}$ and the charge $e_{\alpha}$ of the species.

In Section \ref{para space} we will also extensively refer to the 
momentum corresponding to the group velocity, given
according to \eqref{defmomentum},
by
\begin{equation}
    p_\mathrm{gr} = \gamma_\mathrm{gr} \beta_\mathrm{gr}. \label{pole} 
\end{equation}

This $p_\mathrm{gr}$ appears as a pole in the integrals for $s_\mathrm{d}$ and
$q_\mathrm{d}$ (see Table B2 and B3 in Appendix B). The location of
this pole determines the
magnitude of $s/q$ (see Eq. B23 of Appendix B). Physically, $|s/q|$ is
higher if the pole $p_\mathrm{gr}$ is near the peak of the DF since
then the number of particles interacting with Langmuir waves is
greater, and vice versa.

In the next subsection we discuss under what condition charge separation occurs in the configuration space and how the presence of an iron species component may potentially enhance the charge separation.

\subsection{Charge separation in configuration space}

The slowly varying charge density (in electrostatic units per cubic centimeters) corresponding to the envelope field of Eq. (\ref{NLSE}) is given by (see Eq. A23 of MGP00)
\begin{equation}
    \rho = \mu \; \left( \frac{1}{4\pi k^2 c^2} \right) \; \left( \frac{|e|}{m_\mathrm{e} c^2} \right) \frac{\partial^2 |E|^2 }{\partial \xi'^2 },  \label{slow_charge}
\end{equation}
where
\begin{equation}
    \mu =  \frac{\sum_{\alpha} \mathrm{sgn(\alpha)}\;\varphi_{\alpha}\; \omega^2_\mathrm{p,\alpha} \mathcal{P}\int^{+\infty}_{-\infty} dp \; \frac{1}{(\beta - \beta_\mathrm{gr})} \frac{\partial }{\partial p}\left[ \frac{(\beta - \beta_\mathrm{gr})}{(\beta_\mathrm{ph} - \beta)^2}\frac{\partial f^{(0)}_{\alpha}}{\partial p} \right]   }{ \sum_{\alpha} \omega^2_\mathrm{p,\alpha} \mathcal{P}\int^{+\infty}_{-\infty} p \; \frac{1}{(\beta - \beta_\mathrm{gr})} \frac{\partial f^{(0)}_{\alpha}}{\partial p} } ,  \label{slow_charge_integ} 
\end{equation}
where $\mathrm{sgn}(\alpha)$ is $+$ for positrons and ions, and is $-$ for electrons, and $\varphi_{\alpha} = (|e_{\alpha}|/e) \times ({m_{e}}/{m_{\alpha}})$.

Equation (\ref{slow_charge_integ}) shows that for coinciding electron
and positron DF, the terms pertaining to electrons and positrons in
the numerator of (\ref{slow_charge}) cancel each other. Then, integral
$\mu$ vanishes and there is no charge separation. 
Physically, this effect of charge separation can be understood as follows. The term $ {\partial^2
|E|^2 }/{\partial \xi'^2 } $ represents the ponderomotive/Miller
force.  The Miller force is a pressure force which pushes plasma
particles from regions of strong to low electric fields. The force
is independent of the sign of the charge particles
but depends on the magnitude of charge to mass ratio of the $\alpha-$th plasma species.
For example, in an electron-ion plasma, the Miller force can push an
electron farther away compared to an ion, and hence effective charge
separation can be achieved. In the case of pair plasma, since the
charge to mass ratio is same for both species, there is no such
charge separation possible. Thus, in pulsar relativistic pair plasma
for a coinciding electron-positron DF, no charge separation is possible.
However, it was pointed out by MPG00 that due to flow
of pair plasma along curved magnetic field lines,
the electron and positron DF of pair plasma can separate (\citealt{1977ApJ...212..800C}; \citealt{1998MNRAS.301...59A};
Paper I; also see Appendix F for full derivation) and hence relativistic masses of the electrons and positrons can be unequal.
Thus, the separation of electron-positron DF can produce an effective charge separation in plasma.
As shown in Paper I,
the extent of the separation is determined by the arrangement of
the non- dipolar surface magnetic field. For various arrangements of that field, the separation of the DF remains
nearly constant for around 1000 km above the neutron star surface. In
this context, we can treat the separation of the DF as a free
parameter, and therefore we will consider several
representative values of DF separation in Section 3.

MGP00 also suggested that the presence of an additional
heavier iron ion $^{56}_{26}$ Fe with a high magnitude of charge
component can enhance the charge separation. The PSG model provides an
important motivation for inclusion of an iron ion species as an additional component in the pulsar plasma. One of the goals of the present study is to find out if indeed the presence of an ion species can have appreciable effects on  charge separation.

In the next section, we will evaluate the dimensionless coefficients
of NLSE expressed in Eq. (\ref{GVD}) to Eq. (\ref{NLD}) and
the charge separation integral $\mu$ from equation
(\ref{slow_charge_integ}) as a function of plasma temperature and the
separation of the DF. We also include the contribution of a low-density ion component (see Appendix B3).

\section{Parameter space for NLSE for soliton formation}\label{para space}

The NLSE with NLD can be converted 
into
the dimensionless form  (see Eq. 20 of LMM18) as  
\begin{equation}
    i \frac{\partial u }{\partial t} + \frac{\partial^2 u}{\partial x^2} + Q u \left( |u|^2 + \frac{s}{\pi q} \mathcal{P} \int dx' \; \frac{|u(x',t)|^2}{x - x'}  \right) = 0. \label{dimensionless}
\end{equation}
where
\begin{eqnarray}
&\ u = \frac{E}{E_\mathrm{o}}, \label{u}  \\
&\ x =  \frac{\xi'}{l \theta}, \label{x}\\
&\ t = \frac{\omega_\mathrm{p} G_\mathrm{d}}{\theta^2} \tau' , \label{t}\\ 
&\ Q = \left[ \theta^2  \left( \frac{e}{m_{e} c^2}\right)^2 \left(  \frac{|E_\mathrm{o}|^2}{ 8 \omega^2_\mathrm{p} \gamma } \right)  \right]  \left(2\;\frac{q_\mathrm{d}}{G_\mathrm{d}}\right),  \label{Q} 
\end{eqnarray}
where $u$ is the non-dimensional amplitude of the Langmuir wave envelope, $x$ is the non-dimensional space variable, $t$ is the non-dimensional time variable, $Q$ represents the non-dimensional ratio of of the cubic non-linearity coefficient $q$ to the group velocity dispersion. Here, the characteristic length $l$ of the linear Langmuir waves is given by
\begin{equation}
l = \frac{2\pi}{k},  \label{l}
\end{equation}
where $k$ is the wave number as defined in Eq. (\ref{wave
  number}). 
  The quantity $\theta/2\pi$ is a spatial scaling variable which characterizes the ratio of the spatial extent of the 
    nonlinear 
    wave envelope to the characteristic length $l$
    of linear Langmuir waves. 
    Similarly to LMM18, we will use a
    value $\theta=100$ in the estimates of typical soliton properties,
    which will be presented in subsection \ref{LorentzDF}. For
simplicity, the term in the square brackets in Eq. (\ref{Q}) for
$Q$ 
will be taken to equal 1,
given that $E_\mathrm{o}$ is an
unknown field amplitude. The quantity $Q = 2
q_\mathrm{d}/G_\mathrm{d}$ has to be positive to fulfill the Lighthill
condition (\ref{lighthill condn}). Physically, the typical soliton
formation timescales are on the order of $\sim
\mathcal{O}(1/Q)$. Thus, soliton formation is delayed for smaller $Q$
and vice versa.

Solving Eq. (\ref{dimensionless}) requires us to specify an initial condition. LMM18 represented the initial condition as a combination of the constant electric field component and a random electric field component.  For our analysis, we discount any constant electric field and use only a completely disordered electric field (LMM18):
\begin{equation}
    u (x, 0) = \int^{+\infty}_{-\infty} dk  \; \frac{\hat{w}(k) \exp{[-0.5(k/k_\mathrm{corr})^2 - i k x ]} } {\sqrt{ \sqrt{\pi} k_\mathrm{corr}}}.\label{initial} 
\end{equation}
Here $k_\mathrm{corr}$ is the wave number corresponding to the correlation length $l_\mathrm{corr}$ such that
\begin{equation}
k_\mathrm{corr} = \frac{2 \pi}{l_\mathrm{corr}} ,  \label{corrk}
\end{equation}
and  quantity $\hat{w}(k)$ denotes a white noise field described by
\begin{eqnarray}
&\    \left\langle \hat{w}(k_{1}) \hat{w}(k_{2})\right\rangle = 0 ,  \\
&\   \left\langle \hat{w}^{\star}(k_{1}) \hat{w}(k_{2})\right\rangle = 2 \delta (k_{1} - k_{2}) ,  \label{white_noise} 
\end{eqnarray}
where the angle brackets denote ensemble average.
Let us mention that
increasing $k_\mathrm{corr}$ has the same effect as decreasing $Q$: 
they both increase the time at which solitons emerge; see Table 2 of LMM18.

To solve Eq. (\ref{dimensionless})
numerically, we use the Integrating Factor-Leap-frog method by
\cite{Lakoba2016}. Simulation parameters of the numerical scheme
are summarized in Appendix C.

Next, the maximum dimensionless time for soliton
formation can be estimated as follows. 
The derivation of Eq. (\ref{dimensionless}) assumes that background plasma conditions as captured by the coefficients (\ref{GVD})-(\ref{NLD}) are
steady during the evolution of the wave electric field. For any given separation of the DF, this condition requires that the plasma frequency $\omega_\mathrm{p}$ should not change drastically during the evolution of the wave electric field. 
From Eq. (\ref{plasma_freq}), the change in plasma  frequency $\Delta \omega_\mathrm{p}$ for
segments of $\Delta r $ km  along a field line can be estimated to be $\Delta \omega_\mathrm{p}/ \omega_\mathrm{p} = 1.5 \Delta r / r $. 
Thus, if we choose $\Delta r = 3 $ km
and $r= 500$ km,  the change in plasma frequency is less than 1 $\%$ and can indeed be neglected. Since the outflow is ultra-relativistic, a
typical timescale associated with this spatial length segment is
$\Delta t_\mathrm{OFR} = 3 \; \mathrm{km} / c \approx 10^{-5}$
seconds. We assume that the PFR moves with a Lorentz factor
$\gamma_\mathrm{s} \approx 200$ with respect to OFR. 
Then, the 
typical
timescale in the PFR is $\Delta t_\mathrm{PFR} = \gamma_\mathrm{s} \;
\Delta t_\mathrm{OFR} \approx 2 \times 10^{-3}$ seconds. The MFR moves
relative to OFR in the same direction as PFR (away from the pulsar along the magnetic field lines)
with a typical Lorentz factor $\gamma_\mathrm{gr}\approx p_\mathrm{gr}$
(see \eqref{pole})
with respect to PFR. Combining Lorentz factors for ultra-relativistic co-propagation (see Appendix G), we find that 
the maximum timescale in MFR is:
\begin{equation}
    \tau'_\mathrm{max}  \approx 2 p_\mathrm{gr} \; \Delta t_\mathrm{PFR}. \label{tau_max}
\end{equation}

Next, at
a typical distance of 500 km from the surface we find,  using 
Eq. (\ref{t}), that the maximum dimensionless time $t_\mathrm{max}$ is given by
\begin{equation}
    t_\mathrm{max} \approx 10^{4} \left( \frac{\omega_\mathrm{p}}{10^{8}\;\text{rad s$^{-1}$}} \right) \left( \frac{100}{\theta}\right)^2  \; G_\mathrm{d} \;\tau'_\mathrm{max},  \label{run_time}   
\end{equation}
where we have used that for the typical parameters assumed in this study,
$\omega_\mathrm{p}\sim 10^8$ rad s$^{-1}$; see Eq. (\ref{plasma_freq}).
In subsection \ref{LorentzDF} we will see that in those cases when solitons are formed, one can take 
$p_\mathrm{gr}\approx 6$ 
and $G_\mathrm{d}\lesssim 10$ 
as representative values. 
Then Eq. (\ref{tau_max}) yields 
$\tau'_\mathrm{max} \approx 2\times 10^{-2}$
seconds and Eq.
(\ref{run_time}) yields the following
estimate for the
maximum dimensionless time
$t_\mathrm{max}$ where the \eqref{NLSE}
can be applicable:
\begin{equation}
    t_\mathrm{max} \sim 2\times 10^3 \; \left( \frac{\omega_\mathrm{p}}{10^{8}\;\text{rad s$^{-1}$}} \right) \left( \frac{100}{\theta}\right)^2  \left( \frac{G_\mathrm{d}}{10} \right)  \left( \frac{\tau'_\mathrm{max}}{2\times
    10^{-2} \; \text{sec}}\right).   
\end{equation}
Thus, for the Lorentzian DF, the maximum dimensionless time of the simulation can be restricted to about 2000 units.

In fact, we observed solitons form over dimensionless times that are some two orders of magnitude smaller than the above estimate. This indicates that either solitons can form over distances much less than the above estimate of $\Delta r = 3$ km, or that the factor in
the square brackets in \eqref{Q}, which 
we had assumed to equal 1, can in fact be much smaller (thereby allowing a larger range of values for the dimensional field intensity $|E_\mathrm{o}|^2$, or a combination of both. In other words, a large range of values for the intensity of the initial linear field will be able to lead to soliton formation as long as the condition on $|s/q|$ stated in the
next subsection is fulfilled.

\subsection{Lorentzian DF} \label{LorentzDF}

\begin{figure*}
\begin{tabular}{cc}
\includegraphics[scale=0.5]{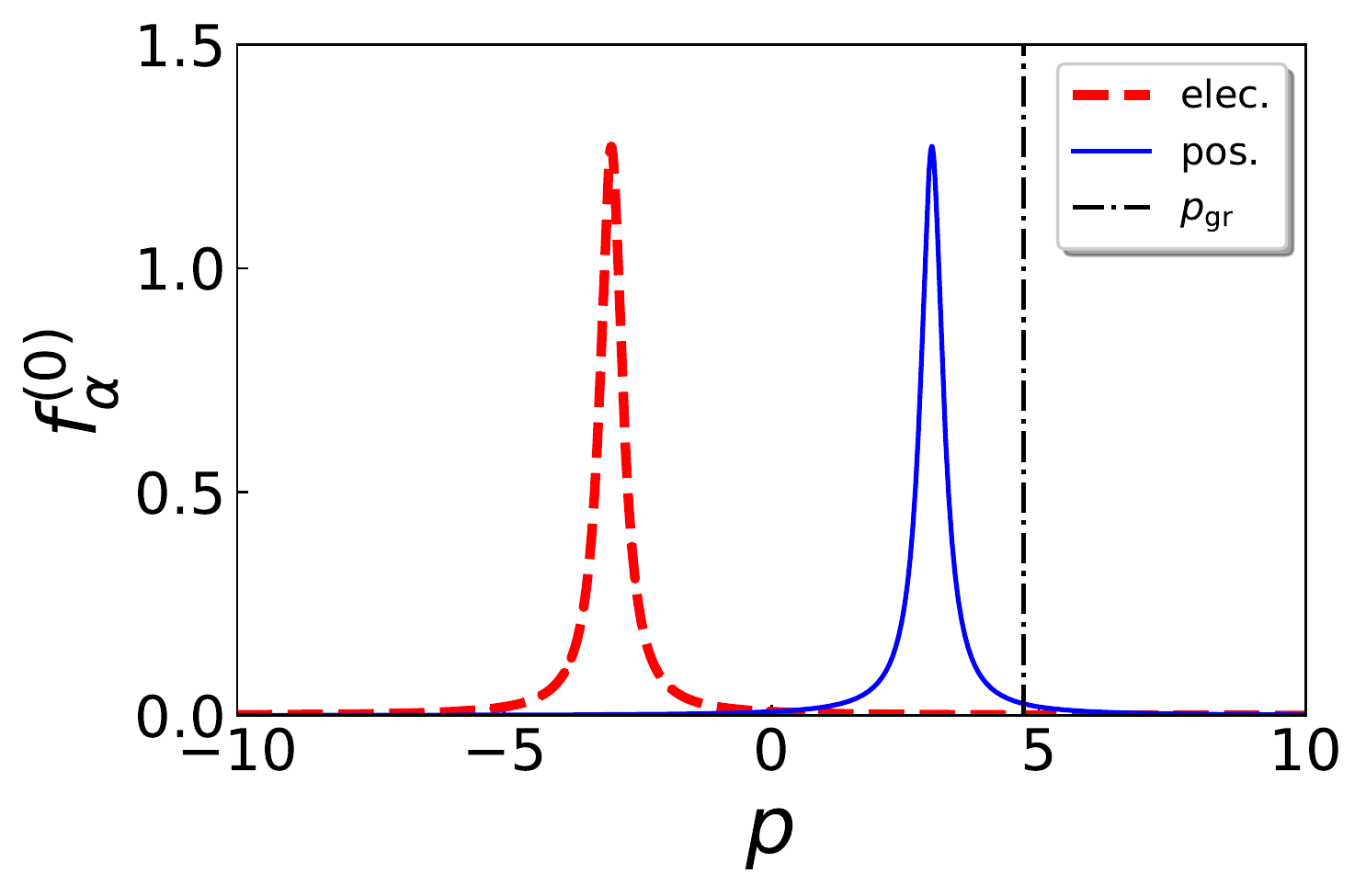} &
\includegraphics[scale=0.5]{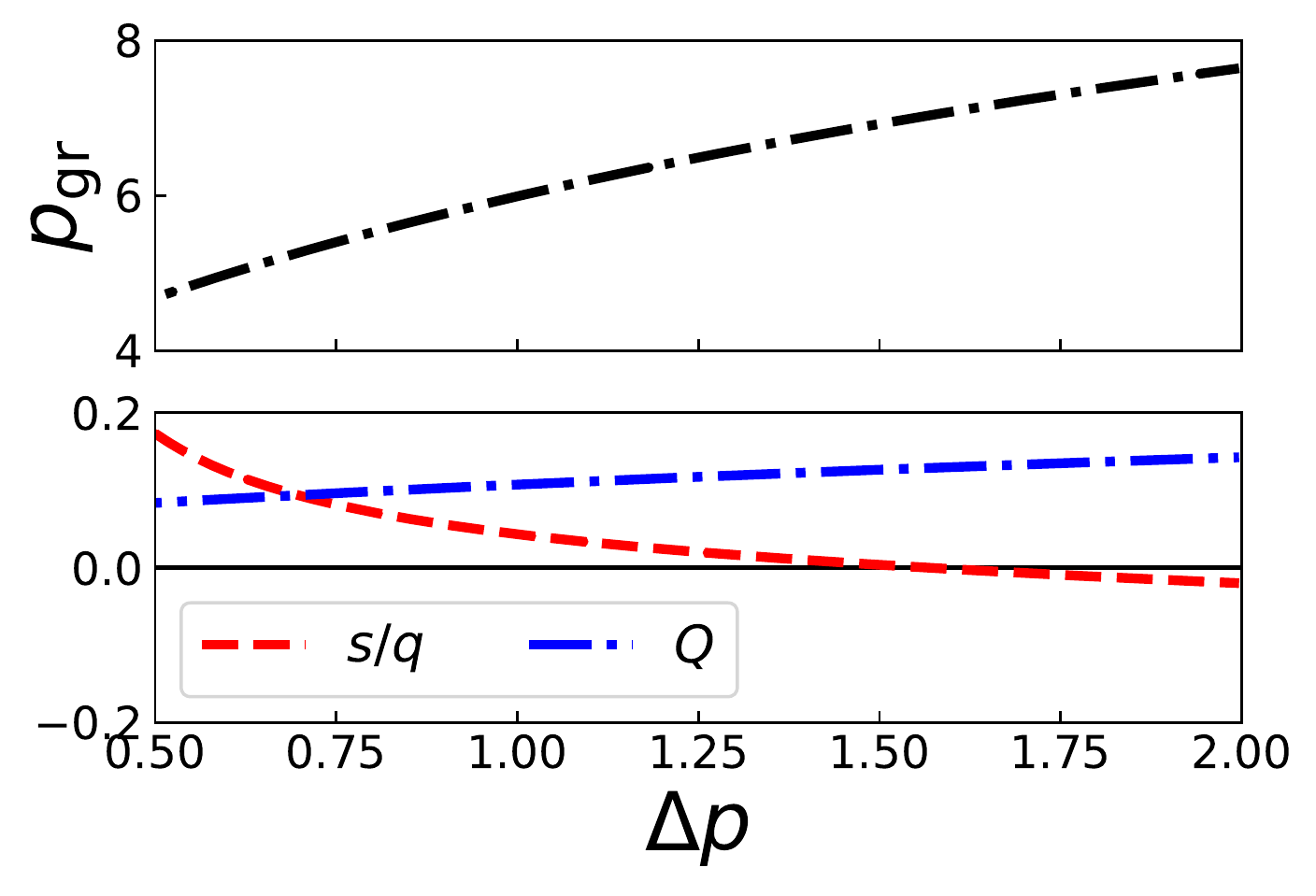} \\ (A) Lorentzian DF
and group velocity pole with ${p}_{\pm}=\pm 3 $ at $\Delta p = 0.5$. 
& (B) Parameter space for Panel (A) as a function of temperature. 
\\ \\ \\ \includegraphics[scale=0.5]{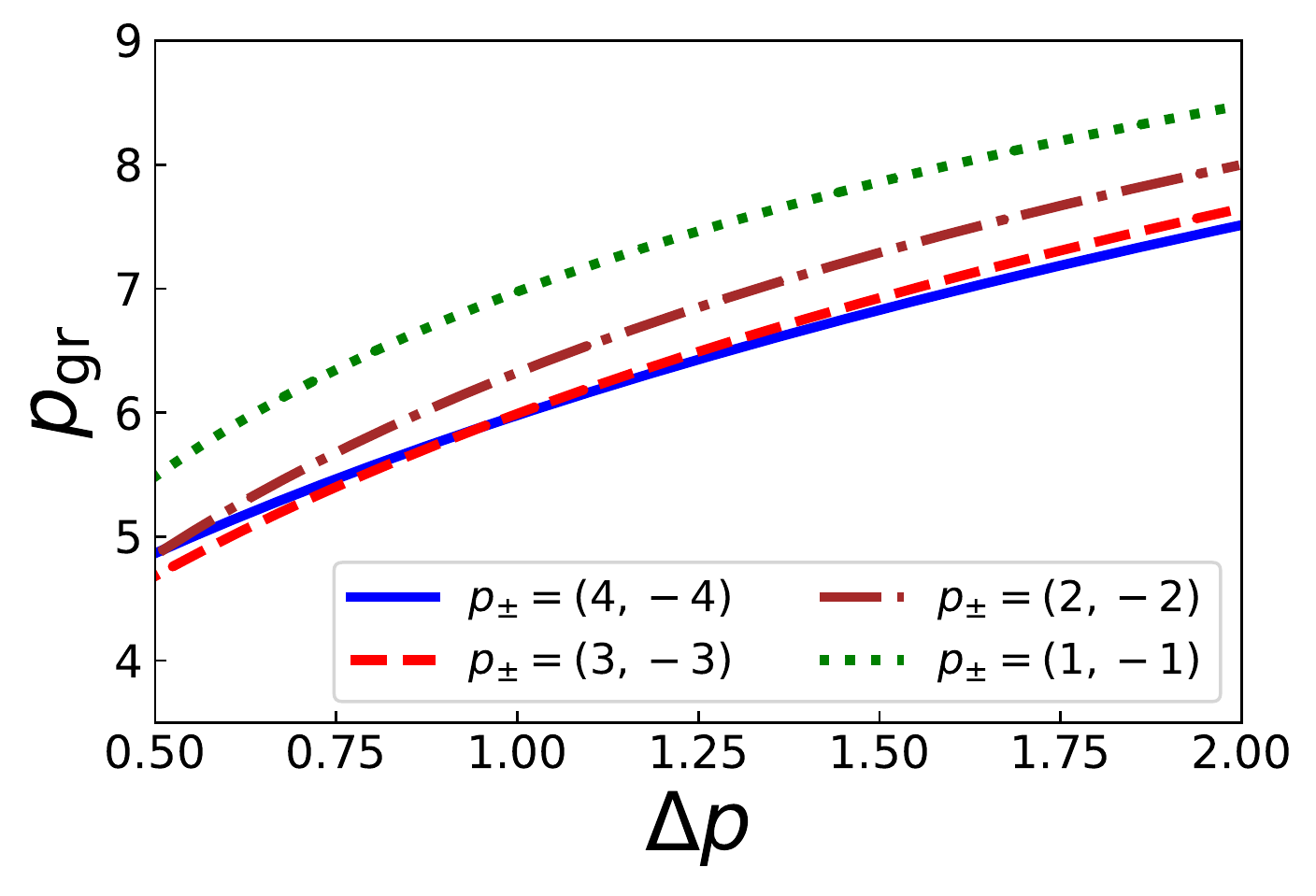} &
\includegraphics[scale=0.5]{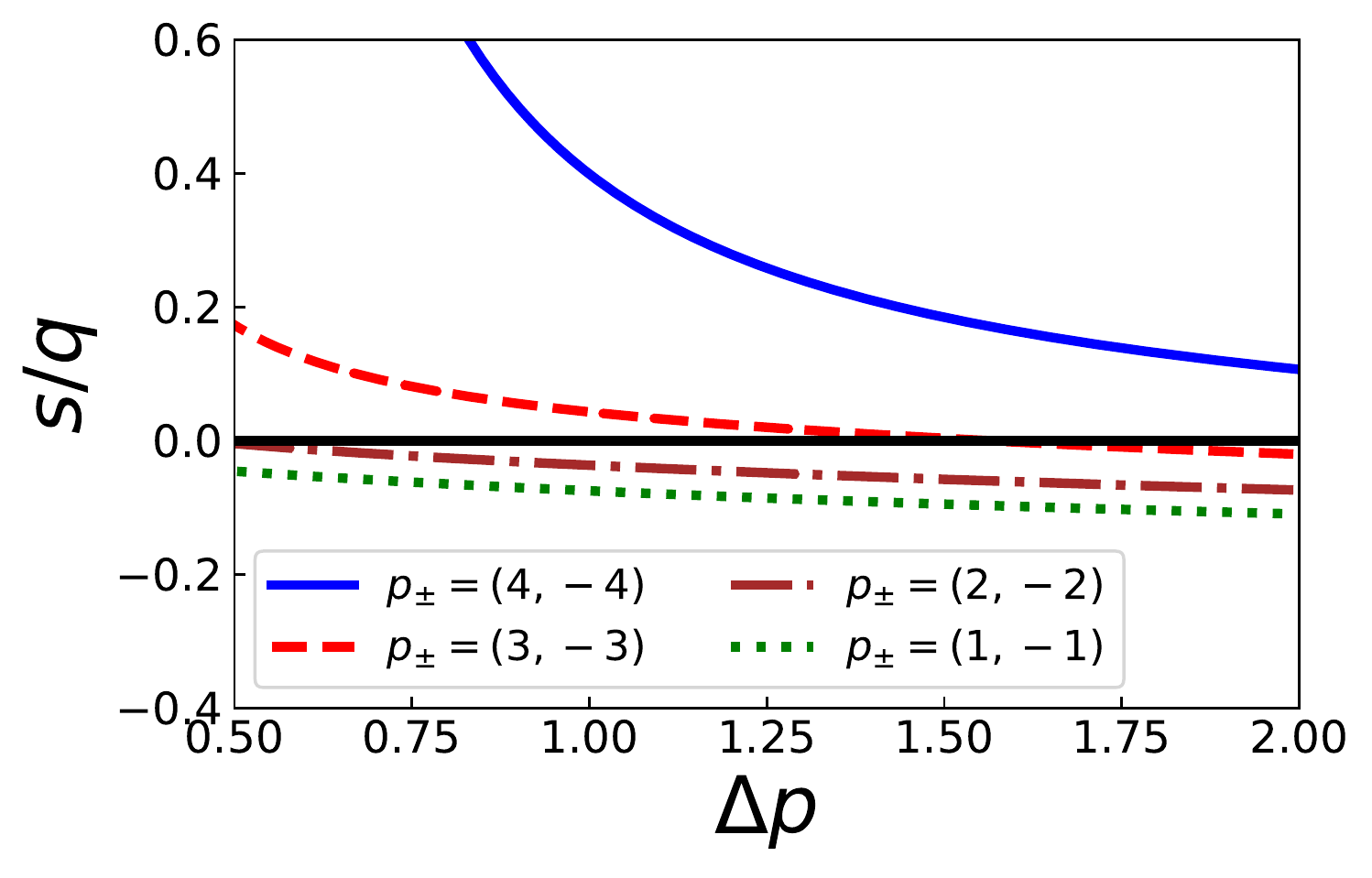} \\ (C) The group velocity pole as a function of temperature and separation of DF.  & (D)  Ratio of NLD to CNL as a function of temperature and separation of DF. \\ \\ \\ \includegraphics[scale=0.5]{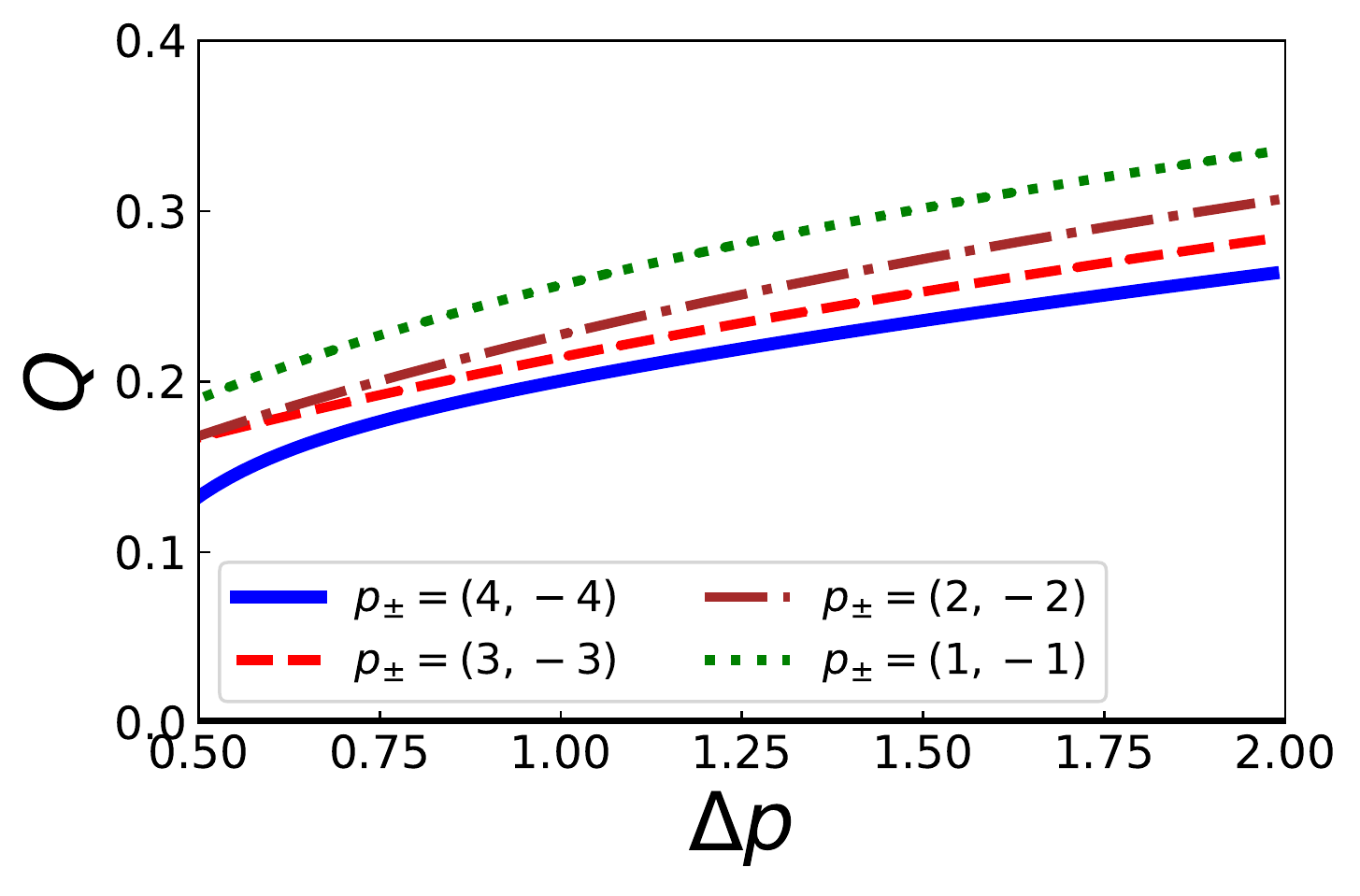} &
\includegraphics[scale=0.5]{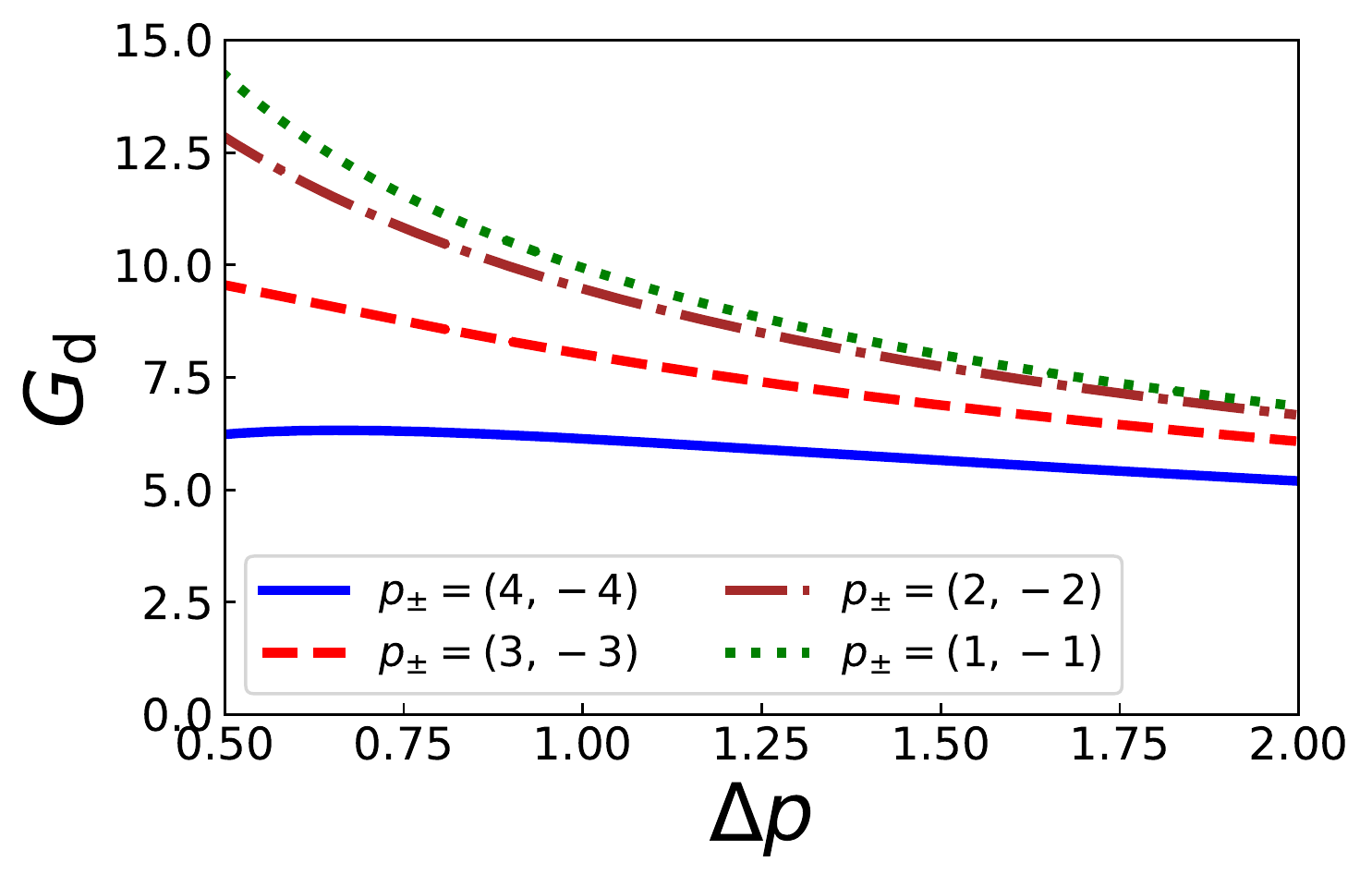} \\ (E) The Lighthill condition as a function of temperature and separation of DF.  & (F)  GVD as a function of temperature and separation of DF.   \\ \\ \\
\end{tabular}
\caption{The figure shows temperature dependence of NLSE coefficients
  for separated Lorentzian DF as defined in equation
  (\ref{Lorentz}). \textbf{Top:} Panel (A) shows a particular example
  of separated electron and positron DF along with the group velocity
  pole $p_\mathrm{gr}$ as defined in Eq. (\ref{pole}) for a
  particular temperature. Panel (B) shows the variation of
  $p_\mathrm{gr}$, $s/q$ and $q_\mathrm{d}/G_\mathrm{d}$ as a function
  of temperature at the same separation of DF as shown in Panel
  (A). \textbf{Middle:} Panel (C) shows that $p_\mathrm{gr}$ at a given
  plasma temperature decreases as the separation of the DF
  increases. Panel (D) shows that the ratio $s/q$ remains tightly
  clustered to values within $0.1$ of zero for moderate separation of
  DF.  Only at sufficiently high separation of DF can the $s/q$ ratio
  increase to values higher than $0.5$. \textbf{Lower:} Panel (E)
  shows that $Q$ stays within the range $(0.1,\,0.3)$ for the
  range of plasma temperatures considered; in particular, 
  the Lighthill condition (\ref{lighthill condn}) is satisfied. Panel (F) shows that the group
  velocity dispersion at any temperature decreases with increasing
  separation.}
\label{Sep Lorentz}
\end{figure*}

\begin{figure*}
\begin{tabular}{c}
\includegraphics[scale = 0.27]{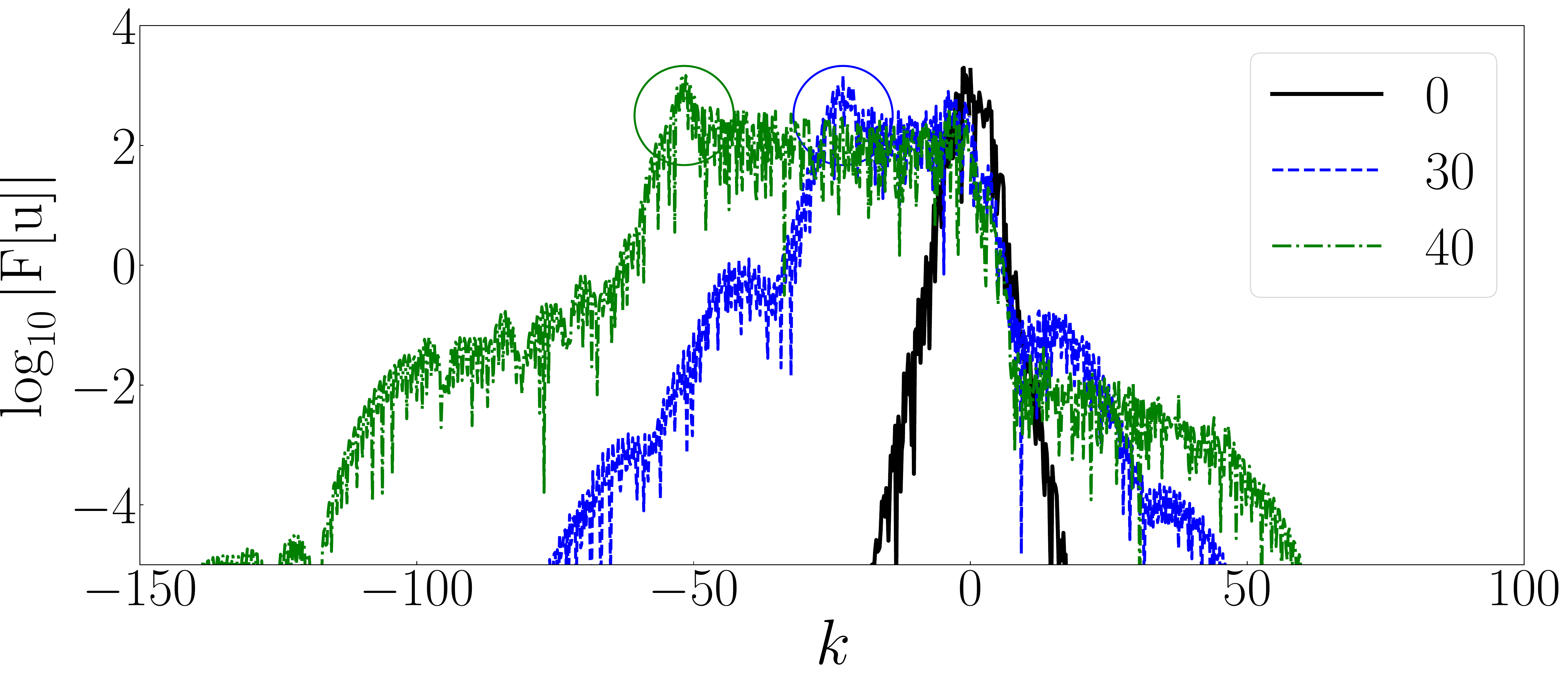}\\
(A) Electric field evolution in the Fourier space\\  
\includegraphics[scale = 0.27]{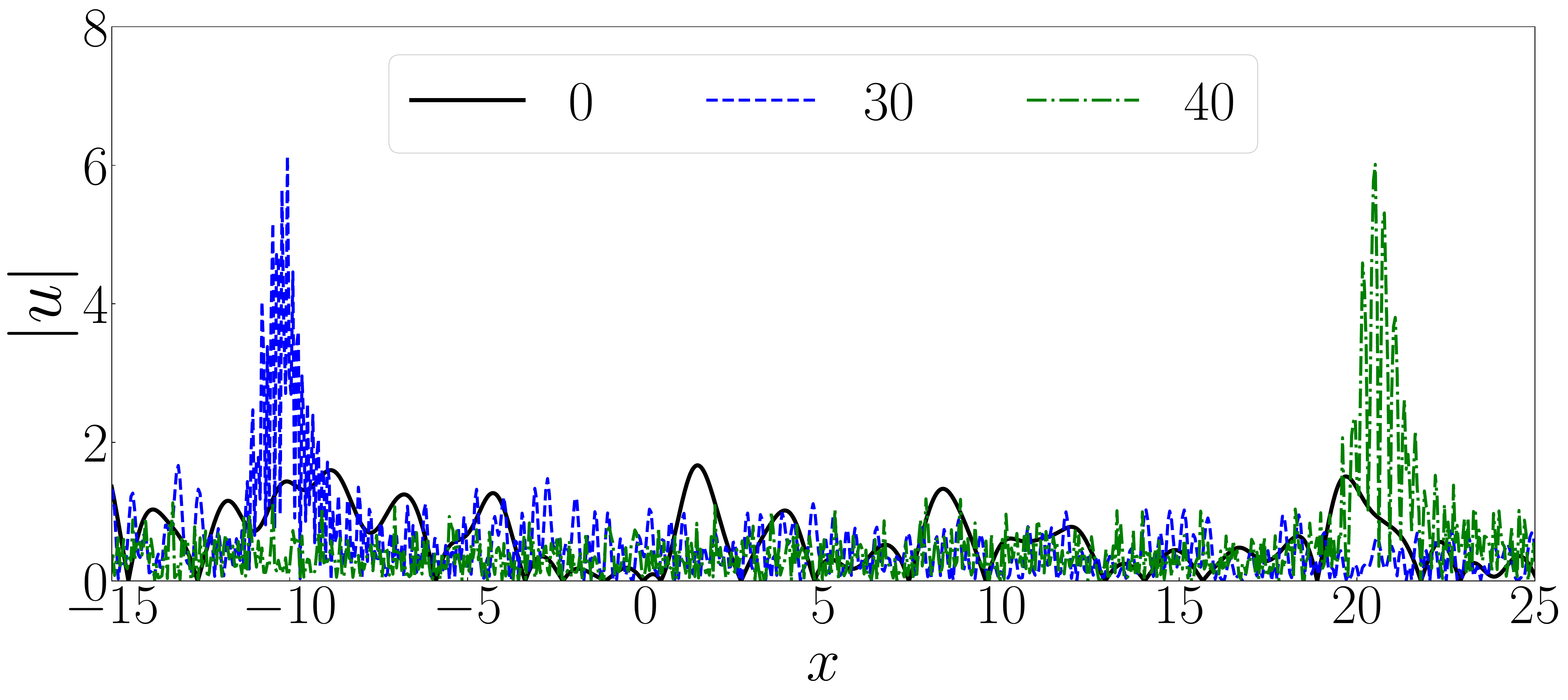} \\ 
(B) Electric field evolution in the configuration space (for a selected range).\\ 
\includegraphics[scale=0.26]{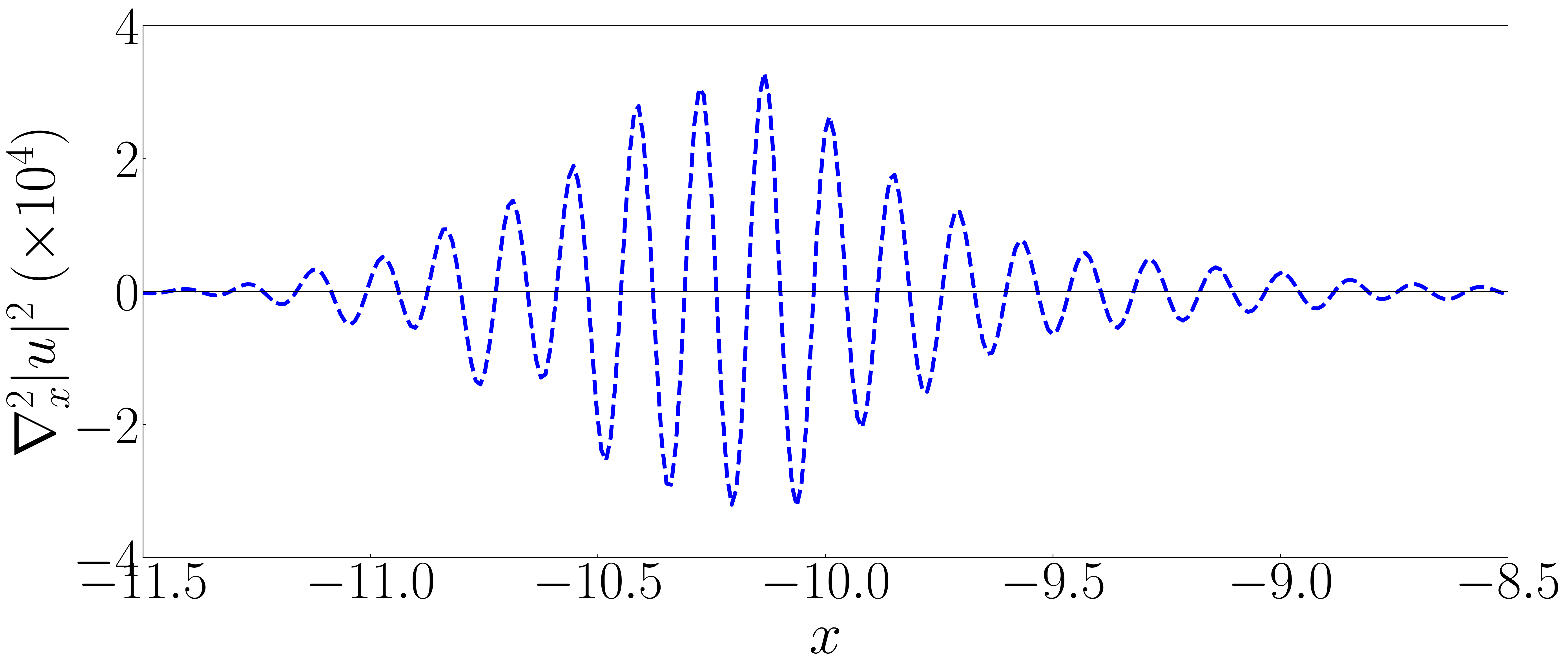}\\
(C) The Miller force associated with soliton electric field at $t = 30$.
\end{tabular}
\caption{Simulation results of soliton formation for Lorentzian DF (
  $Q = 0.25$ and $s/q = 0.1$) by the numerical method described in
  \citet{Lakoba2016}. Panel (A) shows the movement of a prominent
  secondary peak (circled) to $k<0$ at few representative times. Panel
  (B) shows the corresponding soliton formation in configuration space
  in the limited range $(-15,25)$ for clarity. The actual simulation
  box has the range $(-60,60)$. The legends in both panels indicate
  the dimensionless time $t$. The black curve in both panels shows the
  initial condition at $t = 0$. Panel (C) shows the Miller force
  associated with the soliton electric field at $t$ = 30. Here
  `soliton' is the envelope of the pulse with $ \Delta x \sim $ 3 units
  while ripple is what appears to be a ``carrier wave'' with
  wavelength $ \delta x_\mathrm{ripple} \sim $ 0.15 units.   }
\label{soliton_Lorentz1}
\end{figure*}

\begin{figure*}
\begin{tabular}{c}
\includegraphics[scale = 0.27]{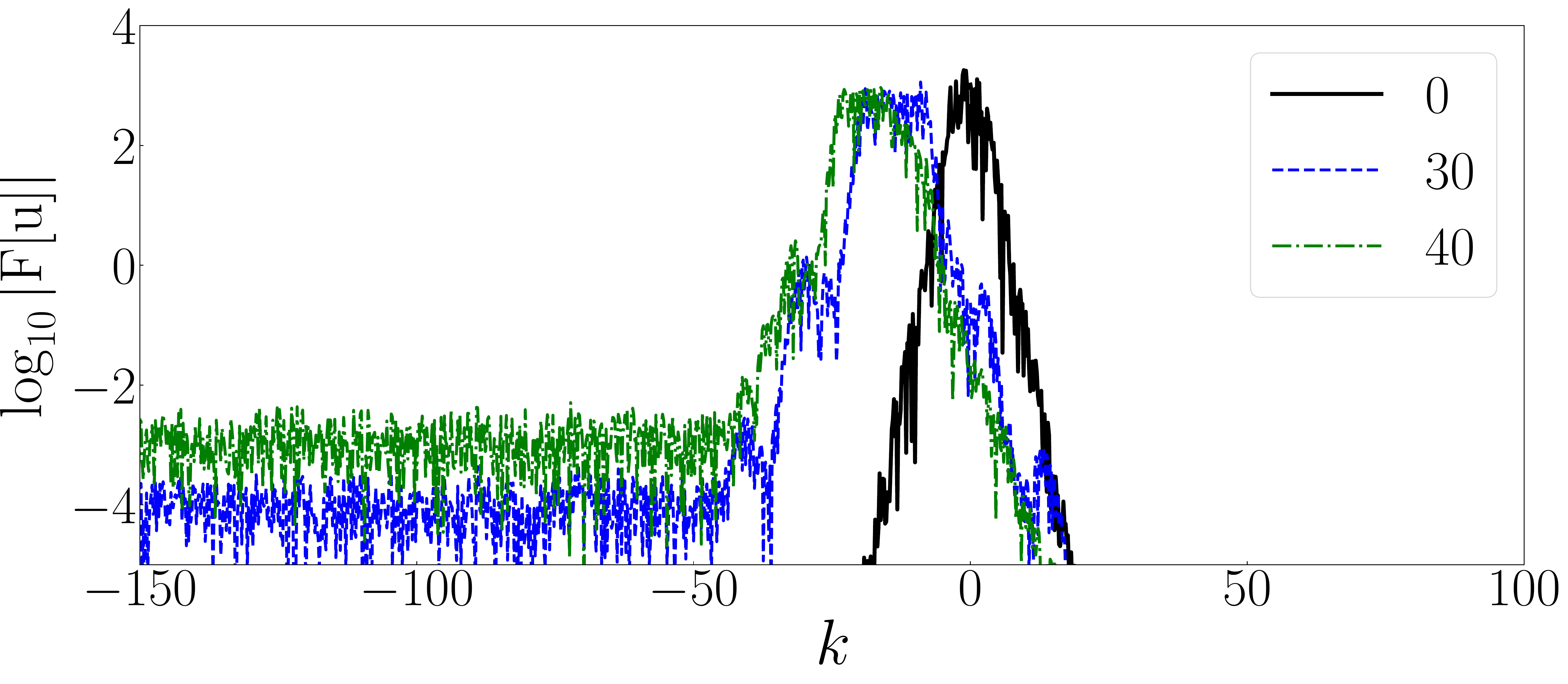}\\
(A) Electric field evolution in the Fourier space.\\ 
\includegraphics[scale = 0.27]{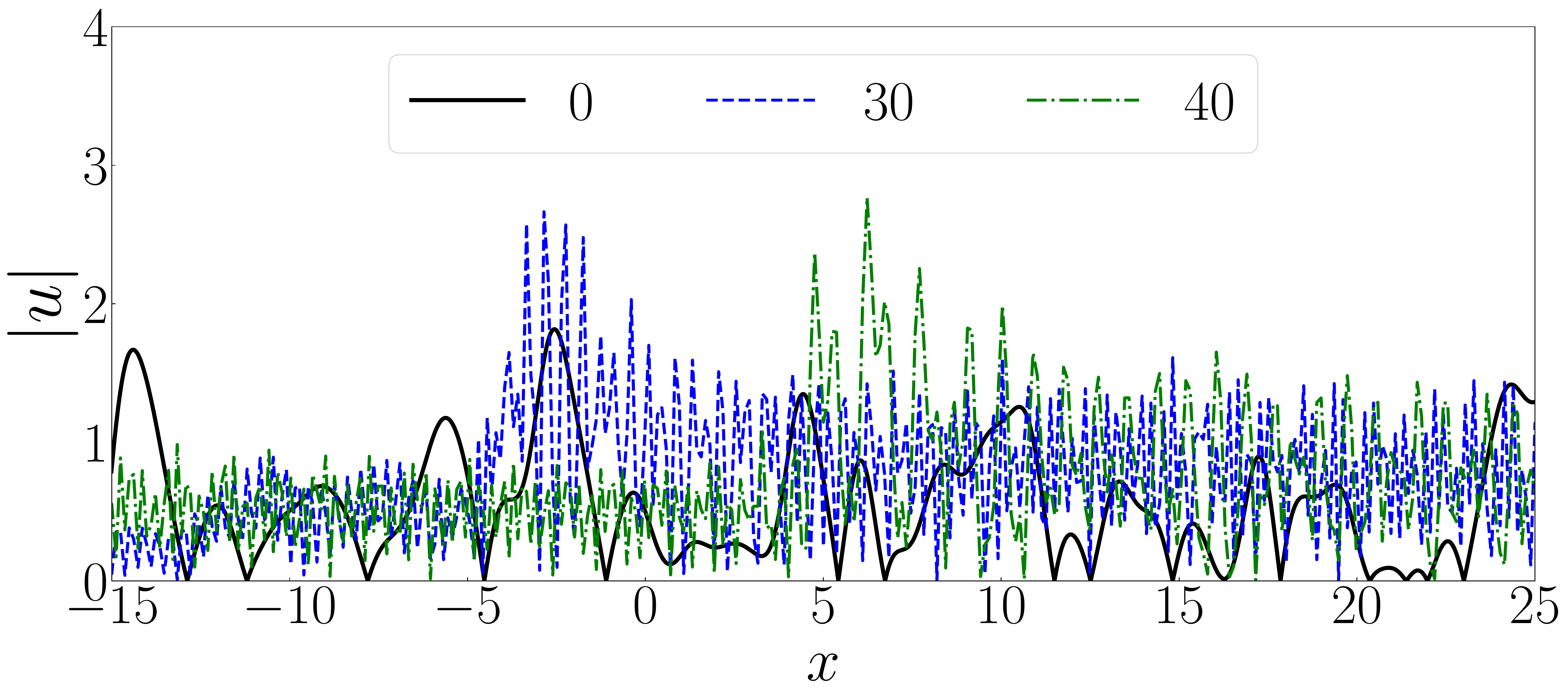} \\ 
(B) Electric field evolution in the configuration space (for a selected range). \\ 
\end{tabular}
\caption{Simulation of pulse evolution for for Lorentzian DF ( $Q =
  0.25$ and $s/q = 0.5$) by the the numerical method described in
  \citet{Lakoba2016}. The black curve in both panels shows the initial
  condition at $t = 0$. Panel (A) shows the absence of a prominent
  secondary peak as was seen in Fig. \ref{soliton_Lorentz1}. Panel (B)
  shows the corresponding wave field evolution in the configuration
  space in the limited range $(-15,25)$ for clarity. The actual
  spatial range of the simulation box is $(-60,60)$. It can be seen
  that the amplitude of the envelope of the pulses at any time does
  not exceed the amplitude of the initial wave electric field. Thus,
  soliton formation is suppressed for $s/q \gtrsim 0.5$.}
\label{soliton_Lorentz2}
\end{figure*}

Let us use the Lorentzian DF to get representative values of the
ratios $s/q$, $q_\mathrm{d}/G_\mathrm{d}$ and the dimensionless group
velocity dispersion $G_\mathrm{d}$. After obtaining these
representative values, we will explore soliton formation numerically
following the method of \cite{Lakoba2016} and LMM18.

The Lorentzian DF for the $\alpha-$th species is given by
\begin{equation}
    f^{(0)}_{\alpha} = \frac{1}{\pi}  \frac{ \frac{1}{2} \; \Delta p }{ \left(p - {p}_{\alpha} \right)^2 + \left(\frac{1}{2} \; \Delta p \right)^2},  \label{Lorentz} 
\end{equation}
where $\Delta p$ is the width of the DF and ${p}_{\alpha}$ is the peak
of the DF. Here $\alpha = \pm$ refers to the positron and the electron
DF, respectively. In this subsection we refer to $\Delta p$ of
Eq. (\ref{Lorentz}), 
which characterizes the spread of particles' momenta,
as ``temperature''. 
In relativistic hot plasma, this spread of the  momenta is assumed to be a significant fraction of the mean momentum. Contrarily, in a cold plasma, the spread is small.
Next, in this study we assume the peak momentum of the particle DF to vary in the range 1 to 3. Thus, to keep the ratio of the width to the peak in the DF to vary from the cold limit to hot limit at all values of peak separation, 
$p_+ - p_-$, the plasma
temperature $\Delta p$ is chosen to be in the range $(0.5,2.0)$ in this study. In dimensional units, this temperature range corresponds to $5 \times 10^{9}$ K to $2 \times 10^{10}$ K.

We evaluate the NLSE coefficients at different separations of the DF
as a function of the plasma temperature
using Eqs. B6, B9 and B12 of Appendix B.
The results are shown in
Fig. \ref{Sep Lorentz} and can be understood physically as
follows. Panel (A) shows for a given separation of the DFs, the pole
due to group velocity $p_\mathrm{gr}$ is at the tail of DF. The upper
sub-panel of Panel (B) shows that the pole $p_\mathrm{gr}$ shifts to
higher values as the temperature of the plasma is increased. Thus, the
number of interacting particles at the group velocity decreases as the
temperature is increased. It is reflected in the lower sub-panel of
Panel (B), which shows that the magnitude of $s/q$ decreases with
increasing plasma temperature. Next, we explore the location of the
pole due to group velocity $p_\mathrm{gr}$ for different separation of
DF. Panel (C) shows that at a given plasma temperature (say $\Delta p
= 1.0$) the pole $p_\mathrm{gr}$ shifts to lower values as the
separation of the DFs increases. It means that with increasing
separation of DF, the pole shifts towards the center of the DF,
thereby increasing the number of plasma particles interacting with the
Langmuir waves, thereby increasing the effect of the nonlinear Landau damping relative to the instantaneous cubic nonlinearity.
Consequently, panel
(D) shows that for moderate separation values,
the magnitude of $s/q$ is clustered within $\sim 0.1$ from zero for a
range of plasma temperatures. However, for larger separation of DF,
the magnitude of $s/q$ 
increases to about
0.5 or even higher, especially for
colder plasma.  Finally, panel (E) shows that the quantity $Q$ is on
the order of $\sim 0.25$ for all separations of the DF across the
range of plasma temperatures. Thus,
we take $s/q=0.1$ and $0.5$ for small/moderate
and for larger DF separation, respectively.
The value of $Q$ can be taken to
have a constant value of $0.25$.

Simulation results for ($Q = 0.25, s/q = 0.1$) 
and the initial condition \eqref{initial}
are shown in
Fig. \ref{soliton_Lorentz1}. 
Following LMM18, we used a representative
value $k_\mathrm{corr}=2$.
In Panel (A) soliton formation can be
clearly identified with the movement of a well-formed secondary
spectral peak from $k=0$ to $k<0$. This peak in the Fourier spectrum
corresponds to a soliton in physical space (LMM18), seen in Panel (B). Panel (C) shows the Miller force associated with the envelope
soliton.

The following remark about identifying 
soliton formation from the field's spectrum
needs to be made. In panel (A) one sees
that the amplitudes of the secondary peak,
corresponding to the soliton in the physical space, and of the spectrum of
the initial field are about the same. 
Yet, the amplitude of the soliton in the 
physical space (panel (B)) is several
times greater than that of the initial
field. Thus, this amplitude increase must
occur via increased coherence of the field
``inside" the secondary spectral peak
compared to the initial fully random field. As was noted in LMM18, this formation
of high-amplitude solitons out of an initial
disordered state is a generic feature that occurs in many (but not all) so-called ``near-integrable but not exactly integrable" nonlinear wave models. (Here, the case $s/q=0$ is that
of the integrable NLSE with purely local cubic nonlinear term; for it, formation of a long-living soliton out of a disordered state will {\em not} occur.) The specific contributions of this study, and earlier of 
LMM18, was to show that this soliton formation does indeed occur for the NLSE 
with a sufficiently small nonlinear Landau damping term,
{\em and} that it occurs within the time
$t_{\max}$ that corresponds to realistic parameters in pulsar plasma.
(In contrast, soliton formation in another model, considered in \citet{JORDAN2001433}, occurred over a
time of many tens of thousands dimensionless units.)

The simulation results for ($Q = 0.25, s/q = 0.5$) are shown in
Fig. \ref{soliton_Lorentz2}. Unlike in Fig. \ref{soliton_Lorentz1}, here no spectral peak is seen to form in Panel (A), and, instead, 
energy gets more uniformly distributed among spectral components of the field. One can
interpret this as the field becoming less coherent for those larger values of $s/q$.
In physical space (Panel (B)), this is manifested by the absence of well-localized, 
long-living and high-amplitude bunches of electric field. It must
be noted that the behaviours, shown in Figs. \ref{soliton_Lorentz1} and \ref{soliton_Lorentz2}, at small and large $s/q$ were found in
LMM18, whereas here we demonstrated that they can actually occur in pulsar plasma.

Next, 
since our assumptions at the beginning of this section
about the strength of the electric field (i.e., parameter
$Q$) and the measure of disorder of the initial field
(i.e., $k_\mathrm{corr}$ in \eqref{initial}) are somewhat
arbitrary, below
we explore the effect of 
these parameters
on soliton formation. 
The simulation setup and technical details of the results are described in Appendix D; here we present only their
gist. 
First, we found that the effect of decreasing $k_\mathrm{corr}$ from 2 to 1 led only to the decrease of
the soliton formation time, in accordance with the statement
at the beginning of this Section; no statistically significant changes were found in the distribution of the amplitude of the long-living solitons that formed. 
Second, we doubled the initial amplitude of $u(x,0)$, 
which is tantamount to quadrupling $Q$. In this case, 
the final amplitude  of the formed solitons was, 
on average, lower than for the original $u(x,0)$;
however, qualitatively, the distribution of the final
soliton amplitudes remained similar to the original case.
(We also found that, in agreement to the statement at the
beginning of this Section, the soliton formation time decreased approximately four-fold.)

To summarize, soliton formation for long-tailed DF can occur for a wide range of plasma temperature for moderate separations of the
electron-positron DF.  Large separation of the DF increases the value
of $s/q$, which necessarily leads to suppression of soliton
formation
via the mechanism explained in our discussion about Fig.~\ref{Sep Lorentz}

\subsection{Gaussian DF}

\begin{figure*}[h]
\begin{tabular}{cc}
\includegraphics[scale=0.5]{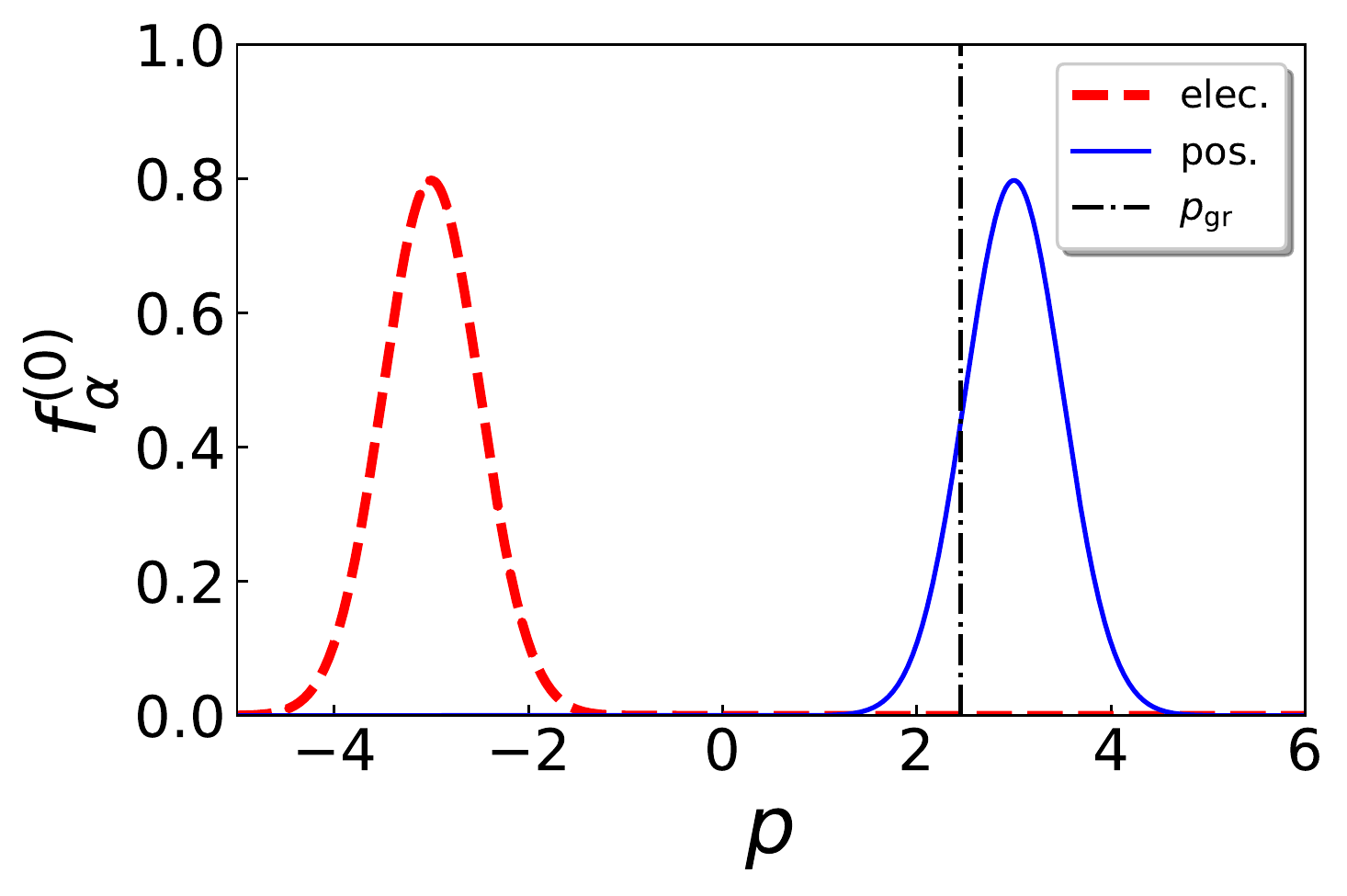} & \includegraphics[scale=0.5]{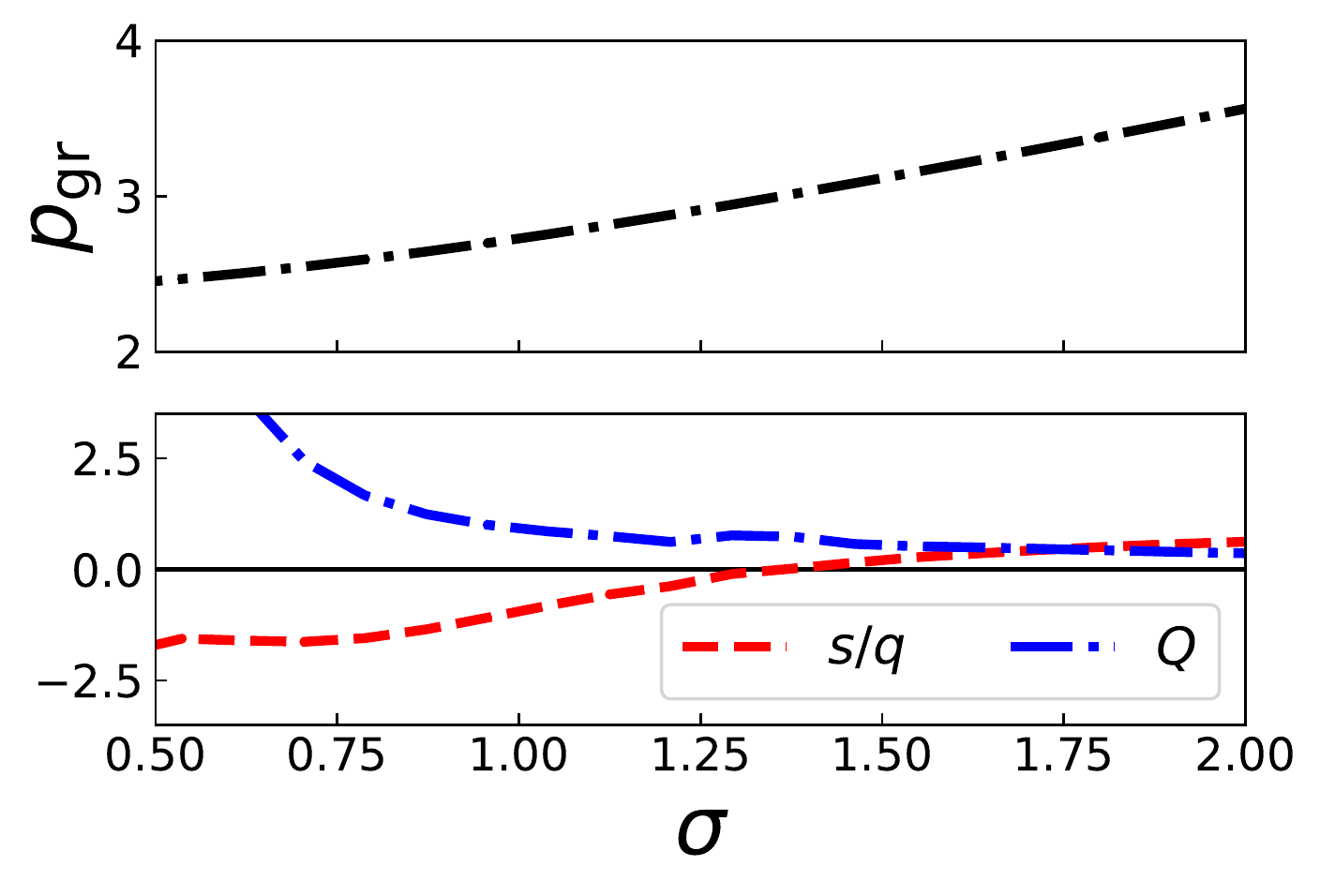} \\
(A) Gaussian DF and group velocity pole with ${p}_{\pm}=\pm 3 $  at $\sigma = 0.5$.   & (B)  Parameter space for Panel (A) as a function of temperature. \\ \\ \\ 
\includegraphics[scale=0.5]{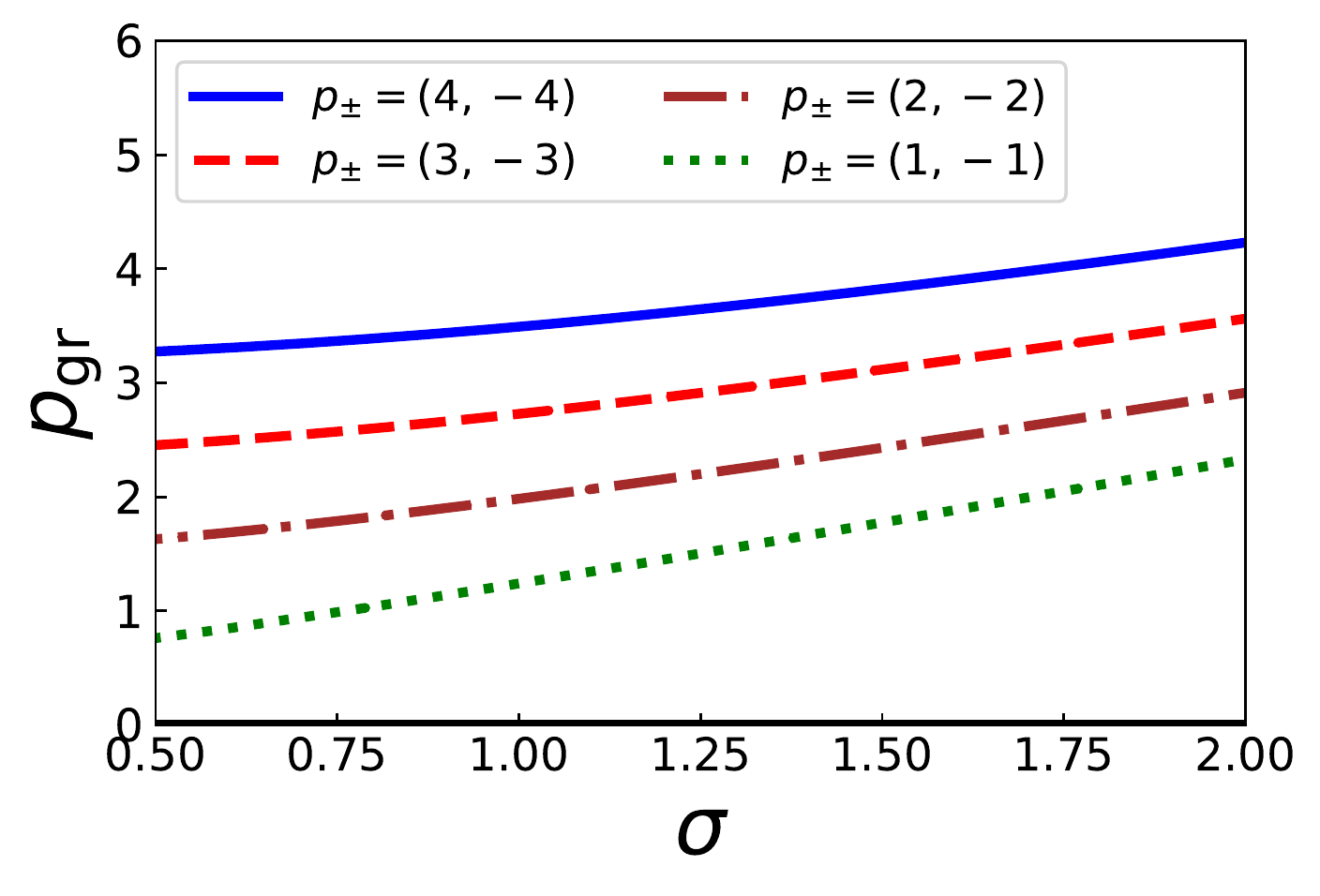} & \includegraphics[scale=0.5]{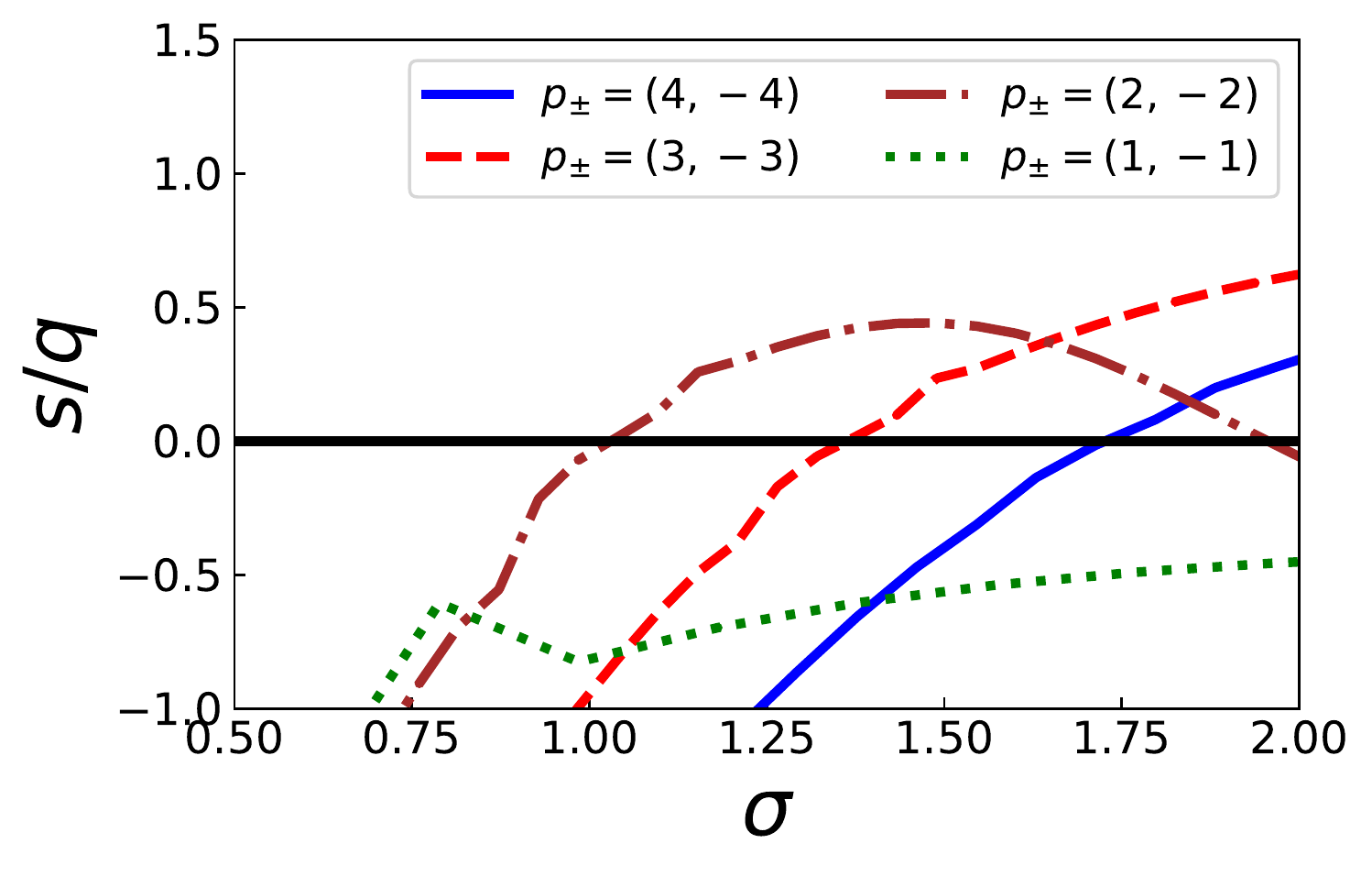} \\
(C) The group velocity pole with temperature and DF separation.   & (D) Ratio of NLD to CNL as a function of temperature and separation of DF. \\ \\ \\ 
\includegraphics[scale=0.5]{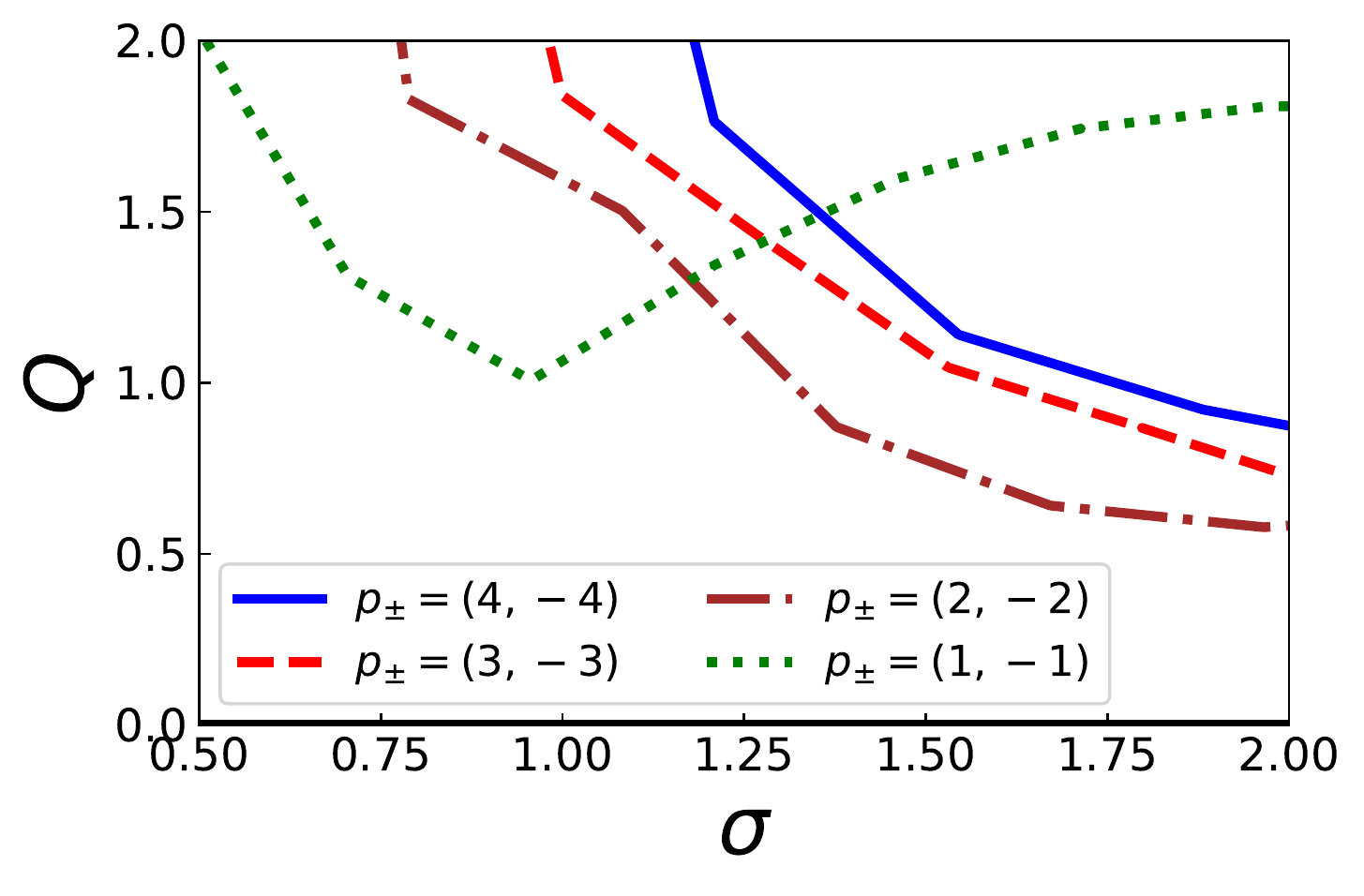} & \includegraphics[scale=0.5]{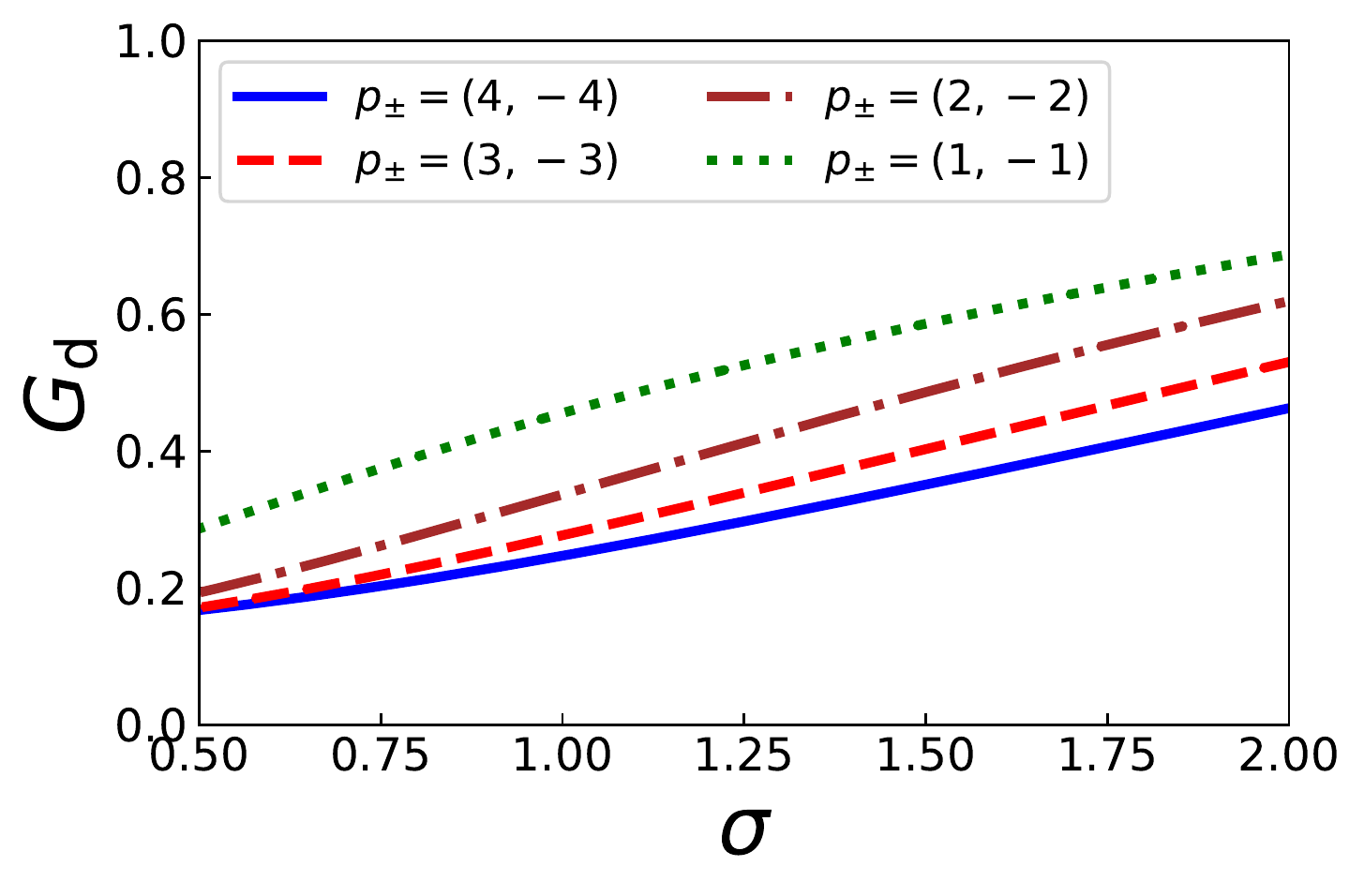} \\
(E) Lighthill condition as a function of temperature and separation of DF.   & (F)  GVD as a function of temperature and separation of DF. \\ \\ \\ 
\end{tabular}
\caption{The figure shows temperature dependence of NLSE coefficients
  for separated Gaussian DF as defined in equation
  (\ref{gauss}). \textbf{Top:} Panel (A) shows a particular example of
  separated electron and positron DF along with the location of the
  group velocity at the same temperature. Panel (B) shows the
  variation of $p_\mathrm{gr}$, $s/q$ and $Q $ as a function of
  temperature at the same separation of DF as shown in Panel
  (A). \textbf{Middle:} Panel (C) shows that the $p_\mathrm{gr}$ for a given
  plasma temperature increases as the separation of the DF
  increases. Panel (D) shows that the ratio $s/q$ remains large for
  all separation of DF. The moderate values of $s/q$ are available
  only near a certain temperature $\sigma_\mathrm{sp}$ where $s/q$
  changes sign. The value of $\sigma_\mathrm{sp}$ decreases with
  decreasing separation of the DF, until below some separation it
  vanishes and the magnitude of $s/q$ ratio settles at around
  0.5. \textbf{Lower:} Panel (E) shows that the Lighthill condition
  (\ref{lighthill condn}) is satisfied across the range of plasma temperatures. Panel
  (F) shows that the group velocity dispersion remains
  clustered around 0.4 for all separation of DF across a wide range of
  plasma temperatures.}
\label{Sep gauss}
\end{figure*}

The Gaussian DF for $\alpha-$th species is given by
\begin{equation}
    f^{(0)}_{\alpha} = \frac{1}{\sqrt{2 \pi} \sigma } \exp{ \Bigg \{ -\frac{(p - {p}_{\alpha})^2}{2 \sigma^2 } \Bigg \} }  , \label{gauss} 
\end{equation}
where $\sigma$ is the width of the DF and $\Bar{p}_{\alpha}$ is the peak of the DF. In this subsection, we will refer to $\sigma$ as the plasma ``temperature''. Like in the previous subsection, in our study the plasma temperature $\sigma$ is restricted to the range $(0.5,2.0)$.

Similar to the previous subsection, we evaluate the NLSE coefficients
for different separations of the Gaussian DF as a function of the
plasma temperature $\sigma$. The results are shown in Fig. \ref{Sep
  gauss} and can be understood physically as follows. Panel (A) shows
that for a given separation of the DF, the pole due to group velocity
$p_\mathrm{gr}$ is near the center of the positron DF. The upper
sub-panel of Panel (B) shows that while pole $p_\mathrm{gr}$ shifts to
higher values with increasing temperature, it still lies very close to
the peak of the positron DF. Thus, the number of particles that can
interact at the group velocity 
of linear Langmuir waves
remains high. This is reflected in the
lower sub-panel of Panel (B), which shows that the magnitude of $s/q$
generally remains high across a range of $\sigma$. Next, we explore
the location of $p_\mathrm{gr}$ for different separations of DF. Panel
(C) shows that for all temperature values considered, the pole $p_\mathrm{gr}$
remains close to the center of the positron DF. Panel (D) shows that
small values of $s/q$ can be obtained only in a very narrow range of $\sigma$ where the quantity $s/q$ changes
sign. The location of this temperature range varies with the DF separation and, in fact, for sufficiently small separation, there is no temperature (in the range considered here) where $s/q$ would be as small as $0.1$. Namely, for $p_{\pm} \sim \pm 1$, one has $s/q \sim -0.5$. 

As shown in
Fig. \ref{soliton_Lorentz2} and in LMM18, higher values of $s/q
\gtrsim 0.5$, observed for most temperature values in the above range,
lead to suppression of soliton formation.

To summarize, the Gaussian DF provides small values of $s/q \lesssim
0.1$ only in a narrow interval of temperatures 
and for moderate separation of DF. As the DF separation decreases, the
interval where $s/q$ remains small, shrinks and eventually vanishes, and the
ratio stays too high: $s/q \approx - 0.5$, for solitons to form. This leads us to conclude that soliton formation for
short-tailed DF can occur only in a very restrictive parameter
regime. As a result, short-tailed DF seems to be unlikely candidates
for sustaining soliton formation under generic hot plasma conditions.

\subsection{Dependence of soliton formation on sign of $s/q$}\label{sign_sq}

\begin{figure*}[h]
    \centering
    \includegraphics[width=\columnwidth]{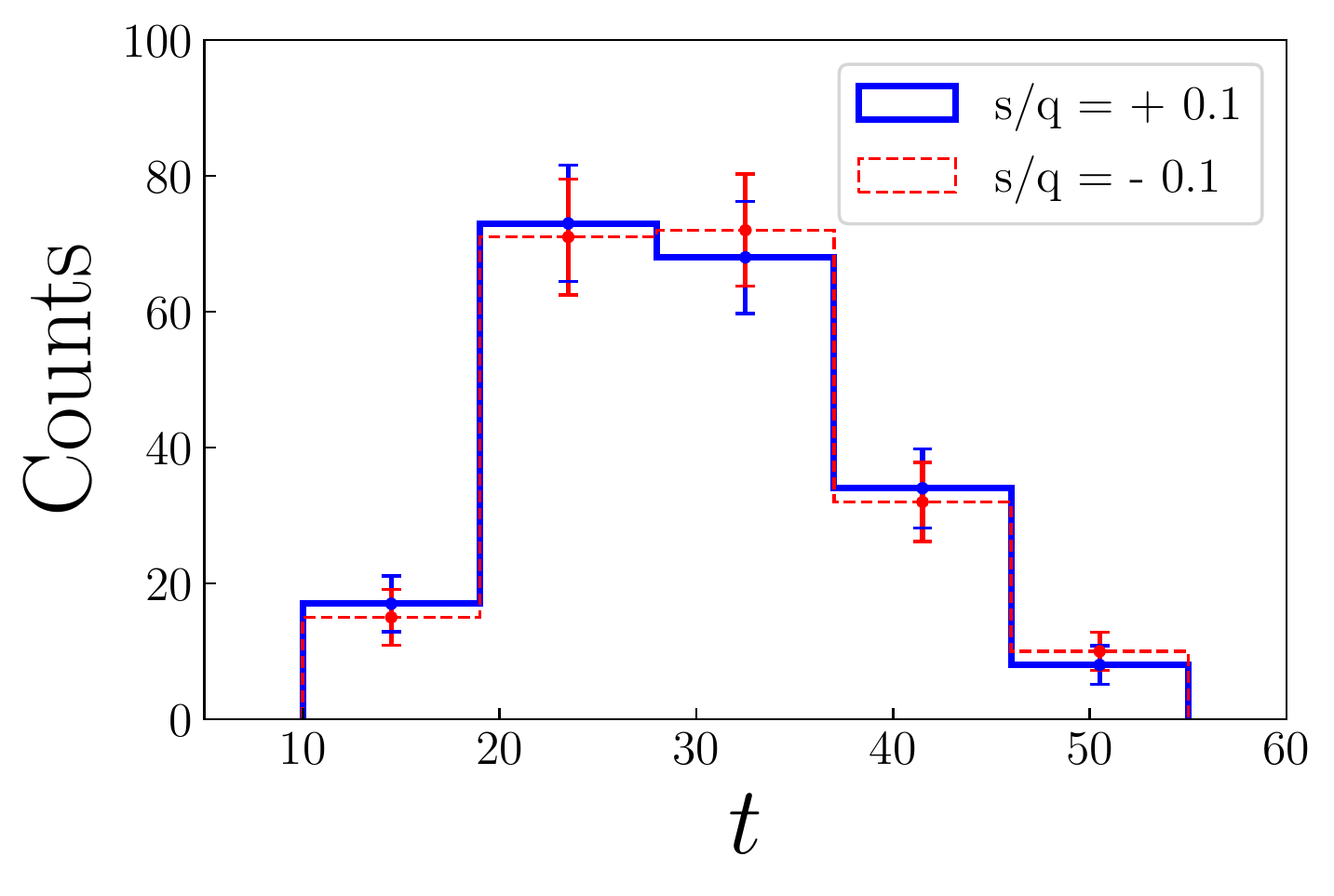}
    \caption{The figure shows the histogram for the earliest soliton
      formation time $t$ (with $Q = 0.25$) which satisfies the
      constraint (\ref{constraint}) for 200 random seeds in equation
      (\ref{white_noise}) for both positive $s/q = + 0.1$ (shown in
      solid blue) and negative $s/q = -0.1$ (shown in dashed red). The
      histogram for $t$ is divided into 5 bins in the range $(10 ,
      55)$ while the error bars $= \sqrt{N}$ where $N$ is the number
      of entries in each bin. It can be seen that the the average time
      for soliton formation is $\sim 30$ units for both signs of
      $s/q$.}
    \label{histogram}
\end{figure*}

It can be seen from Panel (D) of Fig. \ref{Sep Lorentz} and
Fig. \ref{Sep gauss} that the ratio $s/q$ can be both positive and
negative. Physically, the sign of $s/q$ only determines the direction
of the movement of the secondary peak associated with soliton
formation to a higher wave number in the Fourier space (LMM18). For
negative $s/q$, the secondary peak in the Fourier space moves to $k>0$
and vice versa. Physically, there is no difference as the presence of a
secondary peak for both $k > 0$ and $k<0$ gives rise to envelope
solitons in the configuration space. The soliton formation timescale
is not affected in a statistical sense. To show that this is indeed
the case, we simulate soliton formation for $Q=0.25$ for $s/q = 0.1$
and $s/q = -0.1$ for 200 random seed values for the white noise in
Eq. (\ref{white_noise}). Figure \ref{histogram} shows the
histogram for the time $t$ of soliton formation, defined as
\begin{equation}
    \mathrm{max}|u(x,t>0)| \geq 3 \times \mathrm{max}|u(x,0)|.  \label{constraint}
\end{equation}
It can be seen that the statistics of soliton formation times indeed does not depend on the sign of $s/q$.

\subsection{Role of ions in modifying the coefficients of NLSE}

The DF of ions are expected to be near the electron and positron
DF. We treat the location of the ion DF as a free parameter wherein
the maximum contribution to the NLSE coefficients due to ions can only
come if the center of ion DF is near the pole $p_\mathrm{gr}$. The
setup for maximizing the contribution to NLSE coefficients due to ions
is described in Appendix B4. We find that the presence of ions modify
the dimensionless coefficients of NLSE (i.e., $G_\mathrm{d},
q_\mathrm{d}, s_\mathrm{d}$) by less than $10^{-8}$. The result can be understood qualitatively as follows. 
It must be noted in the PSG model (\citealt{2003A&A...407..315G}) the number density of ions is close to 90$\%$ of the Goldreich-Julian co-rotational number density. As defined in the Introduction, $\kappa$ is the ratio of the number density of the pair plasma to the Goldreich-Julian number density. Thus, for simplicity, the ratio of the number density of pair plasma to the number density of ions can be assumed to be $\kappa$.  Then,
the number density of the ions is $\sim 10^4$ times smaller than that of the pair plasma while the mass of the ions is $\sim 10^{4}$ times higher than that of electrons and positrons. 
A combination of these two effects
reduces the contribution of ions to the coefficients of NLSE by the factor $10^{-8}$.
A more expanded discussion of these aspects will be presented in the following section.
We conclude that
ions make negligible contribution in modifying the coefficients of
NLSE.

\begin{figure*}[h]
\begin{tabular}{cc}
\includegraphics[scale=0.5]{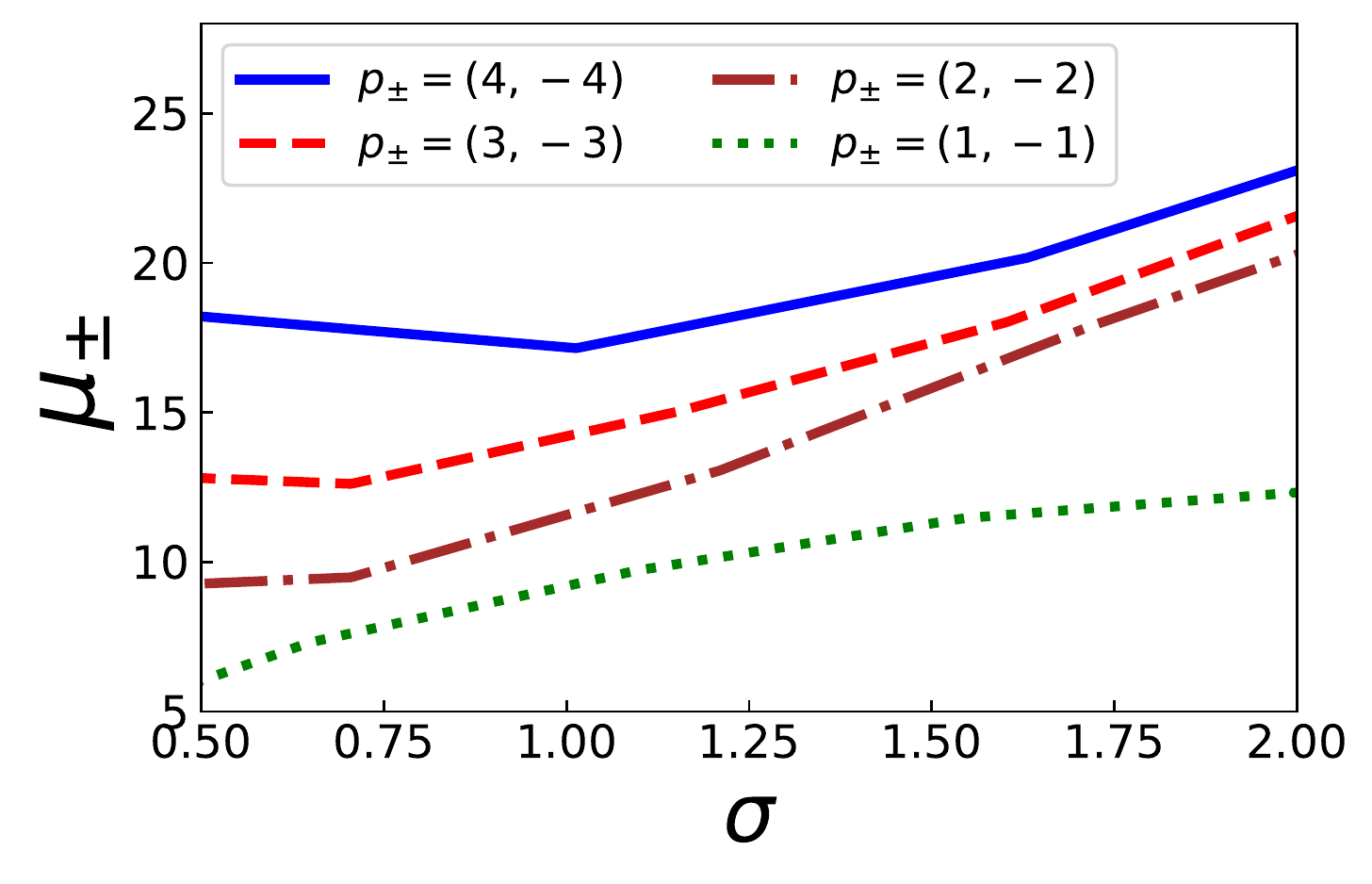} & \includegraphics[scale=0.5]{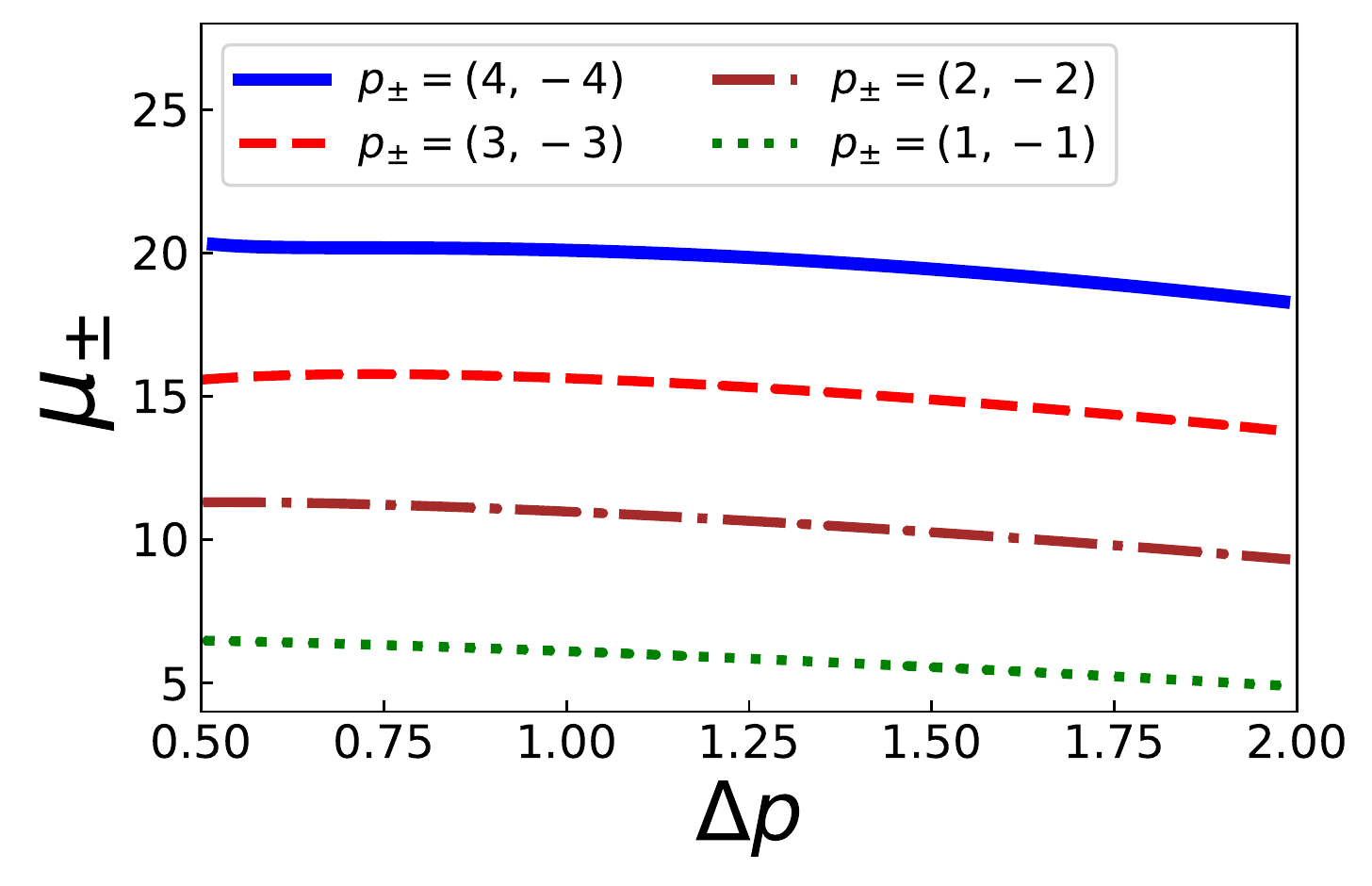} \\
(A) Contribution to $\mu$ from separated electron-positron Gaussian DF.   & (B) Contribution to $\mu$ from separated electron-positron Lorentzian DF.  \\ \\ \\ 
\includegraphics[scale=0.5]{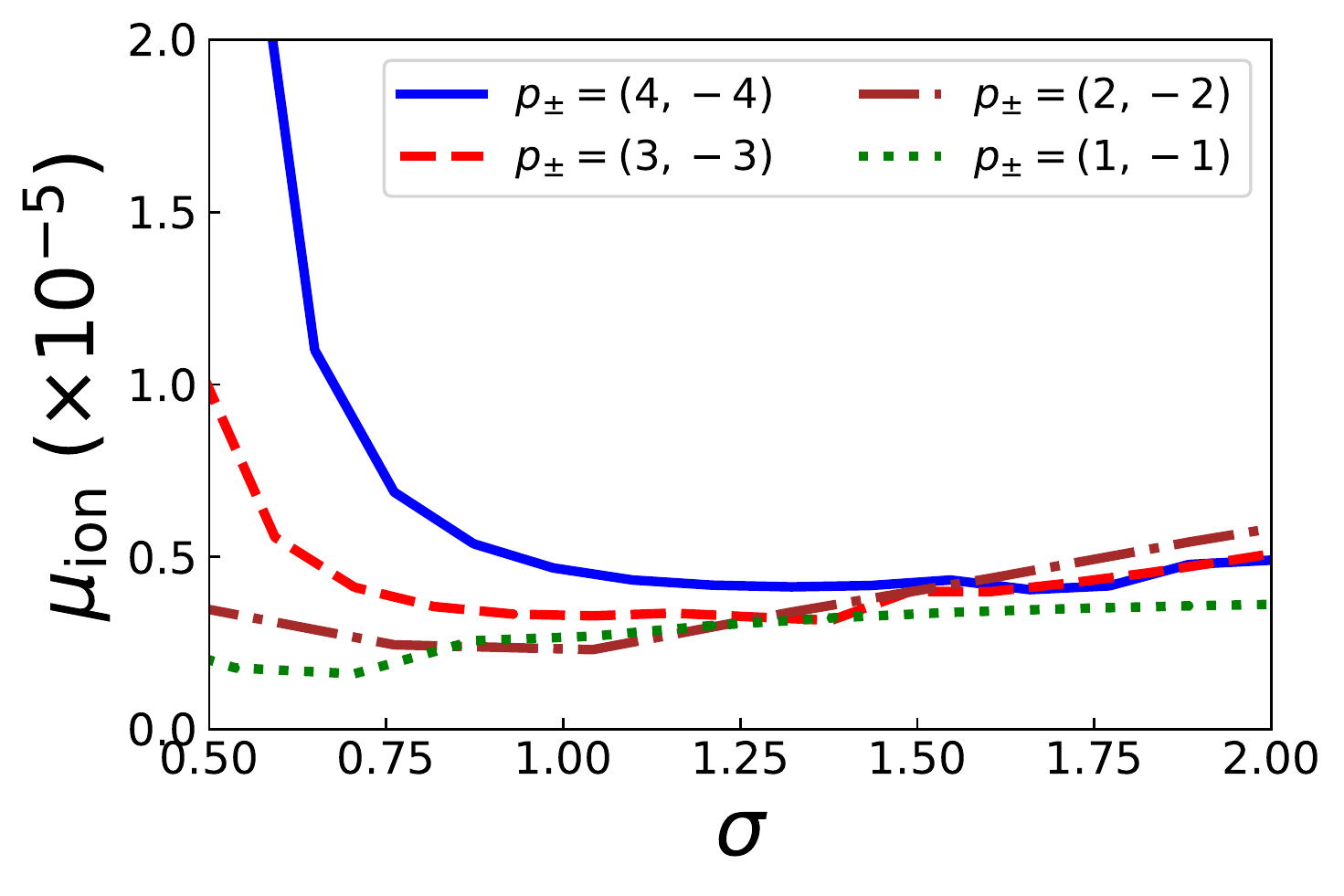} & \includegraphics[scale=0.5]{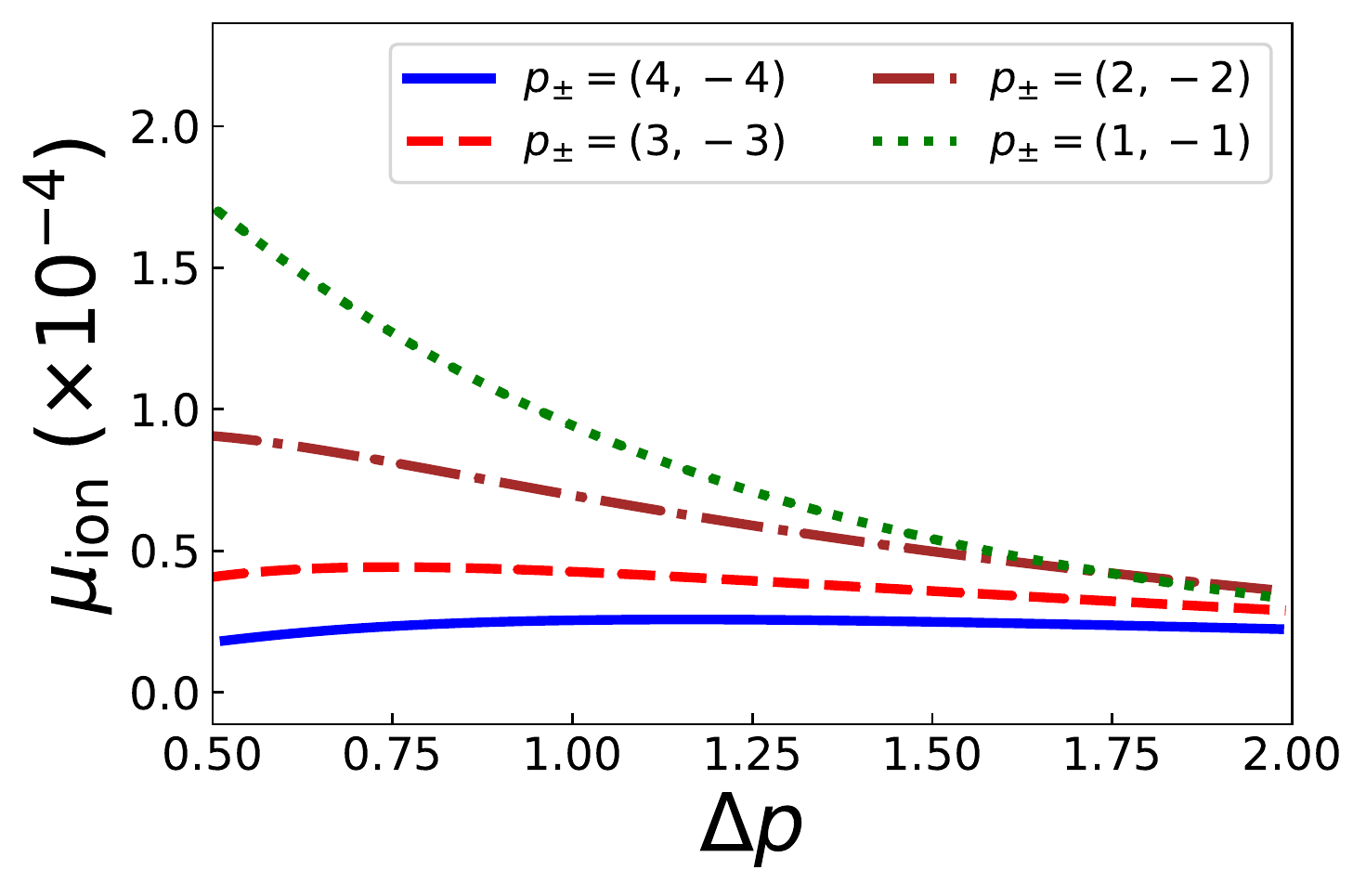} \\
(C) Contribution to $\mu$ for Gaussian ion DF. & (D) Contribution to $\mu$ for Lorentzian ion DF.  \\  \\ \\ 
\end{tabular}
\caption{Variation of the charge separation integral $\mu$ as defined in Eq. (\ref{slow_charge_integ}) for different separation of the DF. Panel (A) and Panel (B) shows that the integral as a function of the plasma temperature for Gaussian and Lorentzian DF respectively. Panel (C) and Panel (D) shows the contribution to the integral due to the presence of an iron ion component. Note the vertical axis scale in Panels (C) and (D) are $10^{-5}$ and $10^{-4}$, respectively.}
\label{mu_var}
\end{figure*} 

\section{Typical properties associated with Langmuir solitons}\label{charge_sep_section}

The typical properties of Langmuir solitons such as spatial extent, structure and charge are crucial in determining if these solitons can be a suitable candidate for the observed coherent radio emission in pulsars. In this section we briefly discuss these aspects.

\subsection{Typical length and ripple associated with the solitons}

In this subsection, following LMM18 we estimate the typical size of the soliton and the ripple
associated with it (in dimensional units) at a distance of $r = 500$
km above the neutron star surface. Using Eq. (\ref{wave number})
and Eq. (\ref{plasma_freq}), the typical Langmuir length scale
$l$ in PFR is given by
\begin{equation}
    l = \frac{2\pi}{k} = \frac{\pi\,c}{ \omega_\mathrm{p} \sqrt{\gamma} } \approx 6\times 10^2 \left( \frac{3}{ \gamma} \right)^{1/2} \text{cm}.
\end{equation}
where $\gamma = \int dp \;  \sqrt{1 + p^2} f^{(0)}_{\alpha} $ is the average Lorentz factor of the electron/positron DF of the pair plasma in PFR and depends on the temperature of the particles. For the cold plasma limit, $\gamma$ equals the Lorentz factor at the peak of the electron/positron DF,
whereas for hot plasma, $\gamma$ can be as much as
twice the Lorentz factor evaluated at DF's peak. For the rest of the analysis, we choose a representative value of $\gamma$ to be $3$.

From panel (C) of Fig. (\ref{soliton_Lorentz1}), the typical size of the soliton envelope $\Delta_\mathrm{PFR}$ in PFR is given as
\begin{equation}
\begin{split}
    & \Delta_\mathrm{PFR} = \frac{\Delta \xi'}{\gamma_\mathrm{gr}} = \frac{l \theta \Delta x}{\gamma_\mathrm{gr}} 
    \\ & \approx 3\times 10^{4}  \left( \frac{6}{\gamma_\mathrm{gr}} \right) \left(\frac{\theta}{100} \right) \left( \frac{\Delta x}{3 }\right) \left( \frac{3}{\gamma} \right)^{1/2} \; \text{cm}, 
\end{split}
\end{equation}
while the typical ripple size $\Delta_\mathrm{ripple,PFR}$ from panel (C) of Fig. \ref{soliton_Lorentz1} associated with the soliton in PFR is given as
\begin{equation}
\begin{split}
    & \Delta_\mathrm{ripple,PFR} = \frac{l \theta \delta x_\mathrm{ripple}}{\gamma_\mathrm{gr}} \\ & \approx 10^{3} \; \left( \frac{6}{\gamma_\mathrm{gr}} \right) \left(\frac{\theta}{200} \right) \left( \frac{\delta x_\mathrm{ripple}}{0.15}\right) \left( \frac{3}{ \gamma } \right)^{1/2} \; \text{cm}.
\end{split}
\end{equation}
Then, in OFR, the typical soliton size is  
\begin{equation}
\begin{split}
    & \Delta_\mathrm{OFR} = \frac{\Delta_\mathrm{PFR}}{\gamma_\mathrm{s}} \\ & \approx 10^2 \; \left( \frac{6}{\gamma_\mathrm{gr}} \right) \left(\frac{\theta}{100} \right) \left( \frac{\Delta x}{3 }\right) \left( \frac{3}{ \gamma} \right)^{1/2} \left( \frac{200}{\gamma_\mathrm{s}}\right) \; \text{cm},
\end{split}
\label{DsolOFR}
\end{equation}
while the ripple size is
\begin{equation}
   \begin{split}
       & \Delta_\mathrm{ripple,OFR} = \frac{\Delta_\mathrm{ripple,PFR}}{\gamma_\mathrm{s}} \\ & \approx  5 \; \left( \frac{6}{\gamma_\mathrm{gr}} \right) \left(\frac{\theta}{100 } \right) \left( \frac{\delta x_\mathrm{ripple}}{0.15}\right) \left( \frac{3}{ \gamma } \right)^{1/2} \left( \frac{200}{\gamma_\mathrm{s}}\right)  \; \text{cm}.
   \end{split}
\label{DripOFR}
\end{equation}
For the case considered in subsection \ref{LorentzDF} where solitons could form (i.e., $s/q=0.1$) and for representative parameters values considered there, 
the typical size for the envelope and the ripple
associated with solitons are about 100 cm and 5 cm, respectively.

 These spatial scales
 correspond to a frequency range from 300 MHz to  6 GHz,
 which spans the observed broad-band frequencies  of curvature radiation. 
 For a typical radius of curvature $r_\mathrm{c} \approx 10^{8}$ cm in the radio emission zone, the characteristic frequency of curvature radiation is $\nu_{c} \approx 3 \gamma^3_\mathrm{s} c / 4 \pi r_\mathrm{c} \approx$ 2 GHz,
 which indeed falls into the above range
 $(0.3,\,6)$ GHz.
  However, the calculation of an actual radiation pattern due to coherent curvature radiation by an {\em ensemble} of such rippled structures,
  as can be expected in pulsar plasma,
  is beyond the scope of this work and will be studied elsewhere.

It must also be noted that the temperature dependence of the size of the solitons is due to the average Lorentz factor $\gamma$ of the plasma particles and the Lorentz factor corresponding to the group velocity of the plasma waves $\gamma_\mathrm{gr}$. As mentioned earlier, $\gamma$ for a high-temperature plasma can be twice as large as the Lorentz factor associated with the peak of the electron/positron DF and enters as a square root dependence in the size estimates \eqref{DsolOFR} and \eqref{DripOFR}. As seen from Panel (C) of Fig. \ref{Sep Lorentz} and \ref{Sep gauss}, the group velocity changes only marginally within the range of temperature considered. Thus, an increase in temperature can decrease the estimates of the soliton size and ripple size by at most $\sim$ 30 $\%$.

It must also be mentioned that the number of ripples within the soliton can vary significantly. Figure \ref{histogram} shows that, for $(Q=0.25,s/q = 0.1, k_{\rm corr} = 2)$, the time of soliton formation (see  \eqref{constraint}) has a significant spread and depends on the particular realization of the random initial condition (\ref{initial}). As a result, location (in Fourier space) of the secondary spectral peak has a wide variation. In Appendix E, we show representative cases of the location of the peak in Fourier space and the Miller force associated with the solitons. We find that while the size of the solitons is roughly the same, the number of ripples within the soliton depends on the location of the secondary peak. In particular, the number of ripples increases as the secondary peak shifts towards higher $k$ values. 
The impact of the variation of the ripple size on radiation pattern will be studied in an upcoming work.

\subsection{Charge separation associated with Langmuir solitons}

The slowly-varying charge density (\ref{slow_charge}) can be re-written using  Eq. (\ref{u}), Eq. (\ref{x}) and Eq. (\ref{l}) as
\begin{equation}
    \rho = \mu \left( \frac{e}{m_\mathrm{e} c^2 } \right) \frac{|E_\mathrm{o}|^2}{4 \pi\; k^2\;c^2\ } \frac{1}{l^2 \theta^2} \frac{\partial |u|^2}{\partial x^2}, \label{eqn1}   
\end{equation}
where the field amplitude $|E_\mathrm{o}|^2$ can be expressed in the form
\begin{equation}
    |E_\mathrm{o}|^2 = \varkappa \; 8 \pi \rho_\mathrm{GJ} \; \kappa c^2   \gamma  , \label{eqn2} 
\end{equation}
where $\rho_\mathrm{GJ}$ is the co-rotational Goldreich-Julian charge density in OFR, 
$\kappa$ is the ratio of the number density of the pair plasma to the co-rotational Goldreich-Julian number density, $\gamma \approx p_{+}$ is the average Lorentz factor of the plasma particles in PFR and 
$\varkappa$ is the ratio of the energy density associated with the envelope field and the particle energy density in PFR.

Using 
the same representative values as above
and a typical value $\varkappa\sim 0.1$ (from MGP00),
Eq. (\ref{eqn1}) and Eq. (\ref{eqn2}) can be combined to give:
\begin{equation}
    \frac{\rho}{\rho_\mathrm{GJ}} \approx \mu \; \left[  \left( \frac{\varkappa}{0.1} \right) \left( \frac{ \gamma }{3} \right) \left( \frac{\kappa}{10^{4}} \right) \left( \frac{100}{\theta} \right)^2 \left( \frac{1}{4 \times 10^{4}} \frac{\partial |u|^2}{\partial x^2} \right) \right], \label{charge_sep}
\end{equation}
where
the quantity $\mu$ defined in Eq. (\ref{slow_charge_integ}) can be expressed in the form
\begin{eqnarray}
   \mu = \mu_{\pm} + \mu_\mathrm{ion}, \label{mu_tot}
\end{eqnarray}
where $\mu_{\pm}$ is the contribution due to separation of
electron-positron DF and $\mu_\mathrm{ion}$ is the contribution due to
iron ion DF near the pole $p_\mathrm{gr}$ (see Appendices B3 and
B4). The variation of $\mu$ with temperature is shown in Fig. \ref{mu_var}. It can be
seen that separation of electron and positron DF leads to $\mu_{\pm}
\sim $ 10. It can also be seen that ions play negligible role in
charge separation since the highest value of $\mu_\mathrm{ion} \approx
10^{-4}$. The result can be understood physically as follows.  The response of $\alpha$-th species to the Miller force ($\nabla^2 |E|^2$) depends on the
mass and density of the species. We find that the very small number density of the ions and their heavier mass leads to this response
being weak. On the other hand, the separation of electron and positron
DF in the pair plasma changes the effective relativistic mass (``inertia'') of the electrons and
positrons. Thus, the
Miller force acts differently on both species to create a spatial
charge separation. Panels (A) shows that $\mu_{\pm}$ for Gaussian DF
varies with temperature while Panel (B) shows that for Lorentzian DF,
$\mu_{\pm}$ remains steady across a wide range of plasma
temperatures. This implies that for the same separation of the DF, the
effective mass is temperature-dependent for short-tailed DF and is
nearly temperature independent for long-tailed DF. For ions, the
nature of the DF determines the number of interacting particles at
$p_\mathrm{gr}$. For ions with large mass $m_\mathrm{ion} = A
m_\mathrm{p}$, where $m_\mathrm{p}$ is the mass of the proton and $A$ is the atomic weight, the
choice of DF has negligible effect on the change in ions' relativistic
mass.

Let us now demonstrate that there is no physically feasible solution where contribution of ions to the  charge density separation could be non-negligible (i.e., comparable to that contribution from electrons and positrons). The expression $\mu_\mathrm{ion}$ can be written from Eq.(\ref{slow_charge_integ}) and Eq. (\ref{mu_tot}) as
\begin{equation}
    \mu_\mathrm{ion} \approx  10^{-4} \left[ \left( \frac{\mathcal{F}}{10^{5}} \right) \;  \left( \frac{10^{4}}{\kappa} \right)  \left( \frac{Z}{26} \right)^3 \left( \frac{56}{A}\right)^2 \right]   ,    
\end{equation}
where $\mathcal{F}$ is the contribution from the integrals involving DF in (\ref{slow_charge_integ}), and $Z$ is the charge of the ions. Firstly, we note that decreasing $\kappa$, while formally increasing $\mu_\mathrm{ion}$, will {\em not} lead to an increased ion's contribution to charge separation, because the latter is proportional to $\mu \kappa$ as seen in Eq. (\ref{charge_sep}). Secondly, considering heavier ions is not an option, either, given that $A \propto Z$ and one need to increase $\mu_\mathrm{ion}$ by a factor $\sim 10^4$ to bring it to the size of $\mu_{\pm}$. Thirdly, decreasing the width of the DF so as to boost $\mathcal{F}$ is also not an option as cold plasma approximation is nonphysical for the ion DF.

\section{Conclusions} \label{conclusions}

As previously shown in LMM18,
soliton formation in the NLSE with NLD
requires small values of 
the ratio of the NLD to the
local cubic nonlinearity,
$|s/q| \lesssim 0.1$, and is suppressed for higher values of $|s/q| \gtrsim 0.5$. In this work, motivated by the PSG model,
we consider an admixture of
electron-positron pairs and ions in the pulsar plasma and derived the NLSE for the envelope of Langmuir waves 
in the plasma. 
We found that  due to the low density of ions compared to the 
density of the 
pair plasma, the ion species contribute negligibly in modifying both
the coefficients of the NLSE and the charge separation.
For subsequent analysis, we neglected the ions and explored the parameter space of  different separation of the electron and positron DF  across a wide range of plasma temperatures, obtaining estimates for the range of $s/q$ values and charge separation. 

We considered two types of DF: a Lorentzian DF with a prominent power law tail and a Gaussian DF with an exponentially decaying tail. The long-tailed Lorentzian  DF provides small values of $|s/q| \sim 0.1$ across a wide range of plasma temperatures for moderate separation of the electron and positron DF. On the other hand, the short-tailed Gaussian DF provides a very restrictive parameter space where small values of  $s/q \lesssim 0.1$ can be 
attained. In
reality, the DF can have a tail in between those of a Gaussian and Lorentzian DF.
However, as long as DF's tail falls off ``sufficiently slowly" for some extended 
range of momenta, soliton formation is feasible in pulsar plasma and thus can be considered as a viable candidate to explain occurrence of CCR charge bunches. The radiation pattern due to curvature radiation under hot plasma conditions will be treated in an upcoming work.

\section*{Acknowledgements}
We thank the anonymous referee for useful comments that improved the quality of the manuscript significantly.
SMR and DM acknowledge the support of the Department of Atomic Energy,
Government of India, under project no. 12-R\&D-TFR-5.02-0700. DM
acknowledges support and funding from the `Indo-French Centre for the Promotion of Advanced Research - CEFIPRA' grant IFC/F5904-B/2018. This work was supported by the grant 2020/37/B/ST9/02215 of the National Science Centre, Poland.
%%%%%%%%%%%%%%%%%%%%%%%%%%%%%%%%%%%%%%%%%%%%%%%%%%
\section*{Data Availability}

Simulation data will be made available upon reasonable request from the corresponding author Sk. Minhajur Rahaman.



\bibliographystyle{mnras}
\bibliography{References} 

\begin{thebibliography}{}
\makeatletter
\relax
\def\mn@urlcharsother{\let\do\@makeother \do\$\do\&\do\#\do\^\do\_\do\%\do\~}
\def\mn@doi{\begingroup\mn@urlcharsother \@ifnextchar [ {\mn@doi@}
  {\mn@doi@[]}}
\def\mn@doi@[#1]#2{\def\@tempa{#1}\ifx\@tempa\@empty \href
  {http://dx.doi.org/#2} {doi:#2}\else \href {http://dx.doi.org/#2} {#1}\fi
  \endgroup}
\def\mn@eprint#1#2{\mn@eprint@#1:#2::\@nil}
\def\mn@eprint@arXiv#1{\href {http://arxiv.org/abs/#1} {{\tt arXiv:#1}}}
\def\mn@eprint@dblp#1{\href {http://dblp.uni-trier.de/rec/bibtex/#1.xml}
  {dblp:#1}}
\def\mn@eprint@#1:#2:#3:#4\@nil{\def\@tempa {#1}\def\@tempb {#2}\def\@tempc
  {#3}\ifx \@tempc \@empty \let \@tempc \@tempb \let \@tempb \@tempa \fi \ifx
  \@tempb \@empty \def\@tempb {arXiv}\fi \@ifundefined
  {mn@eprint@\@tempb}{\@tempb:\@tempc}{\expandafter \expandafter \csname
  mn@eprint@\@tempb\endcsname \expandafter{\@tempc}}}

\bibitem[\protect\citeauthoryear{{Arendt} \& {Eilek}}{{Arendt} \&
  {Eilek}}{2002}]{2002ApJ...581..451A}
{Arendt} Paul~N. J.,  {Eilek} J.~A.,  2002, \mn@doi [\apj] {10.1086/344133},
  \href {https://ui.adsabs.harvard.edu/abs/2002ApJ...581..451A} {581, 451}

\bibitem[\protect\citeauthoryear{{Arumugasamy} \& {Mitra}}{{Arumugasamy} \&
  {Mitra}}{2019}]{2019MNRAS.489.4589A}
{Arumugasamy} P.,  {Mitra} D.,  2019, \mn@doi [\mnras] {10.1093/mnras/stz2299},
  \href {https://ui.adsabs.harvard.edu/abs/2019MNRAS.489.4589A} {489, 4589}

\bibitem[\protect\citeauthoryear{{Asseo} \& {Melikidze}}{{Asseo} \&
  {Melikidze}}{1998}]{1998MNRAS.301...59A}
{Asseo} E.,  {Melikidze} G.~I.,  1998, \mn@doi [\mnras]
  {10.1046/j.1365-8711.1998.01990.x}, \href
  {https://ui.adsabs.harvard.edu/abs/1998MNRAS.301...59A} {301, 59}

\bibitem[\protect\citeauthoryear{{Basu}, {Mitra}, {Melikidze}, {Maciesiak},
  {Skrzypczak}  \& {Szary}}{{Basu} et~al.}{2016}]{2016ApJ...833...29B}
{Basu} R.,  {Mitra} D.,  {Melikidze} G.~I.,  {Maciesiak} K.,  {Skrzypczak} A.,
   {Szary} A.,  2016, \mn@doi [\apj] {10.3847/1538-4357/833/1/29}, \href
  {https://ui.adsabs.harvard.edu/abs/2016ApJ...833...29B} {833, 29}

\bibitem[\protect\citeauthoryear{{Blasi} \& {Amato}}{{Blasi} \&
  {Amato}}{2011}]{2011ASSP...21..624B}
{Blasi} P.,  {Amato} E.,  2011, Astrophysics and Space Science Proceedings,
  \href {https://ui.adsabs.harvard.edu/abs/2011ASSP...21..624B} {21, 624}

\bibitem[\protect\citeauthoryear{{Cheng} \& {Ruderman}}{{Cheng} \&
  {Ruderman}}{1977}]{1977ApJ...212..800C}
{Cheng} A.~F.,  {Ruderman} M.~A.,  1977, \mn@doi [\apj] {10.1086/155105}, \href
  {https://ui.adsabs.harvard.edu/abs/1977ApJ...212..800C} {212, 800}

\bibitem[\protect\citeauthoryear{{Cheng} \& {Ruderman}}{{Cheng} \&
  {Ruderman}}{1980}]{1980ApJ...235..576C}
{Cheng} A.~F.,  {Ruderman} M.~A.,  1980, \mn@doi [\apj] {10.1086/157661}, \href
  {https://ui.adsabs.harvard.edu/abs/1980ApJ...235..576C} {235, 576}

\bibitem[\protect\citeauthoryear{{Daugherty} \& {Harding}}{{Daugherty} \&
  {Harding}}{1982}]{1982ApJ...252..337D}
{Daugherty} J.~K.,  {Harding} A.~K.,  1982, \mn@doi [\apj] {10.1086/159561},
  \href {https://ui.adsabs.harvard.edu/abs/1982ApJ...252..337D} {252, 337}

\bibitem[\protect\citeauthoryear{{Geppert}}{{Geppert}}{2017}]{2017JApA...38...46G}
{Geppert} U.,  2017, \mn@doi [Journal of Astrophysics and Astronomy]
  {10.1007/s12036-017-9460-y}, \href
  {https://ui.adsabs.harvard.edu/abs/2017JApA...38...46G} {38, 46}

\bibitem[\protect\citeauthoryear{{Gil}, {Melikidze}  \& {Mitra}}{{Gil}
  et~al.}{2002}]{2002A&A...388..235G}
{Gil} J.~A.,  {Melikidze} G.~I.,   {Mitra} D.,  2002, \mn@doi [\aap]
  {10.1051/0004-6361:20020473}, \href
  {https://ui.adsabs.harvard.edu/abs/2002A&A...388..235G} {388, 235}

\bibitem[\protect\citeauthoryear{{Gil}, {Melikidze}  \& {Geppert}}{{Gil}
  et~al.}{2003}]{2003A&A...407..315G}
{Gil} J.,  {Melikidze} G.~I.,   {Geppert} U.,  2003, \mn@doi [\aap]
  {10.1051/0004-6361:20030854}, \href
  {https://ui.adsabs.harvard.edu/abs/2003A&A...407..315G} {407, 315}

\bibitem[\protect\citeauthoryear{{Ginzburg}, {Zheleznyakov}  \&
  {Zaitsev}}{{Ginzburg} et~al.}{1969}]{1969Ap&SS...4..464G}
{Ginzburg} V.~L.,  {Zheleznyakov} V.~V.,   {Zaitsev} V.~V.,  1969, \mn@doi
  [\apss] {10.1007/BF00651351}, \href
  {https://ui.adsabs.harvard.edu/abs/1969Ap&SS...4..464G} {4, 464}

\bibitem[\protect\citeauthoryear{{Goldreich} \& {Julian}}{{Goldreich} \&
  {Julian}}{1969}]{1969ApJ...157..869G}
{Goldreich} P.,  {Julian} W.~H.,  1969, \mn@doi [\apj] {10.1086/150119}, \href
  {https://ui.adsabs.harvard.edu/abs/1969ApJ...157..869G} {157, 869}

\bibitem[\protect\citeauthoryear{{Hibschman} \& {Arons}}{{Hibschman} \&
  {Arons}}{2001}]{2001ApJ...560..871H}
{Hibschman} J.~A.,  {Arons} J.,  2001, \mn@doi [\apj] {10.1086/323069}, \href
  {https://ui.adsabs.harvard.edu/abs/2001ApJ...560..871H} {560, 871}

\bibitem[\protect\citeauthoryear{{Ichikawa}}{{Ichikawa}}{1974}]{1974PThPS..55..212I}
{Ichikawa} Y.~H.,  1974, \mn@doi [Progress of Theoretical Physics Supplement]
  {10.1143/PTPS.55.212}, \href
  {https://ui.adsabs.harvard.edu/abs/1974PThPS..55..212I} {55, 212}

\bibitem[\protect\citeauthoryear{{Ichikawa} \& {Taniuti}}{{Ichikawa} \&
  {Taniuti}}{1973}]{1973JPSJ...34..513I}
{Ichikawa} Y.~H.,  {Taniuti} T.,  1973, \mn@doi [Journal of the Physical
  Society of Japan] {10.1143/JPSJ.34.513}, \href
  {https://ui.adsabs.harvard.edu/abs/1973JPSJ...34..513I} {34, 513}

\bibitem[\protect\citeauthoryear{Jordan \& Josserand}{Jordan \&
  Josserand}{2001}]{JORDAN2001433}
Jordan R.,  Josserand C.,  2001, \mn@doi [Mathematics and Computers in
  Simulation] {https://doi.org/10.1016/S0378-4754(00)00292-5}, 55, 433

\bibitem[\protect\citeauthoryear{{Karpman}, {Norman}, {Ter Haar}  \&
  {Tsytovich}}{{Karpman} et~al.}{1975}]{1975PhyS...11..271K}
{Karpman} V.~I.,  {Norman} C.~A.,  {Ter Haar} D.,   {Tsytovich} V.~N.,  1975,
  \mn@doi [\physscr] {10.1088/0031-8949/11/5/006}, \href
  {https://ui.adsabs.harvard.edu/abs/1975PhyS...11..271K} {11, 271}

\bibitem[\protect\citeauthoryear{{Kazbegi}, {Machabeli}  \&
  {Melikidze}}{{Kazbegi} et~al.}{1991}]{1991MNRAS.253..377K}
{Kazbegi} A.~Z.,  {Machabeli} G.~Z.,   {Melikidze} G.~I.,  1991, \mn@doi
  [\mnras] {10.1093/mnras/253.3.377}, \href
  {https://ui.adsabs.harvard.edu/abs/1991MNRAS.253..377K} {253, 377}

\bibitem[\protect\citeauthoryear{{Kijak} \& {Gil}}{{Kijak} \&
  {Gil}}{1997}]{1997MNRAS.288..631K}
{Kijak} J.,  {Gil} J.,  1997, \mn@doi [\mnras] {10.1093/mnras/288.3.631}, \href
  {https://ui.adsabs.harvard.edu/abs/1997MNRAS.288..631K} {288, 631}

\bibitem[\protect\citeauthoryear{{Kijak} \& {Gil}}{{Kijak} \&
  {Gil}}{1998}]{1998MNRAS.299..855K}
{Kijak} J.,  {Gil} J.,  1998, \mn@doi [\mnras]
  {10.1046/j.1365-8711.1998.01832.x}, \href
  {https://ui.adsabs.harvard.edu/abs/1998MNRAS.299..855K} {299, 855}

\bibitem[\protect\citeauthoryear{Lakoba}{Lakoba}{2017}]{Lakoba2016}
Lakoba T.~I.,  2017, Journal of Scientific Computing, 72, 14

\bibitem[\protect\citeauthoryear{{Lakoba}, {Mitra}  \& {Melikidze}}{{Lakoba}
  et~al.}{2018}]{2018MNRAS.480.4526L}
{Lakoba} T.,  {Mitra} D.,   {Melikidze} G.,  2018, \mn@doi [\mnras]
  {10.1093/mnras/sty2152}, \href
  {https://ui.adsabs.harvard.edu/abs/2018MNRAS.480.4526L} {480, 4526}

\bibitem[\protect\citeauthoryear{{Lighthill}}{{Lighthill}}{1967}]{1967RSPSA.299...28L}
{Lighthill} M.~J.,  1967, \mn@doi [Proceedings of the Royal Society of London
  Series A] {10.1098/rspa.1967.0121}, \href
  {https://ui.adsabs.harvard.edu/abs/1967RSPSA.299...28L} {299, 28}

\bibitem[\protect\citeauthoryear{{Lominadze}, {Machabeli}, {Melikidze}  \&
  {Pataraia}}{{Lominadze} et~al.}{1986}]{1986FizPl..12.1233L}
{Lominadze} D.~G.,  {Machabeli} G.~Z.,  {Melikidze} G.~I.,   {Pataraia} A.~D.,
  1986, Fizika Plazmy, \href
  {https://ui.adsabs.harvard.edu/abs/1986FizPl..12.1233L} {12, 1233}

\bibitem[\protect\citeauthoryear{{Manthei}, {Ben{\'a}{\v{c}}ek}, {Mu{\~n}oz}
  \& {B{\"u}chner}}{{Manthei} et~al.}{2021}]{2021A&A...649A.145M}
{Manthei} A.~C.,  {Ben{\'a}{\v{c}}ek} J.,  {Mu{\~n}oz} P.~A.,   {B{\"u}chner}
  J.,  2021, \mn@doi [\aap] {10.1051/0004-6361/202039907}, \href
  {https://ui.adsabs.harvard.edu/abs/2021A&A...649A.145M} {649, A145}

\bibitem[\protect\citeauthoryear{{Melikidze} \& {Pataraia}}{{Melikidze} \&
  {Pataraia}}{1978}]{1978SoGru..90...49M}
{Melikidze} G.~I.,  {Pataraia} A.~D.,  1978, Akademiia Nauk Gruzii
  Soobshcheniia, \href {https://ui.adsabs.harvard.edu/abs/1978SoGru..90...49M}
  {90, 49}

\bibitem[\protect\citeauthoryear{{Melikidze} \& {Pataraia}}{{Melikidze} \&
  {Pataraia}}{1980}]{1980Afz....16..161M}
{Melikidze} G.~I.,  {Pataraia} A.~D.,  1980, Astrofizika, \href
  {https://ui.adsabs.harvard.edu/abs/1980Afz....16..161M} {16, 161}

\bibitem[\protect\citeauthoryear{{Melikidze} \& {Pataraya}}{{Melikidze} \&
  {Pataraya}}{1984}]{1984Ap.....20..100M}
{Melikidze} G.~I.,  {Pataraya} A.~D.,  1984, \mn@doi [Astrophysics]
  {10.1007/BF01005908}, \href
  {https://ui.adsabs.harvard.edu/abs/1984Ap.....20..100M} {20, 100}

\bibitem[\protect\citeauthoryear{{Melikidze}, {Gil}  \& {Pataraya}}{{Melikidze}
  et~al.}{2000}]{2000ApJ...544.1081M}
{Melikidze} G.~I.,  {Gil} J.~A.,   {Pataraya} A.~D.,  2000, \mn@doi [\apj]
  {10.1086/317220}, \href
  {https://ui.adsabs.harvard.edu/abs/2000ApJ...544.1081M} {544, 1081}

\bibitem[\protect\citeauthoryear{{Melrose}}{{Melrose}}{1995}]{1995JApA...16..137M}
{Melrose} D.~B.,  1995, \mn@doi [Journal of Astrophysics and Astronomy]
  {10.1007/BF02714830}, \href
  {https://ui.adsabs.harvard.edu/abs/1995JApA...16..137M} {16, 137}

\bibitem[\protect\citeauthoryear{{Melrose} \& {Gedalin}}{{Melrose} \&
  {Gedalin}}{1999}]{1999ApJ...521..351M}
{Melrose} D.~B.,  {Gedalin} M.~E.,  1999, \mn@doi [\apj] {10.1086/307539},
  \href {https://ui.adsabs.harvard.edu/abs/1999ApJ...521..351M} {521, 351}

\bibitem[\protect\citeauthoryear{{Mitra}}{{Mitra}}{2017}]{2017JApA...38...52M}
{Mitra} D.,  2017, \mn@doi [Journal of Astrophysics and Astronomy]
  {10.1007/s12036-017-9457-6}, \href
  {https://ui.adsabs.harvard.edu/abs/2017JApA...38...52M} {38, 52}

\bibitem[\protect\citeauthoryear{{Mitra} \& {Li}}{{Mitra} \&
  {Li}}{2004}]{2004A&A...421..215M}
{Mitra} D.,  {Li} X.~H.,  2004, \mn@doi [\aap] {10.1051/0004-6361:20034094},
  \href {https://ui.adsabs.harvard.edu/abs/2004A&A...421..215M} {421, 215}

\bibitem[\protect\citeauthoryear{{Mitra}, {Gil}  \& {Melikidze}}{{Mitra}
  et~al.}{2009}]{2009ApJ...696L.141M}
{Mitra} D.,  {Gil} J.,   {Melikidze} G.~I.,  2009, \mn@doi [\apjl]
  {10.1088/0004-637X/696/2/L141}, \href
  {https://ui.adsabs.harvard.edu/abs/2009ApJ...696L.141M} {696, L141}

\bibitem[\protect\citeauthoryear{{Mitra}, {Basu}, {Melikidze}  \&
  {Arjunwadkar}}{{Mitra} et~al.}{2020}]{2020MNRAS.492.2468M}
{Mitra} D.,  {Basu} R.,  {Melikidze} G.~I.,   {Arjunwadkar} M.,  2020, \mn@doi
  [\mnras] {10.1093/mnras/stz3620}, \href
  {https://ui.adsabs.harvard.edu/abs/2020MNRAS.492.2468M} {492, 2468}

\bibitem[\protect\citeauthoryear{{Pataraia} \& {Melikidze}}{{Pataraia} \&
  {Melikidze}}{1980}]{1980Ap&SS..68...49P}
{Pataraia} A.,  {Melikidze} G.,  1980, \apss, \href
  {https://ui.adsabs.harvard.edu/abs/1980Ap&SS..68...49P} {68, 49}

\bibitem[\protect\citeauthoryear{{Rahaman}, {Mitra}  \& {Melikidze}}{{Rahaman}
  et~al.}{2020}]{2020MNRAS.497.3953R}
{Rahaman} S.~M.,  {Mitra} D.,   {Melikidze} G.~I.,  2020, \mn@doi [\mnras]
  {10.1093/mnras/staa2280}, \href
  {https://ui.adsabs.harvard.edu/abs/2020MNRAS.497.3953R} {497, 3953}

\bibitem[\protect\citeauthoryear{{Rahaman}, {Basu}, {Mitra}  \&
  {Melikidze}}{{Rahaman} et~al.}{2021}]{2021MNRAS.500.4139R}
{Rahaman} S. k.~M.,  {Basu} R.,  {Mitra} D.,   {Melikidze} G.~I.,  2021,
  \mn@doi [\mnras] {10.1093/mnras/staa3518}, \href
  {https://ui.adsabs.harvard.edu/abs/2021MNRAS.500.4139R} {500, 4139}

\bibitem[\protect\citeauthoryear{{Ruderman} \& {Sutherland}}{{Ruderman} \&
  {Sutherland}}{1975}]{1975ApJ...196...51R}
{Ruderman} M.~A.,  {Sutherland} P.~G.,  1975, \mn@doi [\apj] {10.1086/153393},
  \href {https://ui.adsabs.harvard.edu/abs/1975ApJ...196...51R} {196, 51}

\bibitem[\protect\citeauthoryear{{Sturrock}}{{Sturrock}}{1971}]{1971ApJ...164..529S}
{Sturrock} P.~A.,  1971, \mn@doi [\apj] {10.1086/150865}, \href
  {https://ui.adsabs.harvard.edu/abs/1971ApJ...164..529S} {164, 529}

\bibitem[\protect\citeauthoryear{{Suvorov} \& {Chugunov}}{{Suvorov} \&
  {Chugunov}}{1973}]{1973Ap&SS..23..189S}
{Suvorov} E.~V.,  {Chugunov} Y.~V.,  1973, \mn@doi [\apss]
  {10.1007/BF00647657}, \href
  {https://ui.adsabs.harvard.edu/abs/1973Ap&SS..23..189S} {23, 189}

\bibitem[\protect\citeauthoryear{{Szary}, {Melikidze}  \& {Gil}}{{Szary}
  et~al.}{2015}]{2015MNRAS.447.2295S}
{Szary} A.,  {Melikidze} G.~I.,   {Gil} J.,  2015, \mn@doi [\mnras]
  {10.1093/mnras/stu2622}, \href
  {https://ui.adsabs.harvard.edu/abs/2015MNRAS.447.2295S} {447, 2295}

\makeatother
\end{thebibliography}

%%%%%%%%%%%%%%%%%%%%%%%%%%%%%%%%%%%%%%%%%%%%%%%%%%

%%%%%%%%%%%%%%%%% APPENDICES %%%%%%%%%%%%%%%%%%%%%

\appendix
\section{Derivation of NLSE with NLD}\label{deriv1}
We follow the procedure by \citealt{1980Afz....16..161M} (MP80), \citealt{1980Ap&SS..68...49P}(PM80), \citealt{1984Ap.....20..100M} (MP84) and \citealt{2000ApJ...544.1081M}(MGP00) with the addition that arbitrary species $\alpha$ can be included as a plasma component. 

We identify two frames of reference. Firstly, we have a Plasma Frame of Reference (hereafter PFR) where the average velocity of the plasma particles is zero. Additionally, we have a Moving Frame of Reference (hereafter MFR) which moves with a velocity $\lambda = v_\mathrm{gr}$ with respect to PFR. Here $v_\mathrm{gr}$ corresponds to the group velocity of the linear plasma waves in PFR. The velocity $\lambda$ corresponds to the Lorentz factor $\gamma_\mathrm{o}$. For simplicity we choose $c=1$. The transformed quantities are shown in table \ref{tab:Lorentz transformation}. 

\begin{table}[h]
    \caption{Transformation of quantities from PFR to MFR }
    \centering
    \begin{tabular}{|c|c|}\hline 
     PFR &  MFR \\ \hline 
      $x$   & $x' = \gamma_\mathrm{o} (x - \lambda t)$    \\
      $t$   & $t' = \gamma_\mathrm{o} (t - \lambda x)$    \\ 
      $\omega$ & $\omega' = \gamma_\mathrm{o}(\omega - k \lambda)$ \\
      $k$   &   $k' = \gamma_\mathrm{o} (k - \lambda \omega)$ \\
      $p$   &   $ p' = \gamma_\mathrm{o} (p - \mathcal{E} \lambda)$ \\
      $v$   &   $v' = \frac{v - \lambda}{1 - \lambda v}$   \\
      $v_\mathrm{gr} = \frac{d \omega}{d k}$ & $v'_\mathrm{gr} = \frac{d \omega'}{d k'} = \frac{v_\mathrm{gr} - \lambda}{1 - \lambda v_\mathrm{gr}}$   \\  \hline 
    \end{tabular}
    \label{tab:Lorentz transformation}
\end{table}

We introduce the stretched variables in MFR as
\begin{eqnarray}
    &\ \xi' = \epsilon x' \\
    &\ \tau' = \epsilon^2 t' 
\end{eqnarray}

The choice of the stretched variables can be motivated as follows. Consider the dispersion relation and group velocity of Langmuir waves in PFR to be defined as
\begin{eqnarray}
    &\ \omega = k (1 - \delta_{1}) \\ 
    &\  \lambda = 1 - \delta_{2}
\end{eqnarray}
where the quantities $(\delta_{1},\delta_{2}) \ll 1$. 

Using the definitions above, the wave quantities $(\omega, k)$ in PFR can be transformed to $(\omega',k')$ in MFR via the Lorentz transformation as
\begin{eqnarray}
    &\ \omega' = \gamma_\mathrm{o}(\omega - k \lambda) = \gamma_\mathrm{o} k (\delta_{2} - \delta_{1}) \\
    &\ k' = \gamma_\mathrm{o} (k - \lambda \omega) = \gamma_\mathrm{o} k (\delta_{2} + \delta_{1})
\end{eqnarray}
Thus, the transformed wave varies much slower in time compared to variation in space. Hence, it justifies the choice of stretched time variable to be second order in $\epsilon$ and the stretched space variable to be first order in $\epsilon$.   

Making use of the stretched variables we have
\begin{eqnarray}
   \label{stretched variable1}&\ \frac{\partial}{\partial t'} \rightarrow \frac{\partial}{\partial t'} + \frac{\partial \tau'}{\partial t'} \frac{\partial}{ \partial \tau'} \rightarrow \frac{\partial}{\partial t'} + \epsilon^2 \frac{\partial}{\partial \tau'} \\  \label{stretched variable2}
  &\ \frac{\partial}{\partial x'} \rightarrow \frac{\partial}{\partial x'} + \frac{\partial \xi'}{\partial x'} \frac{\partial}{\partial \xi'} \rightarrow \frac{\partial}{\partial x'} + \epsilon \frac{\partial}{\partial \xi'} 
\end{eqnarray}

We define the particle distribution function and the electric field in MFR as 
\begin{eqnarray}
  \label{Distri} &\ f_{\alpha}' = f^{(0)'}(p') +  \summa{l}{n}{n} \; f^{(n)'}_{\alpha,l} (p', \xi', \tau') \\
 \label{Elec_field} &\ E = \summa{l}{n}{n} \; E^{(n)}_{l} (\xi', \tau') 
\end{eqnarray}
where the stretched variable amplitudes are given by
\begin{eqnarray}
\label{ampli1}
    &\ f^{(n)'}_{\alpha,l} = \frac{1}{2 \pi} \int d\Omega' \int dK' \; \exp{i(K'\xi' - \Omega' \tau' )} \;  \Tilde{f}^{(n)'}_{\alpha,l} (p \; , \Omega', K') \\ \label{ampli2}
    &\  E^{(n)}_{l} = \frac{1}{2 \pi} \int d\Omega' \int dK' \; \exp{i(K'\xi' - \Omega' \tau' )} \;  \Tilde{E}^{(n)}_{l} (\Omega', K')
\end{eqnarray}
subject to the reality condition
\begin{eqnarray}
    &\ E^{(n) \star }_{l} = E^{(n)}_{-l} \\
    &\ f^{(n)' \star }_{\alpha,l} = f^{(n)'}_{\alpha,-l} 
\end{eqnarray}

Using equation (\ref{stretched variable1}) and equation (\ref{stretched variable2}) the Vlasov Equation in MFR takes the form 
\begin{eqnarray*}
    &\ \frac{\partial f_{\alpha}'}{\partial t'} + v' \frac{\partial f_{\alpha}'}{\partial x'} + \frac{e_{\alpha} E}{m_{\alpha}} \frac{\partial f_{\alpha}'}{\partial p'} = 0 \\ \\ 
    \Rightarrow &\ \left[ \frac{\partial f_{\alpha}'}{\partial t'} + \epsilon^2 \frac{\partial f_{\alpha}'}{\partial \tau'} \right] + v'  \left[ \frac{\partial f_{\alpha}'}{\partial x'}  + \epsilon \frac{\partial f_{\alpha}'}{\partial \xi'} \right] +  \; \left( \frac{e_{\alpha} E}{m_{\alpha}} \right)  \frac{\partial f_{\alpha}'}{\partial p'}= 0 \\ 
\end{eqnarray*}

Using equation (\ref{Distri}) and equation (\ref{Elec_field}) in the equation above the Master equation for Vlasov Equation (hereafter MVE) is
\begin{equation}
  il \left( \omega' - k'v' \right) f^{(n)'}_{\alpha,l} -  \dd{n-2}{l}{\tau'}  -   v' \dd{n-1}{l}{\xi'}   =  \; \frac{e_{\alpha}}{m_{\alpha}} \left[  E^{(n)}_{l} \frac{\partial f^{(0)'}_{\alpha}}{\partial p'} + \sum^{+\infty}_{j=-\infty} \sum^{+\infty}_{m=1}  \dd{m}{j}{p'} \; E^{(n-m)}_{l-j}\right] 
\end{equation}

Using equation (\ref{stretched variable2}) the Poisson's equation in MFR is written as
\begin{eqnarray*}
    &\ \frac{\partial E}{\partial x'}  = 4\pi \sum_{\alpha} e_{\alpha} n_{\alpha}'  \integ{p'}\;  f_{\alpha}'           \\ 
   \Rightarrow &\ \left[  \frac{\partial }{\partial x'}  + \epsilon \frac{\partial }{\partial \xi'} \right]   E  = 4\pi \sum_{\alpha} n_{\alpha}' \; e_{\alpha} \integ{p}\; f_{\alpha}'   
\end{eqnarray*}

 Using equation (\ref{Distri}) and equation (\ref{Elec_field}) in the equation above the Master equation for Poisson's equation (hereafter MPE) in MFR is  
\begin{equation}
il k' \;  E^{(n)}_{l}  +  \frac{\partial E^{(n-1)}_{l}}{\partial \xi'}  = 4 \pi \sum_{\alpha}   n_{\alpha}' \; e_{\alpha} \integ{p'} \; f^{(n)'}_{\alpha,l}
\end{equation}

\subsection{{n=1,l=1 term} }\label{n1l1}
From MVE we get,
\begin{equation}
  f^{(1)'}_{\alpha,1} = \frac{G'_{\alpha}}{i (\omega' - k'v')} E^{(1)}_{1}
\end{equation}
where 
\begin{equation}
 G'_{\alpha} = \frac{e_{\alpha}}{m_{\alpha}} \frac{\partial f^{(0)'}_{\alpha}}{\partial p'} 
\end{equation}
From MPE we have,
\begin{equation}
 k'  + \sum_{\alpha} \omega'^2_\mathrm{p,\alpha} \integ{p'} \; \frac{\partial f^{(0)'}_{\alpha}}{\partial p'} \frac{1}{(\omega' - k'v')} = 0 \label{Dispersion relation}
\end{equation}
where
\begin{equation}
 \omega'^2_\mathrm{p,\alpha} = \frac{4 \pi n'_{\alpha} e^2_{\alpha}}{m_{\alpha}}
\end{equation}

In what follow in each subsection the MPE and MVE term corresponding to a set $(n,l)$ is evaluated. Each term becomes an input to the next term until at $n=3,l=1$ we obtain the NLSE governing the envelope electric field $E^{(1)}_{1}$.

\subsection{{n = 1,l = 0 term}  }
 
Substituting $(n=1,l=0)$ in MVE we have,
\begin{equation}
    E^{(1)}_{0} = 0 
\end{equation}

Substituting $(n=1,l=0)$ in MPE we have,
\begin{equation}
    4\pi \sum_{\alpha}  n'_{\alpha} e_{\alpha} \integ{p'} \; f^{(1)'}_{\alpha,0} = 0  \label{n1l0_expr1}
\end{equation}

\subsubsection{n = 2, l = 0 term}

Substituting $(n=2,l=0)$ in MVE we get,
\begin{equation}
 v' \dd{1}{0}{\xi'} =  -  \frac{e_{\alpha}}{m_{\alpha}} \left[ E^{(2)}_{0} \frac{\partial f^{(0)'}_{\alpha}}{\partial p'} + \Bigg \{ E^{(1)}_{-1} \dd{1}{1}{p'} + E^{(1)}_{1} \dd{1}{-1}{p'} \Bigg \}  \right] \label{n2l0_expr1}
\end{equation}

Now using \ref{n1l1} we have
\begin{equation}
E^{(1)}_{-1} \dd{1}{1}{p'} + E^{(1)}_{1} \dd{1}{-1}{p'} = - \left[ |E^{(1)}_{1}|^2 \frac{\partial}{\partial p'} \Bigg \{ \frac{G'_{\alpha}}{i(\omega' - k'v')}  \Bigg \} - |E^{(1)}_{1}|^2 \frac{\partial}{\partial p'} \Bigg \{ \frac{G'_{\alpha}}{i(\omega' - k'v')} \Bigg \} \right] = 0  \label{n2l0_expr2}
\end{equation}

Substituting equation (\ref{n2l0_expr2})  in equation (\ref{n2l0_expr1})  we get
\begin{equation}
     \dd{1}{0}{\xi'} =  -  \frac{e_{\alpha}}{m_{\alpha}} \frac{1}{v' } \frac{\partial f^{(0)'}_{\alpha}}{\partial p'} E^{(2)}_{0} \label{n2l0_expr3}
\end{equation}

Substituting equation (\ref{ampli1}) in equation (\ref{n2l0_expr3}) we get
\begin{equation}
   \Tilde{f}^{(1)}_{\alpha,0} =  - \frac{1}{i K'}
    \frac{e_{\alpha}}{m_{\alpha}} \frac{1}{v'} \frac{\partial f^{(0)'}_{\alpha}}{\partial p'} \Tilde{E}^{(2)}_{0} \label{n2l0_expr3}
\end{equation}

Using equation (\ref{n2l0_expr3}) in equation (\ref{n1l0_expr1}) we get,
\begin{equation}
    \ 4\pi \sum_{\alpha}  n'_{\alpha} \; e_{\alpha} \integ{p} \; \Tilde{f}^{(1)}_{\alpha,0} = 0 
    \Rightarrow  \Bigg \{ \sum_{\alpha} \omega'^2_{p,\alpha} \integ{p'} \; \frac{1}{v'} \frac{\partial f^{(0)'}_{\alpha}}{\partial p'} \Bigg \}  \Tilde{E}^{(2)}_{0} = 0 
\end{equation}
which gives
\begin{equation}
\Tilde{E}^{(2)}_{0} = 0   \label{n2l0_expr4}
\end{equation}

Using equation (\ref{n2l0_expr4}) in equation (\ref{n2l0_expr3}) we have
\begin{equation}
  \Tilde{f}^{(1)'}_{\alpha,0} = 0  \label{n2l0_expr5}
\end{equation}

Using equation (\ref{n2l0_expr5}) and equation (\ref{n2l0_expr4}) we have
\begin{eqnarray}
&\ E^{(2)}_{0} = 0  \\ 
&\ f^{(1)'}_{\alpha,0} = 0
\end{eqnarray}

\subsection{ {n=1, l$\neq$ 0 term} }
 Substituting $(n=1,l\neq 0)$ in MVE we have
\begin{equation}
 f^{(1)'}_{\alpha,l} =  \frac{G'_{\alpha}}{i\;l (\omega' - k'v')}\; E^{(1)}_{l}   
\end{equation}

Substituting $(n=1,l\neq 0)$ in MPE,
\begin{eqnarray*}
    &\  ilk' \; E^{(1)}_{1} = 4 \pi \sum_{\alpha} n'_{\alpha} \;  e_{\alpha} \;  \integ{p'} \; f^{(1)'}_{\alpha,l} \\
  \Rightarrow &\ \left[ k'  + l^2 \; \sum_{\alpha} \omega'^2_\mathrm{p,\alpha} \integ{p'} \; \frac{\partial f^{(0)'}_{\alpha}}{\partial p'} \frac{1}{(\omega' - k'v')}\right] E^{(1)}_{l} = 0 \\
  \Rightarrow &\  (1 - l^2) \; E^{(1)}_{l} = 0  
\end{eqnarray*}

To summarize we have
\begin{eqnarray}
    &\ E^{(1)}_{l} = 0   \hspace{2cm} \text{for $|l|> 1$} \\
    &\ f^{(1)'}_{\alpha,l} = 0 \hspace{2cm} \text{for $|l|> 1$} 
\end{eqnarray}

\subsection{{n = 2, l = 1 term}}

Substituting $(n=2,l=1)$ in MVE we have,
\begin{equation}
    f^{(2)'}_{\alpha,1} =  \frac{1}{i(\omega' - k' v')} \left[  G'_{\alpha} E^{(2)}_{1} + v'  \dd{1}{1}{\xi'}\right] \label{n2l1_expr1}
\end{equation}

Substituting $(n=2,l=1)$ in MPE we have,
\begin{eqnarray*}
    &\ i k' E^{(2)}_{1} +  \frac{\partial E^{(1)}_{1}}{\partial \xi'} = 4 \pi \sum_{\alpha}  n'_\mathrm{\alpha} \; e_{\alpha}\; \integ{p'} \;  f^{(2)'}_{\alpha,1} \\ \\  
    \Rightarrow &\ i \left[ k' +  \; \sum_{\alpha} \omega'^2_\mathrm{p,\alpha} \integ{p'} \; \frac{\partial f^{(0)'}_{\alpha}}{\partial p'} \frac{1}{(\omega' - k'v')} \right] E^{(2)}_{1} +  \frac{\partial E^{(1)}_{1}}{\partial \xi'}  = 4 \pi \sum_{\alpha}  n'_\mathrm{\alpha} e_{\alpha} \integ{p'} \frac{v'}{i (\omega' - k'v')} \dd{1}{1}{\xi'} \\ \\
    \Rightarrow &\  \left[ 1 + \sum_{\alpha} \omega'^2_\mathrm{p,\alpha}   \integ{p} \; \frac{v'}{(\omega' - k'v')^2}  \frac{\partial f^{(0)'}_{\alpha}}{\partial p'} \right] \frac{\partial E^{(1)}_{1}}{\partial \xi'} = 0 
\end{eqnarray*} 

The condition given above is referred to as the solvability condition. This requires that the term in the square brackets be zero.

We differentiate equation (\ref{Dispersion relation}) with respect to $k'$ to obtain
\begin{equation}
     1 + \sum_{\alpha} \omega'^2_\mathrm{p,\alpha}   \integ{p'} \; \frac{1}{(\omega' - k'v')^2} \bigg\{ - \left( \frac{d \omega'}{d k'} - v' \right) \Bigg \}   \frac{\partial f^{(0)'}_{\alpha}}{\partial p'} = 0 
\end{equation}

Thus, the solvability condition requires the group velocity of the transformed waves in MFR to be zero
\begin{equation}
   v'_\mathrm{gr} = \frac{d \omega'}{d k'} =0    
  \Rightarrow   \frac{v_\mathrm{gr} - \lambda}{1 - \lambda \;  v_\mathrm{gr}} = 0  \Rightarrow
   \lambda = v_\mathrm{gr} = \frac{d \omega}{ d k}
\end{equation}

 Thus, the solvability condition is satisfied since the velocity of MFR has been identified with the group velocity of the linear Langmuir waves in PFR.

\subsection{{n = 2, l = 2 term}  }

 Substituting $(n=2,l=2)$ in MVE and using Eq. A23 we get,
\begin{eqnarray*}
    &\  i 2 (\omega' - k'v') f^{(2)'}_{\alpha,2}  \\
    &\ =  \frac{e_{\alpha}}{m_{\alpha}} \left[ E^{(2)}_{2} \frac{\partial f^{(0)'}_{\alpha}}{\partial p'} + \sum^{+\infty}_{j= -\infty} \sum^{+\infty}_{m=1} \dd{m}{j}{p'} E^{(2-m)}_{2-j}\right] \\  
    &\ =  \frac{e_{\alpha}}{m_{\alpha}} \left[ E^{(2)}_{2} \frac{\partial f^{(0)'}_{\alpha}}{\partial p'} + \sum^{+\infty}_{j= -\infty} \dd{1}{j}{p'} E^{(1)}_{2-j}\right]
    \\ &\  =  \frac{e_{\alpha}}{m_{\alpha}} \left[ E^{(2)}_{2} \frac{\partial f^{(0)'}_{\alpha}}{\partial p'} +  \dd{1}{1}{p'} E^{(1)}_{1} \right] 
\end{eqnarray*}
which gives
\begin{equation}
     f^{(2)'}_{\alpha,2} = \frac{1}{2i} \; \frac{e_{\alpha}}{m_{\alpha} } \; \frac{1}{(\omega' - k'v') } \left[ E^{(2)}_{2} \ddo  + E^{(1)}_{1} \dd{1}{1}{p'}\right]
\end{equation}

Substituting $(n=2,l=2)$ in MPE we have,
\begin{eqnarray*}
    &\  2 i k' \;  E^{(2)}_{2} = 4 \pi \sum_{\alpha}  n'_\mathrm{\alpha} e_{\alpha} \int^{+\infty}_{-\infty} dp' \; f^{(2)'}_{\alpha,2} \\ 
    \Rightarrow &\  \left[ 2i k'   - \frac{1}{2i} \sum_{\alpha} \omega'^2_\mathrm{p,\alpha}  \integ{p'} \; \frac{1}{(\omega' - k' v')} \ddo \right] E^{(2)}_{2} = \frac{1}{2i} \sum_{\alpha} \omega'^2_\mathrm{p,\alpha} \integ{p'} \;  \frac{1}{(\omega' - k' v')} \dd{1}{1}{p'}    \\ 
    \Rightarrow &\  - \frac{3 k' }{2 i} E^{(2)}_{2}  = \frac{1}{2i} \sum_{\alpha} \omega'^2_\mathrm{p,\alpha} \integ{p'} \;  \frac{1}{(\omega' - k' v')} \dd{1}{1}{p'}   \\
    \Rightarrow &\ - \frac{3 k' }{2 i} E^{(2)}_{2}  = \frac{1}{2i} \sum_{\alpha} \omega'^2_\mathrm{p,\alpha} \integ{p'} \;  \frac{1}{(\omega' - k' v')} \frac{\partial}{\partial p'} \Bigg\{  \frac{G'_{\alpha}}{i (\omega' - k'v')}  \Bigg\} E^{ {(1)}^2}_{1} 
\end{eqnarray*}
which finally gives 
\begin{equation}
     E^{(2)}_{2} = \frac{i}{3k'} \; \mathrm{A} \;E^{ {(1)}^2}_{1} \
\end{equation}
where
\begin{equation}
\mathrm{A} = \sum_{\alpha} \omega'^2_\mathrm{p,\alpha} \integ{p'} \;  \frac{1}{(\omega' - k'v')} \frac{\partial}{\partial p'} \Bigg\{  \frac{e_{\alpha}}{m_{\alpha}} \frac{\partial f^{(0)'}_{\alpha}}{\partial p'}\frac{1}{(\omega' - k'v')}  \Bigg\}
\end{equation}

\subsection{ {n = 3, l = 0 term} }
Substituting $(n=3,l=0)$ in MVE we get, 
\begin{eqnarray*}
    &\  v' \; \dd{2}{0}{\xi'}  \\
    &\ = - \frac{e_{\alpha}}{m_{\alpha}}  \; \left[ E^{(3)}_{0} \; \ddo + \sum^{+\infty}_{j=-\infty} \sum^{+\infty}_{m=1} \dd{m}{j}{p} \; E^{(1)}_{-j} \right] \\ 
    &\ = - \frac{e_{\alpha}}{m_{\alpha}} \; \left[ E^{(3)}_{0} \;  \ddo + \sum^{+\infty}_{j=-\infty} \Bigg\{  \dd{1}{j}{p'} \; E^{(2)}_{-j} + \dd{2}{j}{p'} \;  E^{(1)}_{-j} \Bigg \}  \right] \\
    &\ =   - \frac{e_{\alpha}}{m_{\alpha}} \;\left[ E^{(3)}_{0} \; \ddo + E^{(2)}_{-1} \; \dd{1}{1}{p'} + E^{(2)}_{1} \; \dd{1}{-1}{p'} + E^{(1)}_{-1} \; \dd{2}{1}{p} + E^{(1)}_{1} \dd{2}{-1}{p'} \right] 
\end{eqnarray*}
which finally gives 
\begin{equation}
    v' \; \dd{2}{0}{\xi'} =   - \frac{e_{\alpha}}{m_{\alpha}} \;  \left[ E^{(3)}_{0} \; \ddo + E^{(2)}_{-1} \; \dd{1}{1}{p'} + E^{(2) \star}_{-1} \; \dd{1}{-1}{p'} + E^{(1)}_{-1} \; \dd{2}{1}{p'} + E^{(1) \star}_{-1} \dd{2}{-1}{p'} \right] \label{n3l0_expr1}
\end{equation}

From equation (\ref{n2l1_expr1}) we have,

\begin{equation}
    f^{(2)'}_{\alpha,1} =  \frac{1}{i(\omega' - k'v')} \left[  G'_{\alpha} E^{(2)}_{1} + v' \dd{1}{1}{\xi'} \right] =  \frac{G'_{\alpha}}{i(\omega' - k'v')} \; E^{(2)}_{1} + \frac{v'}{i(\omega' - k'v')}\;  \dd{1}{1}{\xi'}  
\end{equation}
which gives
\begin{eqnarray}
    \label{n3l0_expr2}
    &\ E^{(1)}_{-1} \; \dd{2}{1}{p'}  =  \frac{|E^{(1)}_{1}|^2} {E^{{(1)}^2}_{1}} \; E^{(2)}_{1} \; \dd{1}{1}{p'} +  E^{(1)}_{-1} \;\frac{\partial }{\partial p'} \left[ \frac{v'}{i(\omega' - k'v')}\;  \dd{1}{1}{\xi'} \right] \\ \label{n3l0_expr3}
    &\  E^{(1)}_{1} \; \dd{2}{1}{p'}  =  \frac{|E^{(1)}_{1}|^2} {E^{{(1) }^2}_{-1}} \; E^{(2)}_{-1} \; \dd{1}{-1}{p'} -  E^{(1)}_{1} \;\frac{\partial }{\partial p'} \left[ \frac{v'}{i(\omega' - k'v')} \;  \dd{1}{-1}{\xi'}  \right]
\end{eqnarray}

Using equation (\ref{n3l0_expr2}) and (\ref{n3l0_expr3}) in equation (\ref{n3l0_expr1})
\begin{equation} \label{n3l0_expr4}
\begin{split}
    &  v' \; \dd{2}{0}{\xi'}  = - \frac{e_{\alpha}}{m_{\alpha}} \;\left[ E^{(3)}_{0} \; \ddo  + \Bigg \{ \frac{|E^{(1)}_{1}|^2} {E^{{(1)}^2}_{1}} \; E^{(2)}_{1} + E^{(2)}_{-1} \Bigg \} \dd{1}{1}{p'} + \Bigg \{ \frac{|E^{(1)}_{1}|^2} {E^{{(1)}^2}_{1}} \; E^{(2)}_{-1} + E^{(2)}_{1}  \Bigg \} \dd{1}{-1}{p'}  \right] \\
    & - \frac{e_{\alpha}}{m_{\alpha}} E^{(1)}_{-1} \frac{\partial }{\partial p'} \left[ \frac{v'}{i(\omega' - k'v')}\;  \dd{1}{1}{\xi'} \right] + \frac{e_{\alpha}}{m_{\alpha}} E^{(1)}_{1} \frac{\partial }{\partial p'}  \left[ \frac{v'}{i(\omega' - k'v')} \;  \dd{1}{-1}{\xi'}  \right] 
\end{split}    
\end{equation}

Next we simplify 
\begin{eqnarray*}
      &\ \Bigg \{ \frac{|E^{(1)}_{1}|^2} {E^{{(1)}^2}_{1}} \; E^{(2)}_{1} + E^{(2)}_{-1} \Bigg \} \dd{1}{1}{p'} = \left[ E^{(2)}_{1} \; E^{(1)}_{1}  + E^{(2)}_{1} \frac{|E^{(1)}_{1}|^2}{E^{(1)}_{1}} \right] \frac{\partial}{\partial p'} \Bigg \{ \frac{G'_{\alpha}}{i (\omega' - k' v')}\Bigg \} , \\
      &\ \Bigg \{ \frac{|E^{(1)}_{1}|^2} {E^{{(1)}^2}_{1}} \; E^{(2)}_{-1} + E^{(2)}_{1}  \Bigg \} \dd{1}{-1}{p'}  = -  \left[ E^{(2)}_{1} \; E^{(1)\star}_{1}  + E^{(2)\star}_{1} \frac{|E^{(1)}_{1}|^2}{E^{(1)\star}_{1}} \right] \frac{\partial}{\partial p'} \Bigg \{ \frac{G'_{\alpha}}{i(\omega' - k' v')}\Bigg \}  
\end{eqnarray*}
which gives us 
\begin{equation}
     \Bigg \{ \frac{|E^{(1)}_{1}|^2} {E^{{(1)}^2}_{1}} \; E^{(2)}_{1} + E^{(2)}_{-1} \Bigg \} \dd{1}{1}{p'} + \Bigg \{ \frac{|E^{(1)}_{1}|^2} {E^{{(1)}^2}_{1}} \; E^{(2)}_{-1} + E^{(2)}_{1}  \Bigg \} \dd{1}{-1}{p'}  = 0 \label{n3l0_expr5}
\end{equation}

Next we simplify
\begin{equation}
 E^{(1)}_{-1} \frac{\partial X'}{\partial p'} \left[\frac{v'}{i(\omega' - k'v')} \;  \dd{1}{1}{\xi'}\right]   = - \;  \frac{\partial}{\partial p'}\Bigg \{  \frac{v' G'_{\alpha}}{(\omega' - k' v')^2}  \Bigg \}   E^{(1)\star}_{1} \; \frac{\partial E^{(1)}_{1}}{\partial \xi'} = -  \;  I'_{\alpha} \;   \Bigg \{  \frac{\partial }{\partial \xi'}|E^{(1)}_{1}|^2 - \; E^{(1)}_{1} \; \frac{\partial E^{(1)}_{1}}{\partial \xi'}  \Bigg \} \label{n3l0_expr6}
 \end{equation}
where
 \begin{equation}
I'_{\alpha}  = \frac{\partial}{\partial p'}\Bigg \{ \frac{ v'  G'_{\alpha}}{(\omega' - k' v')^2}   \Bigg \}
 \end{equation}

Similarly we have,
\begin{equation}
    E^{(1)}_{1} \frac{\partial Y'}{\partial p'} \left[ \frac{v'}{i(\omega' - k'v')} \;  \dd{1}{-1}{\xi'} \right] =  \; I'_{\alpha} \;  E^{(1)}_{1} \frac{\partial E^{(1) \star}_{1}}{\partial \xi'} \label{n3l0_expr7}
\end{equation}

Substituting equation (\ref{n3l0_expr5}) - (\ref{n3l0_expr7}) in equation(\ref{n3l0_expr4})
\begin{equation}
 \dd{2}{0}{\xi'} = - \frac{1}{v'}   \left[ E^{(3)}_{0} \; G'_{\alpha}  - \frac{e_{\alpha}}{m_{\alpha}} I'_{\alpha} \frac{\partial}{\partial \xi'} |E^{(1)}_{1}|^2 \right] \label{n3l0_expr8} 
\end{equation}

\subsection{{n = 2, l = 0 term}}
 Substituting $(n=2,l=0)$ in MPE we have
\begin{eqnarray*}
   &\ 4\pi \sum_{\alpha}  n'_\mathrm{\alpha} \; e_{\alpha}  \integ{p'} \;  f^{(2)'}_{\alpha,0} = 0 \\
 \Rightarrow  &\ \sum_{\alpha} \omega'^2_\mathrm{p,\alpha} \integ{p'} \; \dd{2}{0}{\xi'} = 0 \\
 \Rightarrow &\  \; \Bigg \{ \sum_{\alpha} \omega'^2_\mathrm{p,\alpha}  \integ{p'} \; \frac{1}{v'} \frac{\partial f^{(0)'}_{\alpha}}{\partial p'} \Bigg \} E^{(3)}_{0} = \Bigg \{   \sum_{\alpha} \omega'^2_{p,\alpha} \integ{p'} \;  \frac{I'_{\alpha}}{v' } \Bigg \} \frac{\partial }{\partial \xi'} |E^{(1)}_{1}|^2 
\end{eqnarray*}
which gives 
\begin{equation}
     E^{(3)}_{0} =  \frac{\zeta'}{\Delta'} \frac{\partial }{\partial \xi' } |E^{(1)}_{1}|^2 \label{n2l0_expr2}
\end{equation}

Substituting equation (\ref{n2l0_expr2}) in  equation (\ref{n3l0_expr8}) we get,
\begin{equation}
     \dd{2}{0}{\xi'} = - \frac{K'}{K' v'} \left[ \frac{\zeta'}{\Delta'} G'_{\alpha} - \left( \frac{e_{\alpha}}{m_{\alpha}} I'_{\alpha} \right) \right] \frac{\partial}{\partial \xi'} |E^{(1)}_{1}|^2 
\end{equation}

Using (\ref{ampli1}) and (\ref{ampli2}) in 
\begin{equation}
    f^{(2)'}_{\alpha,0} = - \frac{1}{v'} \left[ \frac{\zeta'}{\Delta'} G'_{\alpha} - \left( \frac{e_{\alpha}}{m_{\alpha}} I'_{\alpha} \right) \right]  |E^{(1)}_{1}|^2 \label{n3l0_expr9}
\end{equation}

Here $\Delta'$ is given by 
\begin{eqnarray*}
&\ \Delta'  = \lim_{\mu \rightarrow 0} \sum_{\alpha} \omega'^2_\mathrm{p,\alpha} \integ{p'} \;  \frac{K'}{K'v' + i \mu} \frac{\partial f^{(0)'}_{\alpha}}{\partial p'}  \\ 
&\ = \sum_{\alpha} \omega'^2_\mathrm{p,\alpha} \;  \mathcal{P} \integ{p'} \;  \frac{1}{v'} \frac{\partial f^{(0)'}_{\alpha}}{\partial p'} - i \pi \sum_{\alpha} \omega'^2_\mathrm{p,\alpha} \integ{p'} \;  \frac{K'}{|K'|} \; \delta(v') \frac{\partial f^{(0)'}_{\alpha}}{\partial p'} 
\end{eqnarray*}
which gives
\begin{equation}
     \Delta'  =  \mathrm{H} - i  \; \mathrm{U} \; \mathrm{sgn} (K')
\end{equation}
where  
\begin{eqnarray}
 &\ \mathrm{sgn} (K') = \frac{K'}{|K'|} \\
 &\ \mathrm{H} = \sum_{\alpha} \omega'^2_\mathrm{p,\alpha} \;  \mathcal{P} \integ{p'} \; \; \frac{1}{v'} \frac{\partial f^{(0)'}_{\alpha}}{\partial p'} \\ 
 &\ \mathrm{U} = \pi \sum_{\alpha} \omega'^2_\mathrm{p,\alpha} \integ{p'} \; \;  \delta (v') \frac{\partial f^{(0)'}_{\alpha}}{\partial p'} 
\end{eqnarray}
where $\mathcal{P}$ stands for the Principal value integral. Here and in what follows we will assume $\mu$ to be positive following the prescription by \citealt{1974PThPS..55..212I}.

Next we have $\zeta'$ which can be expressed as 
\begin{equation}
    \zeta' = \lim_{\mu \rightarrow 0} \sum_{\alpha}  \omega'^2_\mathrm{p,\alpha} \integ{p'} \; \frac{K'}{K' v' + i \mu} \; \frac{dv'}{dp'}\; \frac{1}{(\omega' - k' v')^2}  G'_{\alpha} 
\end{equation}
which gives
\begin{equation}
    \zeta' = \mathrm{W} - i \; \mathrm{V} \; \mathrm{sgn}(K') 
\end{equation}
where
\begin{eqnarray}
 &\ \mathrm{W} \; = \; \sum_{\alpha} \omega'^2_\mathrm{p,\alpha} \;  \mathcal{P} \integ{p'} \; \frac{1}{v'} \; \frac{dv'}{dp'}\; \frac{1}{(\omega' - k' v')^2}  \Bigg \{ \frac{e_{\alpha}}{m_{\alpha}} \frac{\partial f^{(0)'}_{\alpha}}{\partial p'}\Bigg \} \\
 &\ \mathrm{V} \; = \; \pi \sum_{\alpha} \omega'^2_\mathrm{p,\alpha}  \integ{p'} \; \delta{(v')} \; \frac{dv'}{dp'}\; \frac{1}{(\omega' - k' v')^2}  \Bigg \{ \frac{e_{\alpha}}{m_{\alpha}} \frac{\partial f^{(0)'}_{\alpha}}{\partial p'}\Bigg \} 
\end{eqnarray}

\subsection{{n = 3, l = 1 term}}

Substituting $(n=3,l=1)$ in MVE we have,
\begin{equation}\label{n3l1_1} 
\begin{split}
& f^{(3)'}_{\alpha,1}  =  \frac{1}{i(\omega' - k'v')} \dd{1}{1}{\tau'} + \frac{(v')}{i(\omega' - k'v')} \dd{2}{1}{\xi'}  \\ & +  \frac{1}{i(\omega' - k'v')} \;\frac{e_{\alpha}}{m_{\alpha}} \left[  E^{(3)}_{1} \frac{\partial f^{(0)'}_{\alpha}}{\partial p'} + E^{(2)}_{2} \dd{1}{-1}{p'} + E^{(1)}_{1} \dd{2}{0}{p'} + E^{(1)}_{-1} \dd{2}{2}{p'} \right]
\end{split}
\end{equation}

From MPE we get,
\begin{equation}
     i k' \; E^{(3)}_{1} +  \frac{\partial E^{(2)}_{1}}{\partial \xi'}   = 4\pi \sum_{\alpha}  n'_\mathrm{\alpha} e_{\alpha} \integ{p'} \;  f^{(3)'}_{\alpha,1}, \label{n3l1_2}
\end{equation}    

Substituting equation (\ref{n3l1_1}) in equation (\ref{n3l1_2}) we get,
\begin{equation}
     i k' \; E^{(3)}_{1} +  \frac{\partial E^{(2)}_{1}}{\partial \xi'}  = I_{1} + I_{2} + I_{3} + I_{4} + I_{5} + I_{6}  , \label{n3l1_3}
\end{equation}

\subsection*{Simplification of Integrals ($I_{1} - I_{6}$)}

We can simplify $I_{1}$ as 
\begin{equation*}
     I_{1}   = E^{(3)}_{1} \; \sum_{\alpha} \omega'^2_\mathrm{p,\alpha} \integ{p'} \frac{1}{i(\omega' - k'v')}      \frac{\partial f^{(0)'}_{\alpha}}{\partial p'} = i k'  \; E^{(3)}_{1}, 
\end{equation*}

We can simplify $I_{2}$ using 
\begin{equation}
   I_{2} =   4 \pi \sum_{\alpha}  n'_\mathrm{\alpha} e_{\alpha} \integ{p'} \;  \frac{1}{i(\omega' - k'v')}  \dd{1}{1}{\tau'} =  \left( \frac{2k'}{ \omega'} \right) \; \frac{\partial E^{(1)}_{1}}{\partial \tau'},  
\end{equation}

We can simplify $I_{3}$ as
\begin{equation}
  I_{3}   = 4 \pi \sum_{\alpha}  n'_\mathrm{\alpha} e_{\alpha} \integ{p'} \;  \frac{(v' - v'_\mathrm{gr}) }{i(\omega' - k'v')}  \dd{2}{1}{\xi'}  = \frac{\partial E^{(2)}_{1}}{\partial \xi'} - i \; \frac{1}{2} \frac{d^2 \omega'}{d k'^2} \left( \frac{2 k'}{ \omega'} \right) \frac{\partial^2 E^{(1)}_{1}}{\partial \xi'^2},  
\end{equation}

We can simplify $I_{4}$ as 
\begin{equation}
    I_{4}  = \sum_{\alpha} \omega'^2_\mathrm{p,\alpha}  \integ{p'} \;  \frac{1}{i (\omega' - k'v')}  \Bigg \{  E^{(2)}_{2} \dd{1}{-1}{p'}  \Bigg \}   =  \frac{ i \mathrm{A}^2}{3 k'} |E^{(1)}_{1}|^2  E^{ {(1)}}_{1} , 
\end{equation}

We can simplify $I_{5}$ as 
\begin{equation}
 I_{5}  =  \sum_{\alpha} \omega'^2_\mathrm{p,\alpha} \integ{p'} \; \frac{1}{i (\omega' - k' v')}  \Bigg \{  E^{(1) \star}_{1} \dd{2}{2}{p'}  \Bigg \}  =  -i \; \left( \frac{ \mathrm{A}^2}{6 k'}  - \frac{\mathrm{B} }{2} \right) |E^{(1)}_{1}|^2  E^{ {(1)}}_{1},  
\end{equation}
 where
\begin{equation}
\mathrm{B}  = \sum_{\alpha} \omega'^2_\mathrm{p,\alpha}  \integ{p'}  \frac{1}{(\omega' - k' v')} \; \left(\frac{e_{\alpha}}{m_{\alpha}}\right)^2  \frac{\partial}{\partial p'} \left[ \frac{1}{(\omega' - k' v')} \frac{\partial}{\partial p'} \Bigg \{ \frac{\partial f^{(0)'}_{\alpha} }{\partial p'}  \frac{1}{(\omega' - k'v')} \Bigg \} \right]
\end{equation}

We can simplify $I_{6}$ as 
\begin{eqnarray*}
          I_{6} =    &\ \sum_{\alpha} \omega'^2_\mathrm{p,\alpha} \integ{p'} \;  \frac{1}{(\omega' - k'v')}  \dd{2}{0}{p'}  \\
    &\ =  k' \left(  \frac{\zeta'^2}{\Delta'} - \Lambda' \right) |E^{(1)}_{1}|^2  \\
    &\ = k' \left( \Bigg \{ \frac{  \mathrm{  H (W^2 - V^2) + 2 WVU }  }{ \mathrm{ H^2 + U^2 }  } + \mathrm{C} \Bigg \} \; + \; i \;   \Bigg \{  \frac{ \mathrm{U(W^2 - V^2) - 2WVH }    }{  \mathrm{H^2 + U^2} } + \mathrm{D} \Bigg \} \;  \mathrm{sgn}(K') \right)  |E^{(1)}_{1}|^2  \\
    &\ = k' \; \Theta |E^{(1)}_{1}|^2   + i \; k' \; \Phi \; \Bigg \{ \mathrm{sgn}(K')   |E^{(1)}_{1}|^2 \Bigg \}  \\ 
     &\ = k' \; \Theta |E^{(1)}_{1}|^2   + i \; k' \; \Phi \; \Bigg \{ \mathrm{sgn}(K')  \Bigg \}  |E^{(1)}_{1}|^2  \\ 
    &\ = k' \; \Theta |E^{(1)}_{1}|^2   + i \; k' \; \Phi \; \Bigg \{ \integ{\xi'} \; e^{-iK' \xi'} \; \frac{1}{i \pi \xi'}   \Bigg \} |E^{(1)}_{1}|^2  \\
    &\ = k' \; \Theta |E^{(1)}_{1}|^2  + i k' \; \Phi \;  \Bigg \{ \int^{+\infty}_{-\infty} \; d\xi'\; e^{-iK' \xi'} \; |E^{(1)}_{1}(\xi',\tau')|^2 \; \times \;  \frac{1}{i \pi \xi'} \Bigg \} \\
    &\ =  k' \; \Theta |E^{(1)}_{1}|^2 \; +  k' \; \Phi \;  
\frac{1}{\pi} \left( |E^{(1)}_{1}(\xi', \tau')|^2 \circledast \frac{1}{\xi'} \right)    \\
    &\ =  k' \; \Theta |E^{(1)}_{1}|^2  +  k' \; \Phi \; 
\frac{1}{\pi} \Bigg \{ \mathcal{P} \int^{+\infty}_{-\infty} \; d\xi'' \;  \frac{|E^{1}_{1}(\xi'', \tau')|^2}{\xi' - \xi'' } \Bigg \}  
\end{eqnarray*}
where we have, 
\begin{eqnarray}
     &\ \Theta'(\omega', k') = \Bigg \{ \frac{  \mathrm{  H ( W^2 - V^2 ) + 2 WVU }  }{ \mathrm{ H^2 + U^2 }  } + \mathrm{C} \Bigg \}   \\ 
 &\ \Phi'(\omega',k') =  \Bigg \{  \frac{ \mathrm{U( W^2 - V^2 ) - 2WVH }    }{  \mathrm{H^2 + U^2} } + \mathrm{D} \Bigg \}  
\end{eqnarray}

Substituting $I_1$ to $I_6$ in equation (\ref{n3l1_1})  we have 
\begin{equation}
- \left( \frac{2 k'}{ \omega' } \right)  i \;  \frac{\partial E^{(1)}_{1}}{\partial \tau'} -  \frac{1}{2} \frac{d^2 \omega'}{d k'^2} \left( \frac{2 k'}{\omega'} \right) \frac{\partial^2 E^{(1)}_{1}}{\partial \xi'^2} + \left[ \left( \frac{ \mathrm{A}^2}{6k'} + \frac{ \mathrm{B}}{2}\right) - k' \; \Theta \right]  \; E^{(1)}_{1} \;  |E^{(1)}_{1}|^2   -  k' \; \Phi \; \frac{1}{\pi} \mathcal{P} \int^{+\infty}_{-\infty} d\xi'' \; \frac{|E^{(1)}_{1}(\xi'',\tau)|^2}{\xi' - \xi''}  \; E^{(1)}_{1}  = 0
\end{equation}

Rearranging the terms in equation  gives us, 
\begin{equation}
  i \;  \frac{\partial E^{(1)}_{1}}{\partial \tau'} +   G \; \frac{\partial^2 E^{(1)}_{1}}{\partial \xi'^2} 
  + q \; E^{(1)}_{1} \;  |E^{(1)}_{1}|^2  +  s \; \frac{1}{\pi} \mathcal{P} \int^{+\infty}_{-\infty} d\xi'' \; \frac{|E^{(1)}_{1}(\xi'',\tau)|^2}{\xi' - \xi''}  \; E^{(1)}_{1}  = 0  \label{NLSE2} 
\end{equation}
where the equation is written in MFR while the coefficients are estimated in PFR. 
The integrals being scalars can be estimated in the PFR via Lorentz transformation from MFR. Please note the derivation was done using $c=1$, we have reintroduced $c$ in the appropriate places below. This has been done by identifying $c$ always appears as $m_{\alpha} c$.  
\begin{eqnarray}
  &\ G =  \gamma^3_\mathrm{gr} \; \frac{1}{2} \; \frac{d^2 \omega}{d k^2}  \\ 
&\ q  = - \frac{1}{2} \left( \frac{\omega - k v_\mathrm{gr} } {k} \right)  \; \left[ \left( \frac{ \mathrm{A}^2}{6k} + \frac{ \mathrm{B}}{2}\right) - k  \; \Theta \right]  \\
&\ s =   \frac{\omega - k v_\mathrm{gr}}{2}   \;  \Phi \\  
&\ \mathrm{A} = \frac{1}{c} \; \sum_{\alpha} \omega^2_\mathrm{p,\alpha} \integ{p} \;  \frac{1}{(\omega - kv)} \frac{\partial}{\partial p} \Bigg\{ \left( \frac{e_{\alpha}}{m_{\alpha} c} \right) \frac{\partial f^{(0)}_{\alpha}}{\partial p}\frac{1}{(\omega - kv)}  \Bigg\} \\ 
&\ \hspace*{-1cm} \mathrm{B} = \frac{1}{c} \; \sum_{\alpha} \omega^2_\mathrm{p,\alpha}  \integ{p} \; \frac{1}{(\omega - kv)} \left( \frac{e_{\alpha}}{m_{\alpha} c} \right) \frac{\partial}{\partial p} \left[ \frac{1}{(\omega - kv)} \frac{\partial}{\partial p} \Bigg \{ \left( \frac{e_{\alpha}}{m_{\alpha} c} \right) \frac{\partial f^{(0)}_{\alpha} }{\partial p}  \frac{1}{(\omega - kv)} \Bigg \} \right] \\ 
 &\ \Theta = \Bigg \{ \frac{  \mathrm{  H ( W^2 - V^2 ) + 2 WVU }  }{ \mathrm{ H^2 + U^2 }  } + \mathrm{C} \Bigg \}  \\
 &\ \Phi = \Bigg \{  \frac{ \mathrm{U(W^2 - V^2 ) - 2WVH }    }{  \mathrm{H^2 + U^2} } + \mathrm{D} \Bigg \} \\
 &\ \mathrm{C} = - \frac{1}{c}\; \sum_{\alpha} \omega^2_\mathrm{p,\alpha}  \; \mathcal{P} \integ{p}  \left( \frac{e_{\alpha}}{m_{\alpha} c} \right)^2 \frac{1}{(\omega - kv)^2} \;  \frac{dv}{dp} \; \frac{1}{v - v_\mathrm{gr}} \;  \frac{\partial}{\partial p}\Bigg \{ \frac{ ( v - v_\mathrm{gr} )   }{(\omega - k v)^2} \frac{\partial f^{(0)}_{\alpha}}{\partial p} \Bigg \} \\  
 &\ \mathrm{D} = \pi \; \frac{1}{c}\sum_{\alpha} \omega^2_\mathrm{p,\alpha}  \integ{p}  \left( \frac{e_{\alpha}}{m_{\alpha} c} \right)^2 \frac{1}{(\omega - k v)^2} \;  \frac{dv}{dp} \; \delta(v - v_\mathrm{gr})  \frac{\partial}{\partial p}\Bigg \{ \frac{v - v_\mathrm{gr}}{(\omega - k v)^2} \frac{\partial f^{(0)}_{\alpha}}{\partial p} \Bigg \} \\
 &\ \mathrm{W} \; = \; \frac{1}{c} \; \sum_{\alpha} \omega^2_\mathrm{p,\alpha} \;  \mathcal{P} \integ{p} \; \frac{1}{v - v_\mathrm{gr}} \; \frac{dv}{dp}\; \frac{1}{(\omega - k v)^2}  \Bigg \{ \frac{e_{\alpha}}{m_{\alpha} c} \frac{\partial f^{(0)}_{\alpha}}{\partial p}\Bigg \} \\
 &\ \mathrm{V} \; = \; \pi \; \frac{1}{c} \; \sum_{\alpha} \omega^2_\mathrm{p,\alpha}  \integ{p} \; \delta{(v - v_\mathrm{gr})} \; \frac{dv}{dp}\; \frac{1}{(\omega - k v)^2}  \Bigg \{ \frac{e_{\alpha}}{m_{\alpha} c} \frac{\partial f^{(0)}_{\alpha}}{\partial p}\Bigg \}  \\ 
  &\ \mathrm{H} = \frac{1}{c} \; \sum_{\alpha} \omega^2_\mathrm{p,\alpha} \;  \mathcal{P} \integ{p} \; \; \frac{1}{v - v_\mathrm{gr}} \frac{\partial f^{(0)}_{\alpha}}{\partial p} \\ 
 &\ \mathrm{U} = \pi \; \frac{1}{c} \; \sum_{\alpha} \; \omega^2_\mathrm{p,\alpha} \integ{p} \; \;  \delta (v - v_\mathrm{gr}) \frac{\partial f^{(0)}_{\alpha}}{\partial p}   
\end{eqnarray}

Note  that the expressions for the coefficients are consistent with that of MP80, PM80, MP84 and MGP00 for an electron-positron plasma.
It must be mentioned here that \citet{1973JPSJ...34..513I} used a similar scheme for deriving the NLSE in non-relativistic plasmas. However, their scheme was based on Galilean transformation and is not applicable for ultra-relativistic plasmas as is the case for pulsar plasma. 
It must also be mentioned that an alternative mathematical scheme could have been taken where the MFR would move with arbitrary velocity $\lambda$ with respect to PFR , and both the slow space and time variables retained upto first and second order in $\epsilon$. In that case we would obtain NLSE with additional terms. However, when $\lambda$ is identified with group velocity of Langmuir waves in PFR, then the additional terms vanishes and we recover NLSE with NLD.

\section{Estimating NLSE coefficients}

The linear Langmuir dispersion relation is given by
\begin{equation}
 k  + \sum_{\alpha} \omega^2_\mathrm{p,\alpha} \integ{p} \; \frac{\partial f^{(0)}_{\alpha}}{\partial p} \frac{1}{(\omega - k v)^2} = 0, \label{disp reln}  
\end{equation}
Differentiating the above expression with respect to the wave number $k$ we get,
\begin{equation}
1 -  \sum_{\alpha} \omega^2_\mathrm{p \alpha}\int^{+\infty}_{-\infty} dp \; \frac{\partial f^{(0)}_{\alpha}}{\partial p} \frac{1}{(\omega -  k v)^2} \left [ \frac{d \omega}{d k} - v \right] = 0,  
\end{equation}
where $d \omega / dk $ is the group velocity ($v_\mathrm{gr}$) of the Langmuir waves. Rearranging the equation above we get,
\begin{equation}
    \beta_\mathrm{gr} = \frac{1}{c} \frac{d \omega}{d k} = \frac{1 + \sum_{\alpha} \left( \frac{\omega_\mathrm{p,\alpha}}{kc} \right)^2 \integ{p} \; \frac{\partial f^{(0)}_{\alpha}}{\partial p} \frac{\beta}{ (\beta_\mathrm{ph} - \beta)^2}  }{ \sum_{\alpha} \left( \frac{\omega_\mathrm{p,\alpha}}{k c}\right)^2 \integ{p} \; \frac{\partial f^{(0)}_{\alpha}}{\partial p} \frac{1}{(\beta_\mathrm{ph} - \beta)^2} }
\end{equation}
where equation (\ref{disp reln}) can be re-expressed by performing integration by parts to give the wave number $kc$ as, 
\begin{equation}
    k^2 c^2 = \sum_{\alpha} \omega^2_\mathrm{p,\alpha} \integ{p} \;  f^{(0)}_{\alpha} \frac{1}{\gamma^3 \; (\beta_\mathrm{ph} - \beta)^2},
\end{equation}
The momentum pole corresponding to the group velocity is given by
\begin{equation}
    p_\mathrm{gr} = \gamma_\mathrm{gr} \beta_\mathrm{gr} = \frac{\beta_\mathrm{gr}}{\sqrt{1 - \beta^2_\mathrm{gr}}} , \label{pole2} 
\end{equation}
 
 \subsection{Classification of integrals}

All the integrals are written in the plasma frame of reference (PFR). 

The integrals of the NLSE can be divided into three classes (as shown in Table \ref{class_coeff})
\begin{itemize}
    \item Regular integrals with no poles
    \item Interpolation of delta-function integrals at the group velocity pole
    \item Principal value integrals with group velocity pole
\end{itemize}
The integrals can again be sub-divided into the following categories
\begin{itemize}
    \item Charge-dependent on the plasma particle species
    \item Charge independent of the plasma particle species
\end{itemize}

The integrals can be represented into their dimensionless as shown in Table \ref{dim_coeff}. Using the dimensionless form of the integrals, the coefficients of NLSE ($G,q,s$) can also be represented in their dimensionless form as shown below. 

The group velocity dispersion term $G$ is given by
\begin{equation}
    G  = \gamma^3_\mathrm{o} \; \frac{1}{2} \frac{d^2 \omega}{dk^2} 
     =  \frac{c^2}{\omega_\mathrm{p}} \; \mathrm{G}_\mathrm{d}
\end{equation}
where
\begin{equation}
    \mathrm{G}_\mathrm{d} = \frac{1}{2} \gamma^3_\mathrm{o} \; \Bigg\{ (\beta_\mathrm{gr} - \beta_\mathrm{ph} ) \left( \frac{\omega_\mathrm{p}}{kc}\right)^3 \sum_{\alpha} \; \chi_{\alpha} \;\varphi_{\alpha}  \integ{p} \;  \frac{\partial f^{(0)}_{\alpha} }{\partial p} \frac{(\beta_\mathrm{gr} - \beta)^2}{(\beta_\mathrm{ph} - \beta)^3} \Bigg \} 
\end{equation}

The non-linear co-efficient $q$ is given by
\begin{equation}
    q = \frac{1}{\omega_\mathrm{p}} \left( \frac{e}{m_{e} c}\right)^2   \mathrm{q}_\mathrm{d}, \label{q} 
\end{equation}
where
\begin{eqnarray}
    &\ \mathrm{q}_\mathrm{d}  = - \frac{1}{2} \;  (\beta_\mathrm{ph} - \beta_\mathrm{gr} ) \Bigg \{ \left[ \left( \frac{\omega_\mathrm{p}}{kc} \right) \frac{\mathrm{A}^2_\mathrm{d}}{6} + \mathrm{B}_\mathrm{d}\right] - \left( \frac{kc}{\omega_\mathrm{p}}\right) \Theta_\mathrm{d} \Bigg \} \\ 
    &\ \Theta_\mathrm{d} \; = \; \frac{\mathrm{H}_\mathrm{d} (\mathrm{W}^2_\mathrm{d} - \mathrm{V}^2_\mathrm{d}) + 2 \mathrm{W}_\mathrm{d} \mathrm{V}_\mathrm{d} \mathrm{U}_\mathrm{d}}{\mathrm{H}^2_\mathrm{d} + \mathrm{U}^2_\mathrm{d}} + \mathrm{C}_\mathrm{d} 
\end{eqnarray}

The non-linear Landau damping co-efficient $s$ is given by
\begin{equation}
    s  = \frac{1}{\omega_\mathrm{p}} \left( \frac{e}{m_{e} c} \right)^2 \; \mathrm{s}_\mathrm{d} , \label{s} 
\end{equation}
where
\begin{eqnarray}
 &\  \mathrm{s}_\mathrm{d}  = \frac{1}{2} \left( \beta_\mathrm{ph} - \beta_\mathrm{gr} \right) \left( \frac{kc}{\omega_\mathrm{p}}\right) \Phi_\mathrm{d} \\ 
 &\   \Phi_\mathrm{d} = \frac{ \mathrm{U}_\mathrm{d} ( \mathrm{W}^2_\mathrm{d} - \mathrm{V}^2_\mathrm{d}) - 2 \mathrm{W}_\mathrm{d} \mathrm{V}_\mathrm{d} \mathrm{H}_\mathrm{d}}{ \mathrm{H}^2_\mathrm{d} + \mathrm{U}^2_\mathrm{d}} + \mathrm{D}_\mathrm{d} 
\end{eqnarray}

From equation (\ref{q}) and equation (\ref{s})  we have,
\begin{equation}
    \frac{s}{q} = \frac{s_\mathrm{d}}{q_\mathrm{d}} = - \frac {\left( \frac{kc}{\omega_\mathrm{p}}\right) \Phi_\mathrm{d} }{\left[ \left( \frac{\omega_\mathrm{p}}{kc} \right) \frac{\mathrm{A}^2_\mathrm{d}}{6} + \mathrm{B}_\mathrm{d}\right] - \left( \frac{kc}{\omega_\mathrm{p}}\right) \Theta_\mathrm{d}} ,
\end{equation}
It must be noted that the group velocity $v_\mathrm{gr}$ contribution to $s/q$ is due to  $\Phi_\mathrm{d}$ and $\Theta_\mathrm{d}$ (see Table \ref{class_coeff}).

\subsection{Computing Cauchy Principal Integral due to group velocity pole}

The integrals for the co-efficients of NLSE expressed as equation (\ref{NLSE}) are a function of the particle distribution functions , the number density of the plasma particles and the linear wave dispersion relation of the subluminal Langmuir waves ($\omega,k$) in PFR. The $v$ and $v_\mathrm{gr}$ represent the velocity of the plasma particles and the group velocity ($v_\mathrm{gr}$) of the group velocity of linear Langmuir waves respectively.The integrals which two Cherenkov terms viz., a phase velocity Cherenkov term ($\omega - k v$) and the group velocity Cherenkov term ($\omega - k v_\mathrm{gr}$). In our present study, we assume that the phase velocity of the linear Langmuir waves exceeds the particle velocities ($\omega / k > v$) such that the only pole in the integrals is due to group velocity ($v_\mathrm{gr} = d \omega/ dk$) of the linear Langmuir waves. To ensure this the momentum corresponding to wave phase velocity ($p_\mathrm{ph}$) must be taken much farther from the mean momenta ($p_{\alpha}$) of the particle distribution function of the $\alpha-$th species. The wave phase velocity (notmalized to c)  is $\beta_\mathrm{ph} = p_\mathrm{ph}/ \sqrt{1^2 + p_\mathrm{ph}}$.

Consider a function $f(p)$ with  a pole at  $p = p_\mathrm{gr}$. Consider the Cauchy Principal Value Integral of $f(p)$ given by
\begin{align*}
    I &\  = \mathcal{P} \int^{p_\mathrm{h}}_{p_\mathrm{l}} dp \; f(p) \\
      &\ = \lim _{\epsilon \rightarrow 0} \left[ \int^{p_\mathrm{gr} - \epsilon}_{p_\mathrm{l}} dp \; f(p) + \mathrm{Re} (I_\mathrm{o}) +    \int^{p_\mathrm{h}}_{p_\mathrm{gr} - \epsilon} dp \; f(p)  \right] \\
      &\ =  I_{-} + I_\mathrm{pole} + I_{+}   \numberthis 
\end{align*}

Next we consider a semi-circular contour centred at $p_\mathrm{gr}$ and of radius  $\epsilon$ in the complex $\tilde{p}$ plane
\begin{align*}
&\ \tilde{p} = p_\mathrm{gr} + \epsilon \exp{(i \theta)} \\
&\ d\tilde{p} = i \epsilon \exp{(i \theta)} \; d\theta \\ 
\end{align*}

Then we have the integral
\begin{align*}
    I_\mathrm{o} = \oint^{p_\mathrm{gr} + \epsilon}_{p_\mathrm{gr} - \epsilon} d\tilde{p} \; f(\tilde{p}) = i \epsilon \int^{\theta = 0}_{\theta = \pi} d\theta \; e^{i\theta} f(p_\mathrm{gr} + \epsilon \; e^{i\theta} )
\end{align*}

Finally we have
\begin{align*}
    I_\mathrm{pole} &\ = \mathrm{Re} \left[ \lim_{\epsilon \rightarrow 0} I_\mathrm{o} \right] \\
    &\ = \mathrm{Re} \left[ \lim_{\epsilon \rightarrow 0} i \epsilon \int^{\theta = 0}_{\theta = \pi} d\theta \; e^{i\theta} f(p_\mathrm{gr} + \epsilon \; e^{i\theta} ) \right]
\end{align*}

\subsection{Introduction of $\alpha-$th species}

\begin{table}
    \centering
    \caption{For $\alpha$-th species in the plasma, $\varphi_\mathrm{\alpha}$ is the ratio of charge to mass in units of ($e/m_{e}$)   and $\chi_{\alpha}$ is the ratio of number density of the  plasma to the number density of pair plasma. The subscript $\pm$ stands for positrons and electrons in the pair plasma respectively. The subscript $\mathrm{ion}$ stands for the ion contribution.}
    \begin{tabular}{|c|c|} \hline
          Quantity  & Value \\\hline 
         $\varphi_{+}$ & 1  \\
         $\varphi_{-}$ & 1  \\
         $\varphi_\mathrm{ion}$ & 1.42$\times $ 10$^{-2}$ \\
         $\chi_{+}$ & 1 \\
         $\chi_{-}$ & 1  \\
         $\chi_\mathrm{ion}$ & 10$^{-4}$ \\ \hline 
    \end{tabular}
    \label{ions}
\end{table}

We introduce the following notation
\begin{eqnarray}
    &\ \varphi_{\alpha} = Z_{\alpha} \times \frac{m_{e}}{m_{\alpha}}, \label{varphi}\\
    &\ \chi_{\alpha} = \frac{Z_{\alpha}}{A_{\alpha}} \times \frac{n_{\alpha}}{n_\mathrm{p}}, \label{chiphi}\\
    &\ \frac{\omega^2_\mathrm{p,\alpha}}{\omega^2_\mathrm{p}} = \chi_{\alpha}\; \varphi_{\alpha} \label{plasma_freq2},   
\end{eqnarray}
where
\begin{eqnarray}
     &  Z_{\alpha} = \Bigg | \frac{e_{\alpha}}{e} \Bigg | ,   \\ 
     & \omega^2_\mathrm{p} = \frac{4 \pi \; n_\mathrm{s} \; e^2}{m_{e}}, 
\end{eqnarray}

For ions we have,
\begin{equation}
     \omega^2_\mathrm{ion} = \Bigg \{ \frac{Z^2_\mathrm{ion}}{A_\mathrm{ion}} \times \frac{m_\mathrm{e}}{m_\mathrm{p}} \times \frac{n_\mathrm{ion}}{n_\mathrm{s}}  \Bigg \} \; \omega^2_\mathrm{p} = \varphi_\mathrm{ion} \; \chi_\mathrm{ion} \;  \omega^2_\mathrm{p}, \numberthis
\end{equation}
For iron ions we have,
\begin{eqnarray*}
    &\ Z_\mathrm{ion} = 26 \\
    &\ A_\mathrm{ion} = 56 \\
    &\ \frac{m_{p}}{m_{e}} = 1836 \\
    &\ \frac{n_\mathrm{ion}}{n_\mathrm{s}} = \frac{1}{\kappa} \approx 10^{-4}
\end{eqnarray*}
Typical estimates for quantities defined in equation (\ref{varphi}) and equation (\ref{chiphi}) are presented in Table \ref{ions}.

\subsection{Setup for maximizing the contributions due to ions}

\begin{figure}[h]
    \centering
    \includegraphics[scale=0.5]{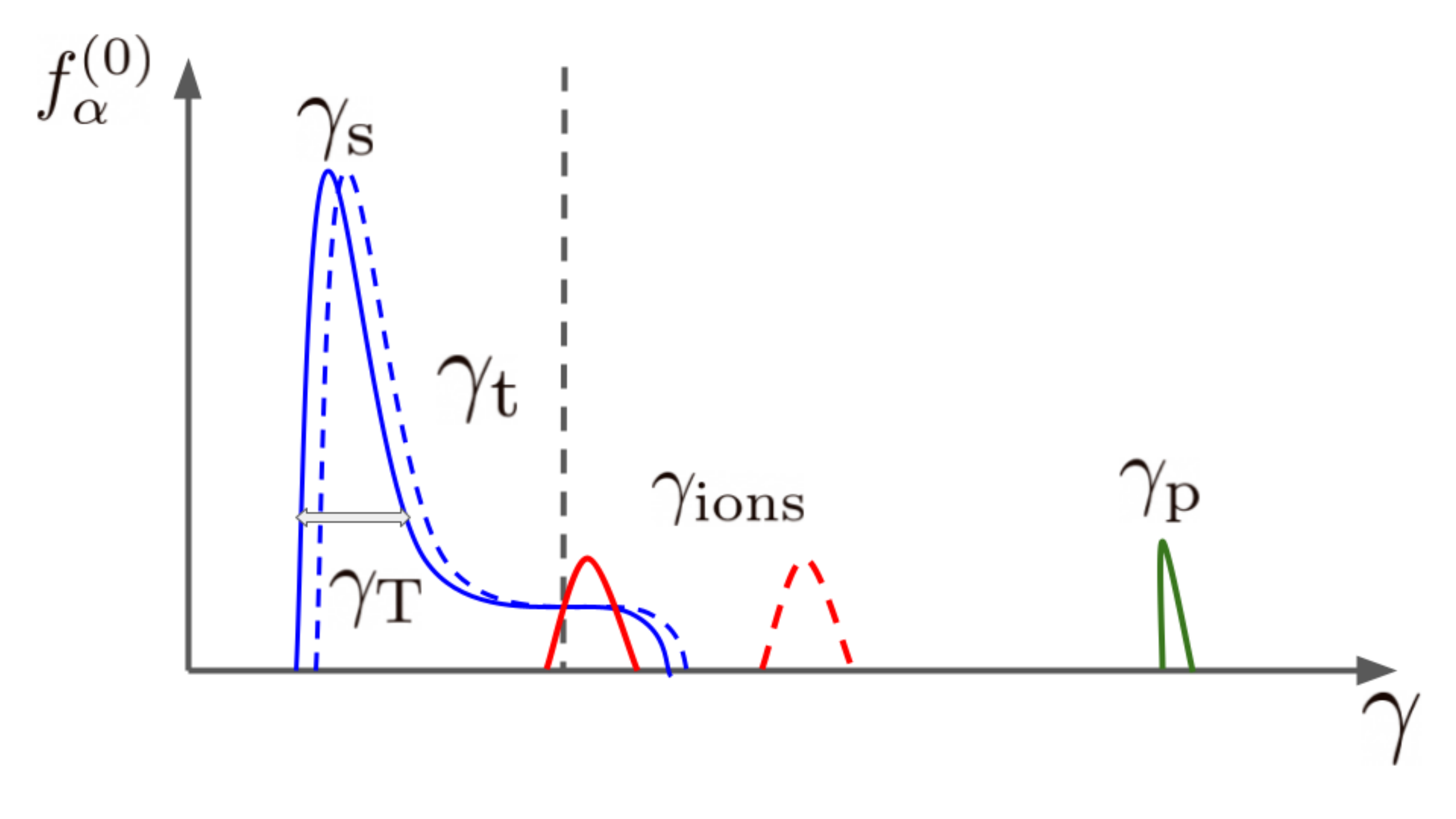}
    \caption{ A schematic of the particle distribution function and the pole of the group velocity (shown as horizontal dashed line). Only particles in the immediate neighbourhood of the pole due to group velocity can contribute to resonant interaction that characterizes the dimensionless coefficients $s_\mathrm{d}$ and $q_\mathrm{d}$ respectively. The electron and positron DF of the pair plasma is shown in solid blue and dashed line respectively. The DF of the positron beam (shown in solid green) is far-removed from the pole. The DF due to iron can contribute only if located near the pole (as shown in solid red) and will fail do so if away from the pole (as shown in dashed red). It must be noted that the peaks of the electron and positron DF are separated due to plasma outflow along curved magnetic field lines as shown in Appendix \ref{long_drift}.   }
    \label{Distribution function}
\end{figure}

The integrals as shown in Table \ref{class_coeff} depend on the derivative of the particle distribution function around the group velocity pole as shown in Fig. \ref{Distribution function}. 
Ions have an unique advantage in that location of the ion distribution function is close to the electron and positron distribution function. In order to maximise the contribution of an ion distribution function to the integrals presented in Table \ref{dim_coeff}, , the peak of the ion distribution function ($\bar{p}_\mathrm{ion}$) (for both Lorentzian and Gaussian distribution function) is chosen to be at 
\begin{eqnarray}
    & \bar{p}_{\alpha} = \sigma \pm p_\mathrm{gr}, \hspace{2cm} \text{For Gaussian DF}\\ 
    & \bar{p}_{\alpha} = \Delta p \pm p_\mathrm{gr}, \hspace{2cm} \text{For Lorentzian DF}
\end{eqnarray}
so to maximize the derivative of the ion distribution function at $p_\mathrm{gr}$.

We find that the contribution of ions in modifying the dimensionless coefficients is very small $(\delta G_\mathrm{d}, \delta s_\mathrm{d}, \delta q_\mathrm{d}) \leq 10^{-8}$.

\begin{table*}
    \hskip-2.0cm
    \caption{Classification of integrals for the coefficients in NLSE. The contribution of the group velocity $v_\mathrm{gr}$ in the  Interpolation and the Principal value integrals must be noted. }
    \begin{tabular}{l|l|l}\hline 
        Type      & Charge-dependence & Integral     \\ \hline 
        Regular   & Yes               & $ \mathrm{A} = \frac{1}{c} \sum_{\alpha} \omega^2_\mathrm{p,\alpha} \integ{p} \;  \frac{1}{(\omega - k v)} \frac{\partial}{\partial p} \Bigg\{  \left( \frac{e_{\alpha}}{m_{\alpha} c} \right) \frac{\partial f^{(0)}_{\alpha}}{\partial p}\frac{1}{(\omega - k v)}  \Bigg\} $           \\ 
                  & No     & $\mathrm{B} = \frac{1}{c} \; \sum_{\alpha} \omega^2_\mathrm{p,\alpha}  \integ{p} \; \frac{1}{(\omega - k v)} \frac{\partial}{\partial p} \left[ \left( \frac{e_{\alpha}}{m_{\alpha} c} \right) \; \frac{1}{(\omega - k v)} \frac{\partial}{\partial p} \Bigg \{ \left( \frac{e_{\alpha}}{m_{\alpha} c} \right) \frac{\partial f^{(0)}_{\alpha} }{\partial p}  \frac{1}{(\omega - k v)} \Bigg \} \right]$            \\
                        &      &      \\  \hline 
        Interpolation       & Yes     & $\mathrm{V} \; = \; \pi \; \frac{1}{c} \; \sum_{\alpha} \omega^2_\mathrm{p,\alpha}  \integ{p} \; \delta{(v - v_\mathrm{gr}  )} \; \frac{dv}{dp}\; \frac{1}{(\omega - k v)^2}  \Bigg \{ \left( \frac{e_{\alpha}}{m_{\alpha} c} \right)  \frac{\partial f^{(0)}_{\alpha}}{\partial p}\Bigg \}$           \\
                            & No      & $ \mathrm{D} = \pi \; \frac{1}{c} \sum_{\alpha} \omega^2_\mathrm{p,\alpha}  \integ{p}  \left( \frac{e_{\alpha}}{m_{\alpha} c} \right)^2 \frac{1}{(\omega - k v)^2} \;  \frac{dv}{dp} \; \delta(v - v_\mathrm{gr})  \;  \frac{\partial}{\partial p}\Bigg \{ \frac{(v - v_\mathrm{gr})}{(\omega - k v)^2} \frac{\partial f^{(0)}_{\alpha}}{\partial p} \Bigg \}$           \\ 
                            &  No  & $ \mathrm{U} = \pi \; \frac{1}{c}  \sum_{\alpha} \omega^2_\mathrm{p,\alpha} \integ{p} \; \;  \delta {\left(v - v_\mathrm{gr} \right ) } \frac{\partial f^{(0)}_{\alpha}}{\partial p} $  \\ 
                            &      &   \\ \hline 
        Principal value     & Yes     & $\mathrm{W} \; = \;\frac{1}{c} \; \sum_{\alpha} \omega^2_\mathrm{p,\alpha} \;  \mathcal{P} \integ{p} \; \frac{1}{(v - v_\mathrm{gr}  )} \; \frac{dv}{dp}\; \frac{1}{(\omega - k v)^2}  \Bigg \{ \left( \frac{e_{\alpha}}{m_{\alpha} c} \right) \frac{\partial f^{(0)}_{\alpha}}{\partial p}\Bigg \} $           \\ 
                            & No      &  $ \mathrm{C} \; = \;  - \frac{1}{c} \; \sum_{\alpha} \omega^2_\mathrm{p,\alpha}  \mathcal{P} \integ{p}  \left( \frac{e_{\alpha}}{m_{\alpha} c} \right)^2 \frac{1}{(\omega - k v)^2} \;  \frac{dv}{dp} \; \frac{1}{(v - v_\mathrm{gr} ) } \;  \frac{\partial}{\partial p}\Bigg \{ \frac{(v - v_\mathrm{gr}) }{(\omega - k v)^2} \frac{\partial f^{(0)}_{\alpha}}{\partial p} \Bigg \}$ \\
                            & No      &  $\mathrm{H} \; = \;  \frac{1}{c} \;  \sum_{\alpha} \omega^2_\mathrm{p,\alpha} \;  \mathcal{P} \integ{p} \; \; \frac{1}{( v - v_\mathrm{gr}) } \frac{\partial f^{(0)}_{\alpha}}{\partial p}$            \\ \hline  
    \end{tabular}
    \label{class_coeff}
\end{table*}

\begin{table*}
    \caption{Conversion of the integrals from dimensional form to dimensionless form. The quantities $\beta_\mathrm{ph}$, $\beta_\mathrm{gr}$ and $\beta$ are wave phase velocity, wave group velocity and particle velocity normalised to the speed of light $c$. The quantities $\varphi_{\alpha}$ and $\chi_{\alpha}$ are defined in equation (\ref{varphi}) and equation (\ref{chiphi}) respectively. In the charge dependent integrals the quantity $\mathrm{sign}_{\alpha}$ is + 1 for positrons and ions and -1 for electrons.}
    \begin{tabular}{c|c}\hline 
        Dimensional & Dimensionless\\ \hline 
        $\mathrm{A} =  \frac{1}{c} \; \left( \frac{|e|}{m_\mathrm{e} c} \right) \sum_{\alpha} \mathrm{A}_\mathrm{d,\alpha} =   \frac{1}{c} \; \left( \frac{|e|}{m_\mathrm{e} c} \right)  \mathrm{A}_\mathrm{d}$ & $ \mathrm{A}_\mathrm{d,\alpha} = \left( \frac{\omega_\mathrm{p}}{k c} \right)^2  \mathrm{sign}_{\alpha} \; \varphi^2_{\alpha} \; \chi_{\alpha} \; \integ{p} \; \frac{1}{(\beta_\mathrm{ph} - \beta)} \frac{\partial }{\partial p} \Bigg\{ \frac{\partial f^{(0)}_{\alpha}}{\partial p} \frac{1}{(\beta_\mathrm{ph} - \beta )}\Bigg\} $ \\  \\ 
        
        $\mathrm{B} =   \frac{1}{c} \; \left( \frac{e}{m_\mathrm{e} c} \right)^2 \frac{1}{\omega_\mathrm{p}} \sum_{\alpha} \mathrm{B}_\mathrm{d,\alpha} = \frac{1}{c} \; \left( \frac{e}{m_\mathrm{e} c} \right)^2 \frac{1}{\omega_\mathrm{p}}  \mathrm{B}_\mathrm{d} $   &  $\mathrm{B}_\mathrm{d,\alpha} = \left( \frac{\omega_\mathrm{p}}{k c} \right)^3   \varphi^3_{\alpha} \; \chi_{\alpha} \;  \integ{p} \; \frac{1}{(\beta_\mathrm{ph} - \beta )} \frac{\partial}{\partial p} \left[  \frac{1}{(\beta_\mathrm{ph} - \beta)} \frac{\partial }{\partial p} \Bigg\{ \frac{\partial f^{(0)}_{\alpha}}{\partial p} \frac{1}{(\beta_\mathrm{ph} - \beta )} \Bigg\} \right]$  \\ \\ \hline 
         
         $\mathrm{V} =  \frac{1}{c} \; \left( \frac{|e|}{m_{e} c} \right) \sum_{\alpha} \mathrm{V}_\mathrm{d, \alpha} = \frac{1}{c} \; \left( \frac{|e|}{m_{e} c} \right) \mathrm{V}_\mathrm{d}$       &     $ \mathrm{V}_\mathrm{d,\alpha} = \pi \; \left( \frac{\omega_\mathrm{p}}{kc} \right)^2 \mathrm{sign}_{\alpha} \; \varphi^2_{\alpha} \;\chi_{\alpha} \; \integ{p} \; \delta(p - p_\mathrm{gr}) \;  \frac{1}{(\beta_\mathrm{ph} - \beta)^2} \frac{\partial f^{(0)}_{\alpha}}{\partial p}$ \\ \\ 
         
         $\mathrm{D}  = \left( \frac{e}{m_{e} c} \right)^2 \frac{1}{\omega^2_\mathrm{p}} \;  \sum_{\alpha} \mathrm{D}_\mathrm{d,\alpha} = \left( \frac{e}{m_{e} c} \right)^2 \frac{1}{\omega^2_\mathrm{p}} \; \mathrm{D}_\mathrm{d}$  &  $\mathrm{D}_\mathrm{d,\alpha} =  \pi \; \left( \frac{\omega_\mathrm{p}}{kc} \right)^4 \; \varphi^3_{\alpha}\; \chi_{\alpha} \;    \integ{p} \; \frac{1}{(\beta_\mathrm{ph} - \beta)^2} \delta(p - p_\mathrm{gr}) \frac{\partial}{\partial p} \Bigg \{ \frac{\partial f^{(0)}_{\alpha}}{\partial p} \; \frac{\beta - \beta_\mathrm{gr}}{(\beta_\mathrm{ph} - \beta)^2} \Bigg \}$ \\ \\ 
         
         $ \mathrm{U} = \frac{\omega^2_\mathrm{p}}{c^2} \;  \sum_{\alpha} \mathrm{U}_\mathrm{d,\alpha} = \frac{\omega^2_\mathrm{p}}{c^2} \; \mathrm{U}_\mathrm{d}  $   &    $ \mathrm{U}_\mathrm{d,\alpha} =  \pi \;  \varphi_{\alpha} \; \chi_{\alpha}   \integ{p} \; \gamma^3 \; \delta{\left( p - p_\mathrm{gr} \right)} \; \frac{\partial f^{(0)}_{\alpha}}{\partial p}$     \\  \\ \hline 
         
         $\mathrm{W} = \frac{1}{c} \; \left( \frac{|e|}{m_{e} c} \right) \sum_{\alpha} \mathrm{W}_\mathrm{d,\alpha} = \frac{1}{c} \; \left( \frac{|e|}{m_{e} c} \right)  \mathrm{W}_\mathrm{d} $   &     $ \mathrm{W}_\mathrm{d,\alpha} = \left(\frac{\omega_\mathrm{p}}{kc} \right)^2 \mathrm{sign}_{\alpha} \; \varphi^2_{\alpha} \; \chi_{\alpha} \;  \mathcal{P} \integ{p} \; \frac{1}{(\beta - \beta_\mathrm{gr}  )} \; \frac{1}{\gamma^3}\; \frac{1}{(\beta_\mathrm{ph} - \beta)^2} \frac{\partial f^{(0)}_{\alpha}}{\partial p} $  \\ \\
         
         $ \mathrm{C}  =  \left( \frac{e}{m_{e} c}\right)^2 \; \frac{1}{\omega^2_\mathrm{p}} \;    \sum_{\alpha} \mathrm{C}_\mathrm{d,\alpha} = \left( \frac{e}{m_{e} c}\right)^2 \; \frac{1}{\omega^2_\mathrm{p}} \;  \mathrm{C}_\mathrm{d} $   &     $\mathrm{C}_\mathrm{d,\alpha} =  - \left( \frac{\omega_\mathrm{p}}{kc} \right)^4 \; \varphi^3_{\alpha} \; \chi_{\alpha} \;  \;  \mathcal{P} \integ{p} \; \frac{1}{(\beta_\mathrm{ph} - \beta)^2} \;  \frac{1}{\gamma^3} \; \frac{1}{(\beta - \beta_\mathrm{gr}) } \;  \frac{\partial}{\partial p}\Bigg \{ \frac{(\beta - \beta_\mathrm{gr})  }{(\beta_\mathrm{ph} - \beta )^2} \frac{\partial f^{(0)}_{\alpha}}{\partial p} \Bigg \} $  \\ \\ 
         
         $\mathrm{H}= \frac{\omega^2_\mathrm{p}} {c^2} \sum_{\alpha} \mathrm{H}_\mathrm{d,\alpha} = \frac{\omega^2_\mathrm{p}} {c^2} \mathrm{H}_\mathrm{d}$  &  $\mathrm{H}_\mathrm{d,\alpha} = \varphi_{\alpha} \; \chi_{\alpha} \; \mathcal{P} \integ{p} \; \; \frac{1}{(\beta - \beta_\mathrm{gr})} \frac{\partial f^{(0)}_{\alpha}}{\partial p}  $    \\   \\ \hline 
    \end{tabular}\label{dim_coeff}

\end{table*}

\section{Parameters for the numerical simulation}
To solve NLSE with NLD numerically we use the integrating factor and Leap frog method (IF-LF) numerical method as \cite{Lakoba2016} and \cite{2018MNRAS.480.4526L}.

 While an expanded discussion is already available in the previously mentioned works, some essential steps are reproduced below for the sake of completion.  

The dimensionless non-linear Schrodinger equation (NLSE) with non-linear Landau damping (NLD) term is
 \begin{equation}
      i \frac{\partial u}{\partial t} +   \frac{\partial^2 u}{\partial x^2 } + Q u \left( |u|^2 + \frac{s}{\pi q} \mathcal{P} \int^{+\infty}_{-\infty} dx' \frac{|u(x',t)|^2}{x-x'} \right) = 0 \label{Eqn1}
 \end{equation}

We define 
\begin{equation}
    \hat{u}(k,t) = \frac{1}{\sqrt{2\pi}} \int^{+\infty}_{-\infty} dx \; u(x,t) \; e^{-ikx}
\end{equation}
such that
\begin{equation}
    u(x,t) = \frac{1}{\sqrt{2\pi}} \int^{+\infty}_{-\infty} dx \; \hat{u}(k,t) \; e^{ikx}
\end{equation}

We can reduce equation (\ref{Eqn1}) to the form
\begin{equation}
    i u_{t} + \mathcal{L} u + \mathcal{N} = 0 \label{Eqn2}
\end{equation}
where
\begin{eqnarray}
    & \mathcal{L} = \frac{\partial^2 }{\partial x^2} \\
    & \mathcal{N} = Q u \left( |u|^2 + \frac{s}{\pi q} \mathcal{P} \int^{+\infty}_{-\infty} dx' \frac{|u(x',t)|^2}{x-x'} \right) 
\end{eqnarray}

Taking Fourier transform $\mathbf{FT}$ of equation (\ref{Eqn2}) we get, 
\begin{equation}
    i \hat{u}_{t} + \hat{\mathcal{L}} \hat{u} + \hat{\mathcal{N}} = 0 \label{Eqn3}
\end{equation}
where $\hat{\mathcal{L}} = - k^2$. 

Solution of equation (\ref{Eqn3}) from time $t_{2}$ to time $t_{1}$ has the form
 \begin{equation}
e^{-i \hat{\mathcal{L}} t_2} \; \hat{u}(t_2) - e^{-i \hat{\mathcal{L}} t_{1}} \; \hat{u}(t_1) = \int^{t_2}_{t_1} e^{-i \hat{\mathcal{L}} t'} \; i\hat{\mathcal{N}}(t') dt' \label{Eqn4}
\end{equation}
where 
\begin{eqnarray}
    & \hat{\mathcal{N}}(k,t) =  \mathbf{FT}[(\; Q |u|^2 + V(x) \;)u]    \\ 
    & \hspace{-2cm} V(x) = \chi Q \; \frac{1}{2 \pi} \int^{+\infty}_{-\infty} dk \; e^{ikx} \left[\frac{i}{\sqrt{2\pi}} \mathrm{sgn}(k) \times \int^{+\infty}_{-\infty} dx \; e^{-ikx} |u|^2\right] = \frac{\chi Q}{2} \; \mathcal{H}[|u|^2] 
\end{eqnarray}
where the the Hilbert transform $\mathcal{H}$ of $|u|^2$ is defined as
\begin{equation}
 \mathcal {H}[|u(x,t)|^2] = \frac{1}{\pi} \mathcal{P} \int^{+\infty}_{-\infty} dx' \frac{|u(x',t)|^2}{x-x'}\\
= \frac{1}{\pi} \lim_{\epsilon \rightarrow 0^{+}} \left ( \int^{x - \epsilon}_{x - 1/\epsilon} dx' \frac{|u(x',t)|^2}{x-x'} + \int^{x + 1/\epsilon}_{x + \epsilon} \frac{|u(x',t)|^2}{x-x'} \right)
\end{equation}

Next, we employ the three-point numerical scheme where
\begin{eqnarray}
    & t' = t_\mathrm{n} = n  \Delta t \\ 
    & t_{2} = t_\mathrm{n+1} = (n+1) \Delta t  \\
    & t_{1} = t_\mathrm{n-1} = (n-1) \Delta t 
\end{eqnarray}

The application of the 3-point numerical scheme reduces equation (\ref{Eqn4}) to the form
\begin{equation}
e^{i k^2 \Delta t} \;\hat{u}(t_\mathrm{n+1}) - e^{- i k^2 \Delta t} \;\hat{u}(t_\mathrm{n-1}) =  2 i \; \Delta t \; \hat{\mathcal{N}}(t_\mathrm{n}) \label{Eqn5}  
\end{equation}

Next we define, 
\begin{equation}
    \hat{v}_\mathrm{n} = \hat{u}_\mathrm{n} \; e^{i k^2\; t_n} = \hat{u}_\mathrm{n} \; e^{i k^2\; n \Delta t} \label{Eqn6}
\end{equation}

Substituting equation (\ref{Eqn6}) in equation (\ref{Eqn5}) we get
\begin{equation}
\hat{v}_{n+1} - \hat{v}_{n-1} = 2i \; \Delta t \; e^{i k^2 t_n} \; \hat{\mathcal{N}} (t_n)
\end{equation} 

For checking numerical stability we need to evaluate the following quantities after every eight time steps
\begin{eqnarray}
& \hspace{-1cm} \hat{\bar{v}}_n = a_0 \; \hat{v}_{n} + a_1 \; (\hat{v}_{n-1} + \hat{v}_{n+1}) + a_2 \; (\hat{v}_{n-2} + \hat{v}_{n+2}) + a_3 \; (\hat{v}_{n-3} + \hat{v}_{n+3}) \\
& \hspace{-1cm} \hat{\bar{v}}_{n-1} = a_0 \; \hat{v}_{n-1} + a_1 \; (\hat{v}_{n-2} + \hat{v}_{n}) + a_2 \; (\hat{v}_{n-3} + \hat{v}_{n+1}) + a_3 \; (\hat{v}_{n-4} + \hat{v}_{n+2})
\end{eqnarray}
where  $a_0 = \frac{11}{16} ; a_1 = \frac{15}{64} ; a_2 = - \frac{3}{32} ; a_3 = \frac{1}{64}$. 

After every eight time steps,  we use $\hat{\bar{v}}_n$ and $\hat{\bar{v}}_{n-1}$ to restart equation (\ref{Eqn6}) by replacing, on the l.h.s., $\hat{v}_{n-1}$ with $\hat{\bar{v}}_{n-1}$ and, on the r.h.s., $\hat{v}_{n}$  with $\hat{\bar{v}}_{n-1}$. As shown in \cite{Lakoba2016} this procedure suppresses the parasitic solution at the edges of the spectral window for long time of evolution and requires that the interval $\Delta t$ to be smaller than a critical time interval given by 
\begin{equation}
    \Delta t < \Delta t_\mathrm{crit} = \frac{\pi}{k^2_\mathrm{max}} ,
\end{equation}

In what follows we briefly describe the simulation box parameters. For a box of length $L$ (normalised to $\theta$) with $N$ grid points, the spacing interval is given by
\begin{equation}
    \Delta x = \frac{L}{N},
\end{equation}
The spacing in the Fourier domain is given by
\begin{equation}
    dk = \frac{2 \pi}{L} , 
\end{equation}
The highest value of wave number in the Fourier domain is given by
\begin{equation}
    k_\mathrm{max} = \frac{2 \pi}{2 \Delta x } = \frac{ \pi N}{L}
\end{equation}
In the IF-LF method, the parasitic solutions are suppressed by choosing the discrete time interval $\Delta t$ to be smaller than a critical time interval given by 
\begin{equation}
    \Delta t_\mathrm{crit} = \frac{\pi}{k^2_\mathrm{max}} ,
\end{equation} in order to suppress parasitic solution. 

The simulation is terminated if the power ($\log_{10}|\mathrm{F(u)}|$) at $k_\mathrm{max} \approx 300$ exceeds a threshold value ($10^{-2}$). For the numerical simulations in the main text, the simulation parameters are summarized in Table (\ref{simbox}).

\begin{table*}[]
    \centering
    \caption{Parameters of the box used for simulation. The quantity $L$ denotes the length of the simulation box, the quantity $N$ denotes, the number of grid-points, the highest wave number $k_\mathrm{max}$ in the Fourier domain, the quantity $\Delta t_\mathrm{crit} = \pi/ k^2_\mathrm{max} $ is the time interval above which parasitic solutions leads to rapid saturation of power at higher harmonics and $\Delta t < \Delta t_\mathrm{crit}$ is the time interval used for the simulation, $t_\mathrm{max}$ is the maximum run time of the simulation and  Th$_{300}$ is value of log$_{10}(F[u])$ at $k = 300$ beyond which the simulation is stopped.}
    \begin{tabular}{c|c|c|c|c|c|c}\hline 
         $L$ & $N$ & $k_\mathrm{max}$ & $\Delta t_\mathrm{crit}$ & $\Delta t$ & $t_\mathrm{max}$ & Th$_{300}$\\ \hline 
         40$\pi$ &  12288  & 308 & 3e-5 & 1e-5 & 100 & 1e-2 \\  \hline  
    \end{tabular}
    \label{simbox}
\end{table*}

\section{Impact of the initial injected spectrum on soliton formation} \label{sol form}

\begin{figure*}[h]
\begin{tabular}{cc}
\includegraphics[scale=0.5]{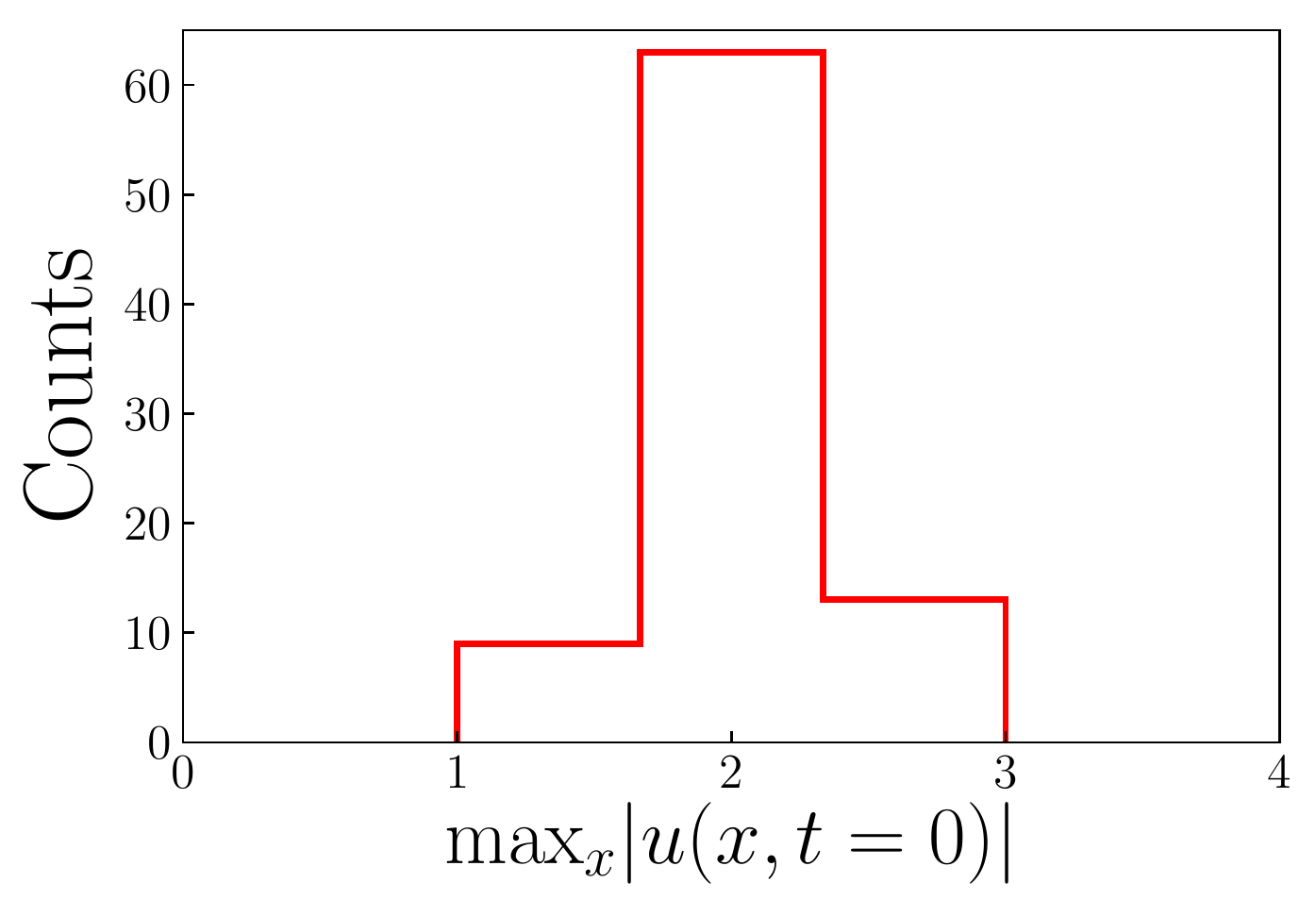} &
\includegraphics[scale=0.5]{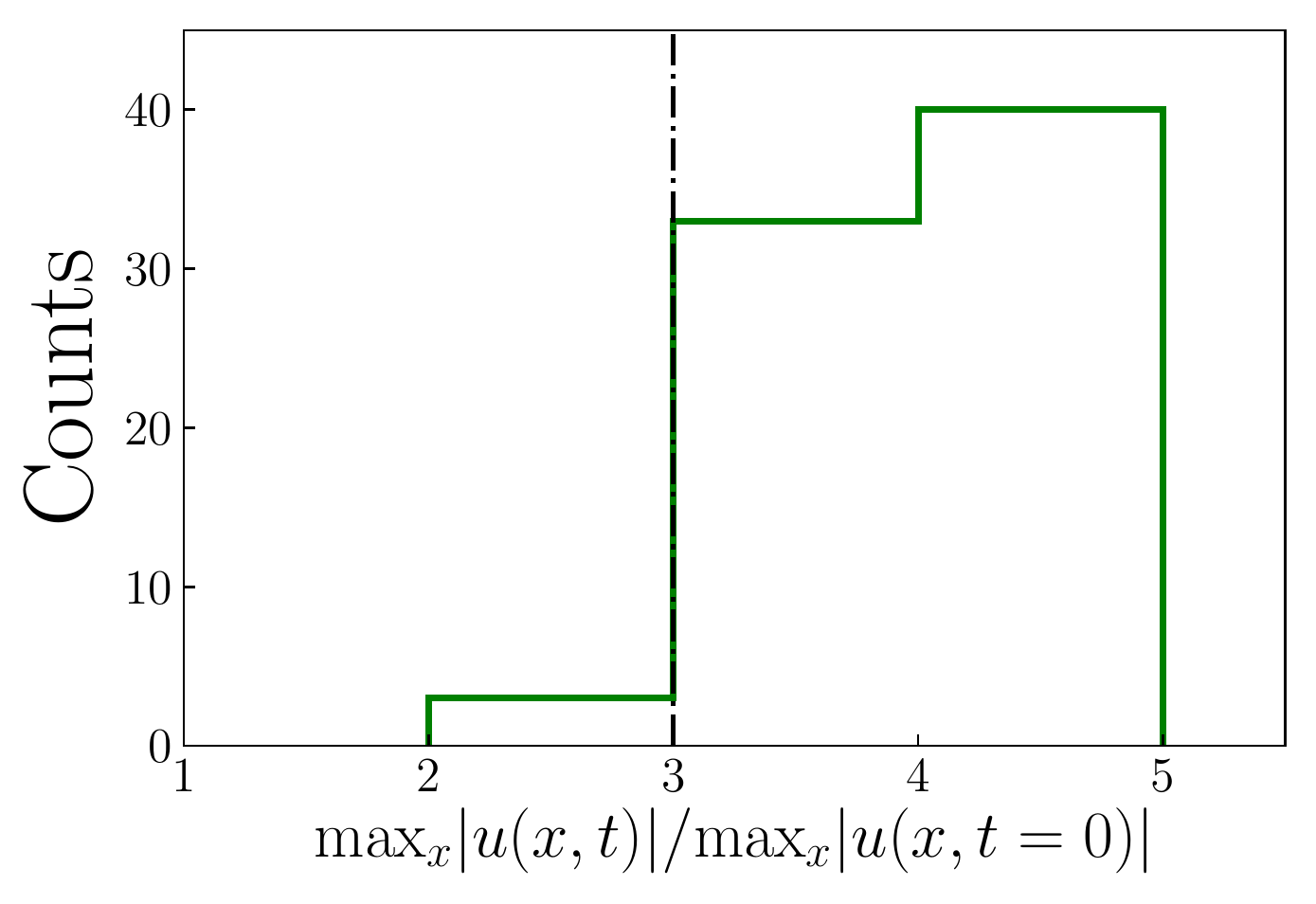} \\ (A) Initial amplitude for case (I).
& (B) Ratio of the final to initial amplitude for case (I).
\\ \\ \\ \includegraphics[scale=0.5]{Histogram_ampli1_initial.pdf} &
\includegraphics[scale=0.5]{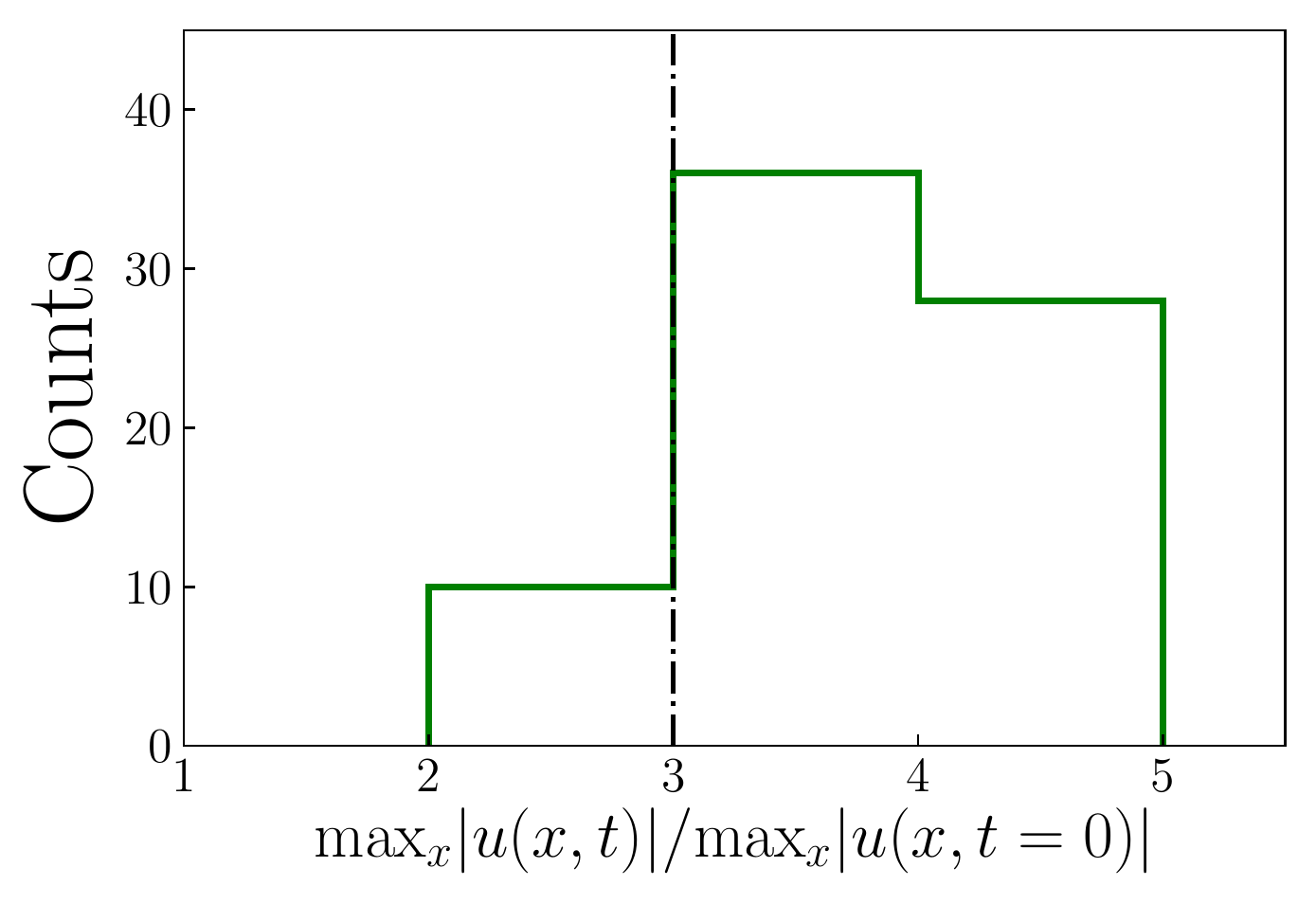} \\ (C) Initial amplitude for case (II).  & (D) Ratio of final to initial amplitude for case (II). \\ \\ 
\includegraphics[scale=0.5]{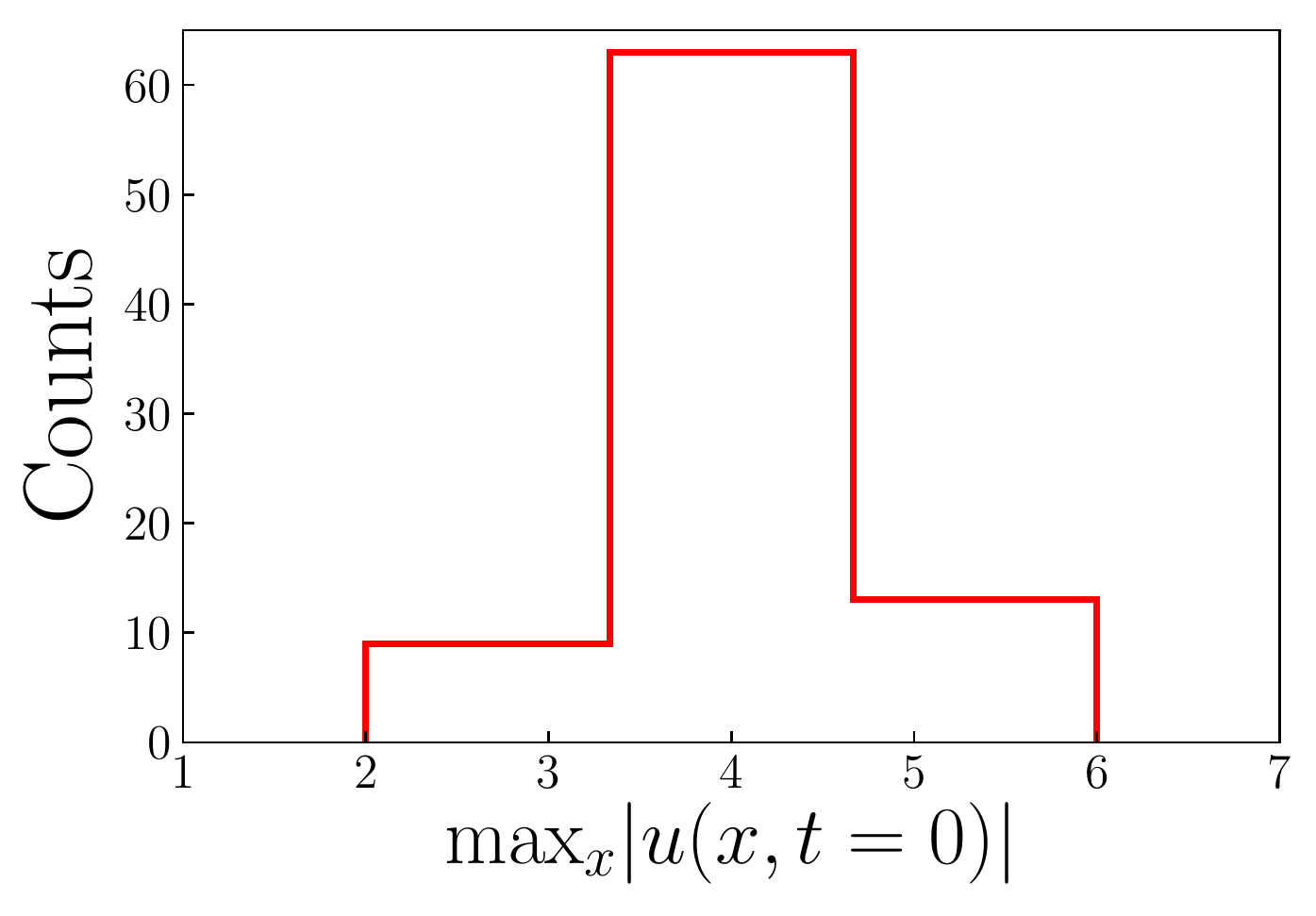} &
\includegraphics[scale=0.5]{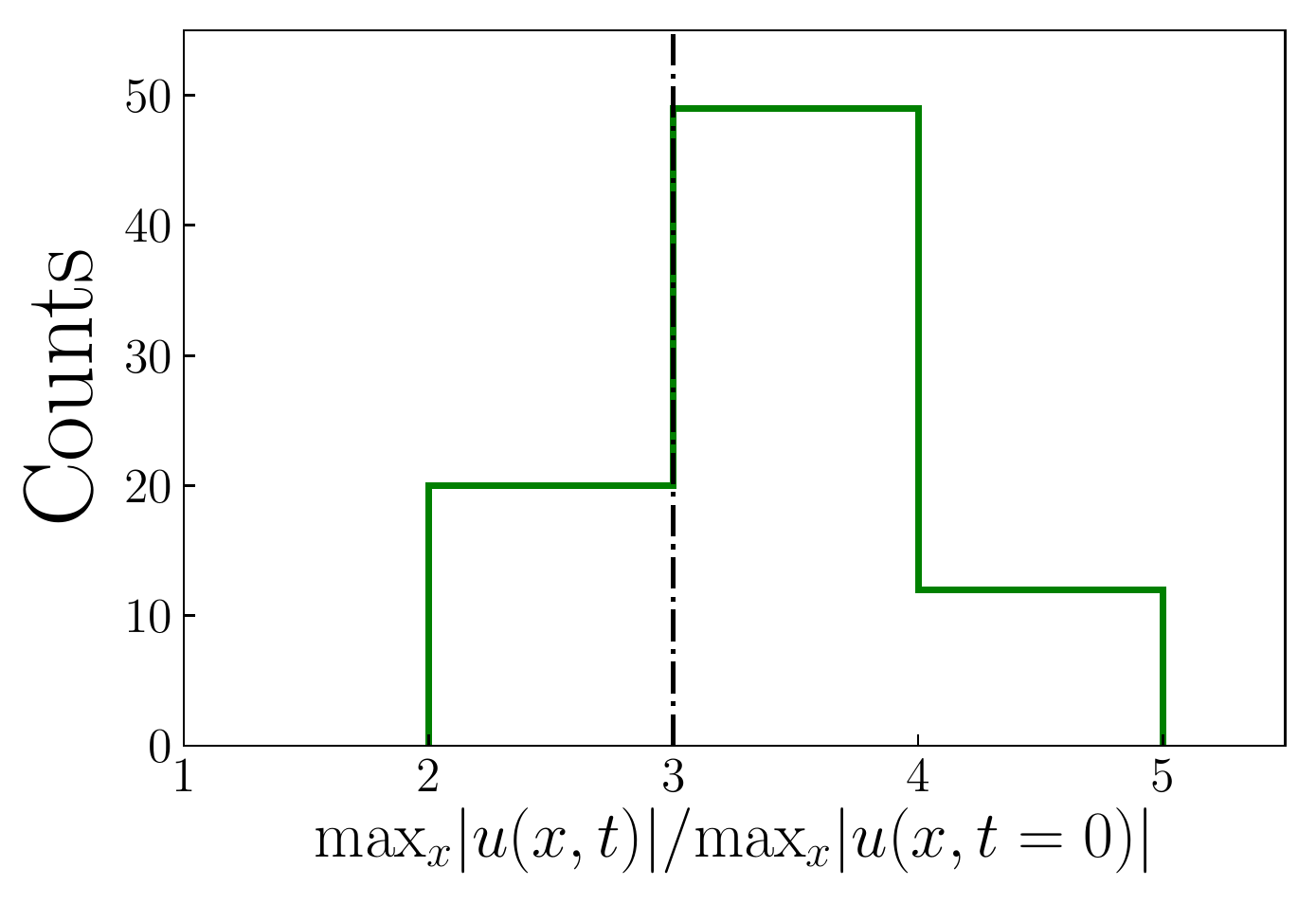} \\ (E) Initial amplitude for case (IV).  & (F) Ratio of final to initial amplitude for case (IV). 
\end{tabular}
\caption{The histograms for the cases (I) to (III) (see \ref{sol form} ) for details. The vertical black dashed lines on all the right panels represents the threshold of 3.}
\label{Spectrum}
\end{figure*}
 
First we choose a random number generators for the white noise field $w(k)$ as defined in Eq. (24) and Eq. (25) of the main text. Then, for each of the random number generator, the simulation runs till the secondary peak in the Fourier space reaches $k = -100$. The final amplitude of $|u|$ is then measured in the configuration space. For well-formed solitons, we put the criterion that the amplitude of the soliton $|u|$ must be three times or higher than the amplitude of the initial field. We construct two histograms. The first histogram shows the amplitude of the initial field. The second histogram shows the ratio between the final amplitude and the initial amplitude. The histograms are shown in Fig. \ref{Spectrum}. 

\begin{itemize}
    \item Case I: In the first simulation we have $Q = 0.25, s/q = 0.1, k_\mathrm{corr} = 2$. In this case the final amplitude of the field does exceed the initial value by 3 times.
    \item Case II: In the first simulation we have $Q = 0.25, s/q = 0.1, k_\mathrm{corr} = 1$. We take hundred random number generators and show that the for most random number generators the final amplitude of the field does exceed the initial value by 3 times. 
    \item Case III: In the first simulation we have $Q = 0.25, s/q = 0.1, k_\mathrm{corr} = 2$, 
    \  \ initial amplitude doubled compared to Case I.
\end{itemize}

\section{On the variation of ripple sizes in solitons}

\begin{figure*}
\begin{tabular}{cc}
\includegraphics[scale=0.15]{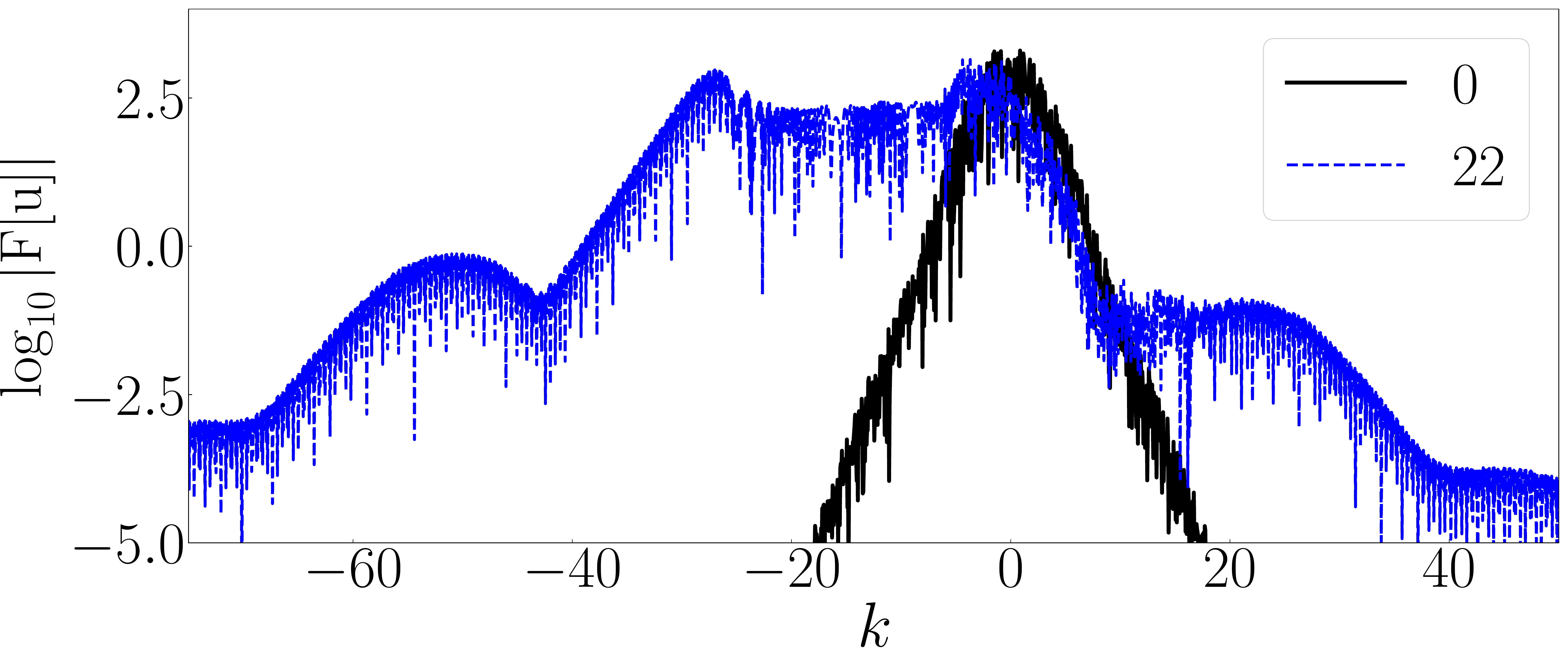} &
\includegraphics[scale=0.15]{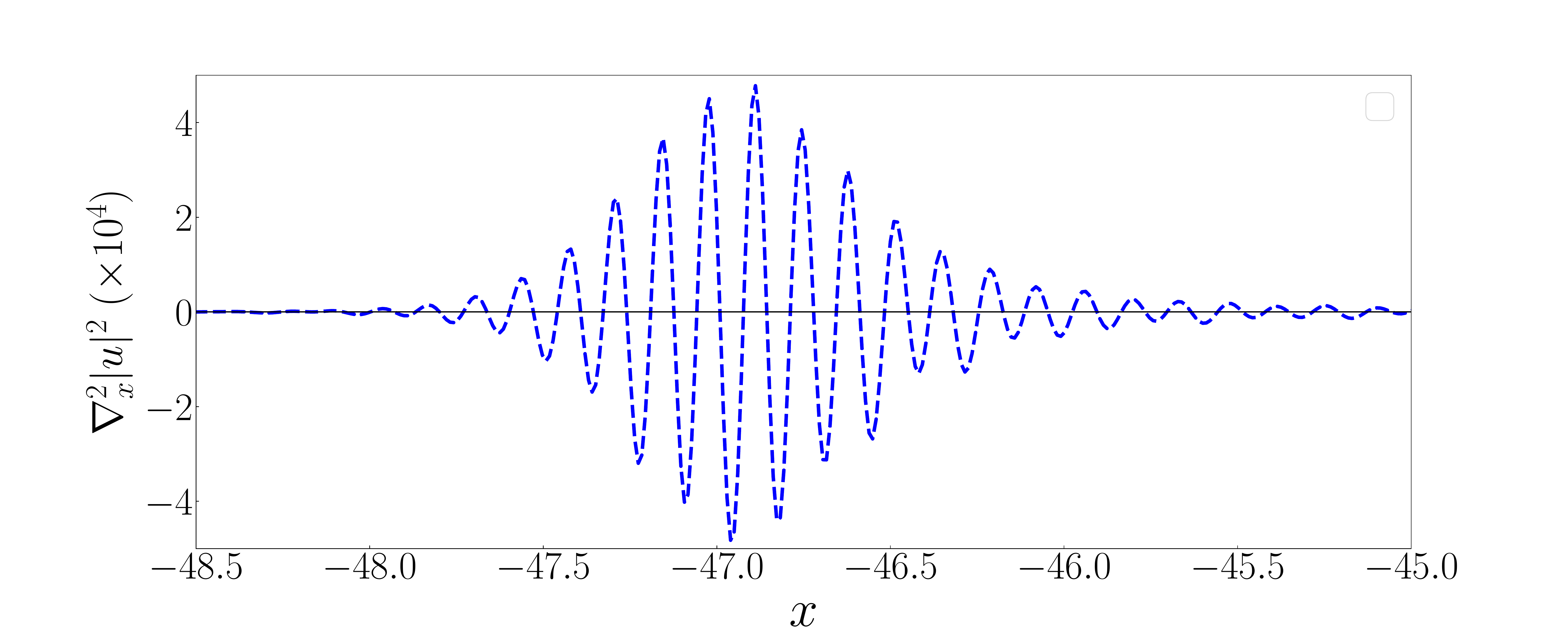} \\ (A) Fourier space for random seed 1.
& (B) Miller force corresponding to Panel (A).
\\ \\ \\ \includegraphics[scale=0.15]{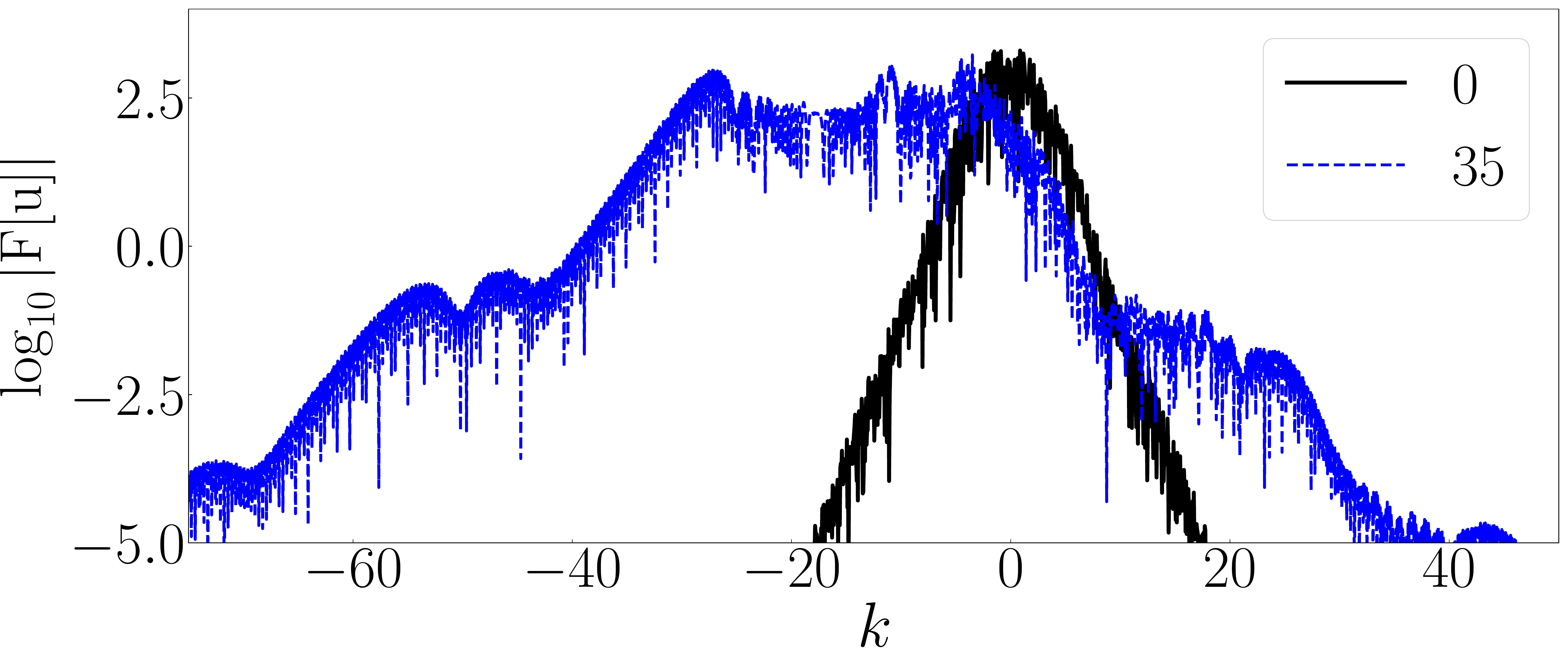} &
\includegraphics[scale=0.15]{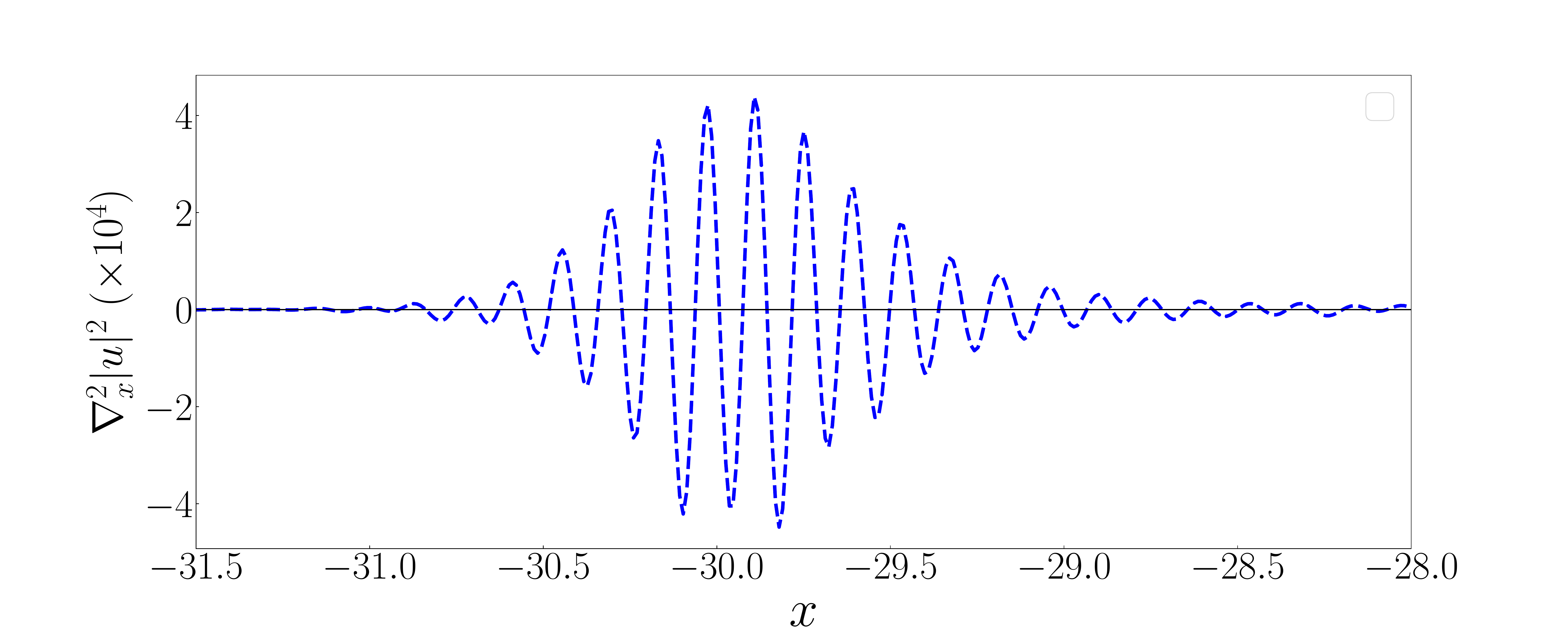} \\ (C) Fourier space for random seed 2.  & (D) Miller force corresponding to Panel (C). \\ \\ 
\includegraphics[scale=0.15]{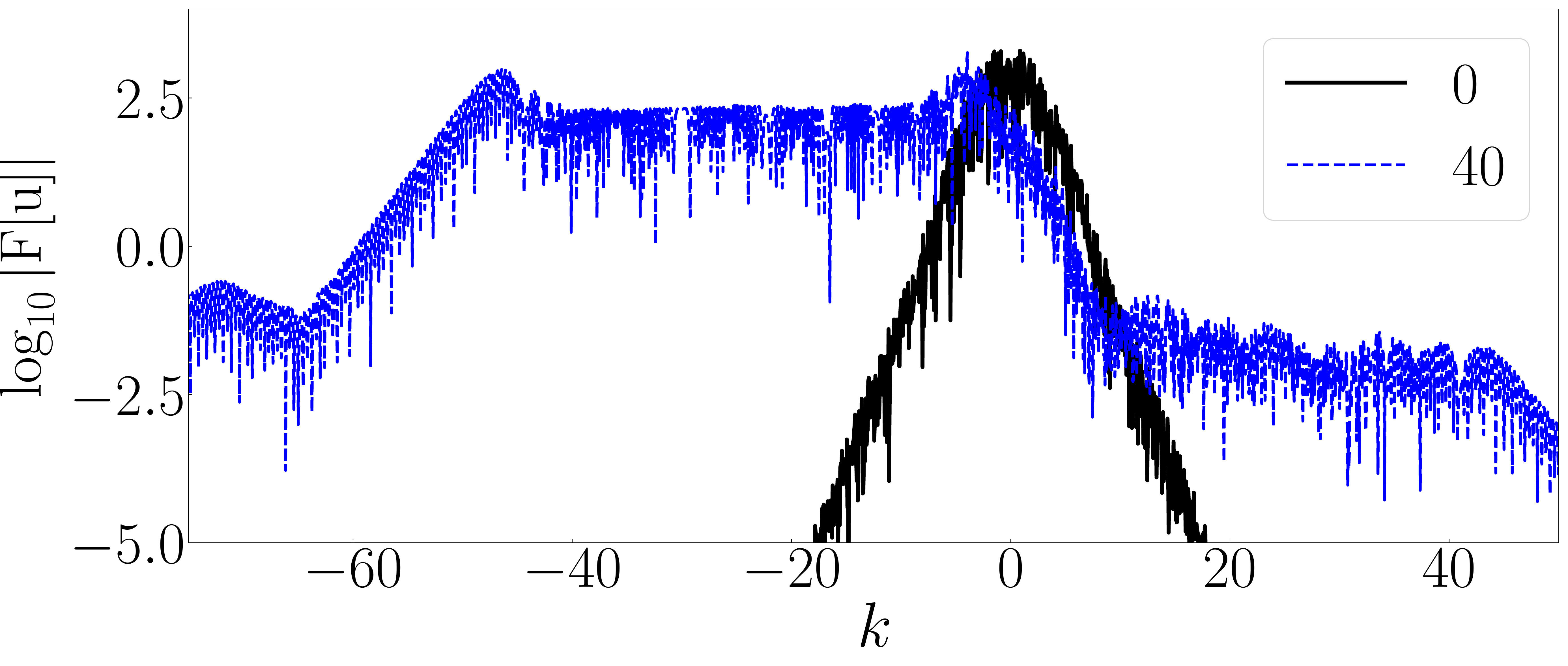} &
\includegraphics[scale=0.15]{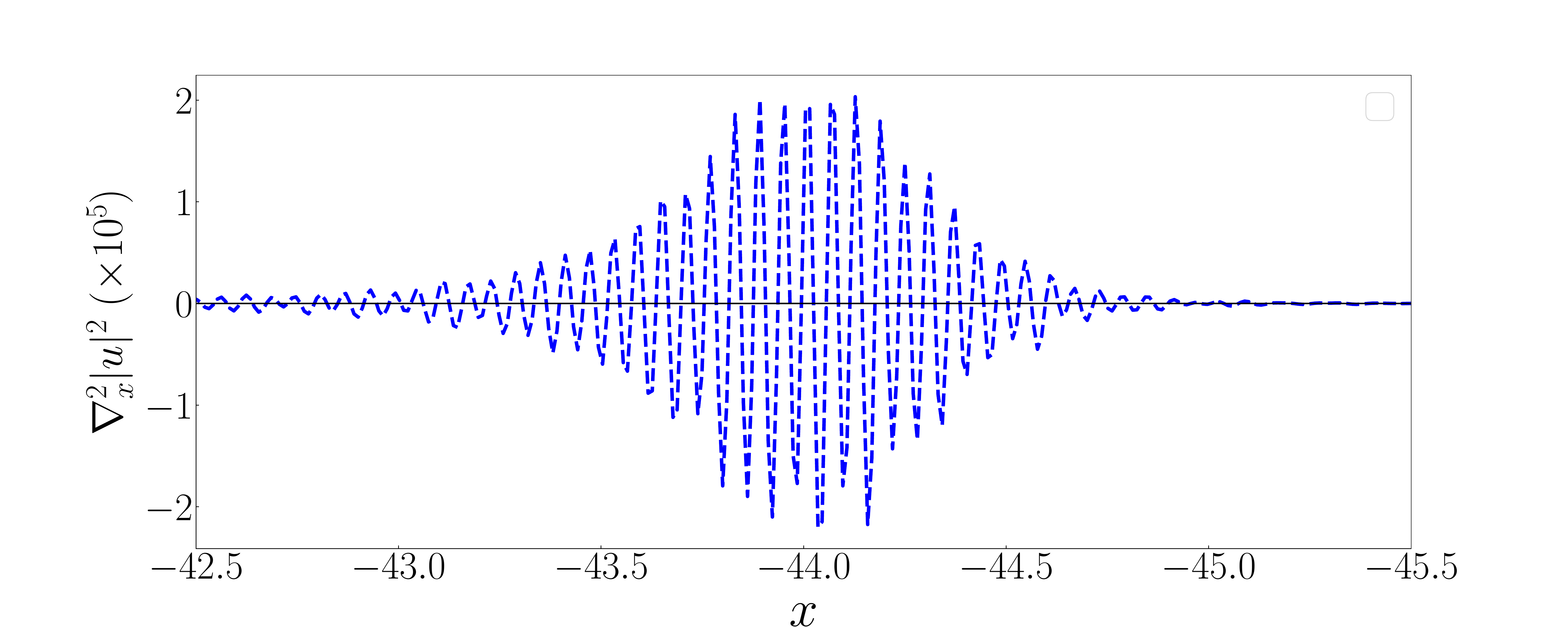} \\ (E) Fourier space for random seed 3.  & (F) Miller force corresponding to Panel (E). 
\end{tabular}
\caption{The figure shows the Fourier space and the configuration space for fixed $(Q=0.25,s/q=0.1,k_\mathrm{coor} = 2)$ for 3 different random seed values at times (shown in legends on the left panel) when the amplitude of the solitons exceeds the initial field amplitude by at least three times. It is seen that while the length of the soliton nearly remains constant, the number of ripples in the soliton structure is more if the secondary peak is located at a higher $k$. The number of ripples are similar if the secondary peak is located at similar $k$.   }
\label{Ripple size}
\end{figure*}

The criterion that we have used for a well-formed solitons is that the amplitude of the soliton must be at least three times higher than the amplitude of the initial field strength. In Fig. \ref{Ripple size} we show that for a fixed $(Q=0.25, s/q = 0.1,k_\mathrm{corr} = 2)$ we show the location of the secondary peak in the Fourier space (shown on left panels) and the Miller force (on the right panels) for three different random seed values. The time of soliton formation is indicated on the legends of the left panels.  It can be seen that when the location of the secondary peak is similar (Panels(A) and (C)), the number of ripples in the Miller force is similar. As shown in Panel (E) if the peak is located at a higher $k$ then as shown in Panel (F) the number of ripples also increases. It can be seen that the  number of ripples can change as much as $50 \%$. This variation has implications for the coherent curvature radiation pattern which will be explored in an upcoming work.

\section{Separation of distribution functions in curved magnetic field lines}\label{long_drift}

The following mechanism was proposed by \cite{1977ApJ...212..800C} (hereafter CR77). Let $\vec{\Omega}_\mathrm{Rot.}$ be the angular velocity of the pulsar and $\vec{B}$ be the local magnetic field at any distance $r$ on a given magnetic field line. The condition for the pulsar magnetosphere to co-rotate with the pulsar requires charge particles to maintain the co-rotational Goldreich-Julian value $\rho_\mathrm{GJ} = \vec{\Omega}_\mathrm{Rot}.\vec{B} f/(2 \pi c)$. Along the open magnetic field lines $\rho_\mathrm{GJ}$ is provided by the one-dimensional flow  of the charged high energy beams and the quasi-neutral plasma pair plasma. Magnetically induced pair creation cascades gets quenched at a distance of around $r_\mathrm{O} = 1.02 \;  R_\mathrm{NS}$. Beyond $r_\mathrm{O}$, no new particles are created. This implies that the total current across any cross-section of the flux tube formed by the open field lines remains constant beyond $r_\mathrm{O}$. At $r_\mathrm{O}$, the charged beams contribute exclusively to $\rho_\mathrm{GJ}$. The number density of the particles decreases as the strength of the field ($\propto 1/r^3$) while  Goldreich-Julian value varies as $\rho_\mathrm{GJ} \propto \cos \alpha / r^3$ where $\alpha$ is the angle between $\vec{B}$ and $\vec{\Omega}_\mathrm{Rot.}$. For the curved open field lines $ \alpha$ changes further away from the neutron star. The divergence less nature of the current flow  for $r \geq R_\mathrm{O}$ insures that the charged beams cannot completely provide for $\rho_\mathrm{GJ}$ all along a curved field line. The pair plasma provides the offset charge density ($\delta \rho = \rho_\mathrm{GJ} - \rho_\mathrm{b}$) by acquiring a net charge density. It requires the separation  of the bulk velocities of electrons and positrons in the pair plasma. We re-derive the expression for bulk-separation from CR77.

In the steady-state for which $\vec{E}.\vec{B} = 0$ , the charge density at any point is given by Goldreich-Julian charge density
\begin{align*}\label{GJ density}
\rho_\mathrm{GJ} \;=\; - \frac{\vec{\Omega}_\mathrm{Rot}.\vec{B}\;f}{2\pi c}   \numberthis
\end{align*}
where 
\begin{align*}
f \;=\; 1 + \mathcal{O} \left(\frac{\Omega_\mathrm{Rot}^2r^2}{c^2} \right)
\end{align*} is the contribution due to rotation.

The contribution to the total charge density is a summation of charge density due to each species `$\alpha$'-th such that 
\begin{align*}\label{charge density}
\rho_\mathrm{GJ} \;=\; \sum_{\alpha} \; \rho_{\alpha} 
\end{align*}
At any arbitrary point the species present are beam (positrons , ions or both), and electrons and positrons of pair plasma such that
\begin{align*}
\rho_\mathrm{GJ} \;=\; \rho_\mathrm{b} \;+\; \rho_{-} \;+\; \rho_{+}  \numberthis
\end{align*}

At the injection point `$\mathrm{O}$'  the entire contribution comes from the beam such that using \ref{GJ density} and \ref{charge density} we get
\begin{align*}\label{GJ at O}
\rho_\mathrm{GJ,O} \;=\; \rho_\mathrm{b,O} =  - \frac{\left(\vec{\Omega}_\mathrm{Rot}.\vec{B} \;f\right)_\mathrm{O}}{2\pi} \numberthis 
\end{align*}

The current density at any point ($\vec{r}$) is given by
\begin{align*}
\vec{J} \;=\; \rho_\mathrm{GJ}\;\left( \vec{\Omega}_\mathrm{Rot} \times \vec{r}\right) + \vec{J}_{\parallel} 
\end{align*}
where $\vec{J}_{\parallel}$ is the current in the direction of the local magnetic field $\vec{B}$ such that
\begin{align*}\label{parallel current density}
\vec{J}_{\parallel} \;=\; \sum_{\alpha} \;\vec{v}_{\parallel,\alpha}\; \rho_{\alpha} \;=\; \sum_{\alpha} \; \vec{J}_{\parallel,\alpha} \numberthis
\end{align*}

In some portion of the field line where there is no source term ( meaning there is no particle production and dissipation ) we can write
\begin{align*}
\nabla . \vec{J}_{\parallel,\alpha} \;=\; 0   
\end{align*} 
for each component. 

The solution of the above equation is given by
\begin{align*}\label{current density for ith component}
\vec{J}_{\parallel,\alpha} \;=\; - \frac{\Phi_{\alpha}\;\Omega_\mathrm{Rot}}{2\pi} \; \vec{B} \numberthis
\end{align*} 
such that $\Phi_{\alpha}\; \Omega_\mathrm{Rot}$ is an invariant along any given field line.

Using equation \ref{parallel current density} and \ref{current density for ith component} we get
\begin{align*}\label{alternative current density for ith component}
&\ \vec{J}_\mathrm{\parallel,b} \;=\; \rho_\mathrm{b} \;c \; \hat{B} \;=\; -\frac{\Phi_\mathrm{b}\; \Omega_\mathrm{Rot}}{2\pi}\; \vec{B} \\\\ 
&\ \vec{J}_{\parallel,\pm} \;=\;  \rho_{\pm}\;v_{\pm} \; \hat{B} \;=\; -\frac{\Phi_{\pm} \;\Omega_\mathrm{Rot}}{2\pi}\; \vec{B}   \numberthis
\end{align*} 
where the ultra-relativistic nature of the beam particles  ($v \sim c$) has been taken account.

At point `$\mathrm{O}$' using Eq \ref{GJ at O} and \ref{alternative current density for ith component} we get
\begin{align*}\label{alpha for beam}
&\ \rho_\mathrm{b,O} \;=\; - \frac{\left(\Omega_\mathrm{Rot} \;B\right)_\mathrm{O}}{2\pi c}\; \Phi_\mathrm{b} \;=\; - \frac{( \vec{\Omega}_\mathrm{Rot}.\vec{B} \; f)_\mathrm{O}}{2\pi c} \\\\
&\ \Rightarrow \Phi_\mathrm{b} \;=\;   \; \left( \hat{\Omega}_\mathrm{Rot}.\hat{B}\; f\right)_\mathrm{O} \numberthis
\end{align*}

Similarly for $e^{\pm}$ of pair plasma using  equation \ref{alternative current density for ith component} at point `$O$' we get
\begin{align*}
&\ \Phi_{\pm} \; \frac{(\Omega_\mathrm{Rot} \;\vec{B})_\mathrm{O}}{2\pi} \;=\; \left[ \rho_{\pm}\; v_{\pm} \; \hat{B} \right]_\mathrm{O} \\\
&\ \Rightarrow \Phi_{\pm} \;=\; -2\pi \; \left[ \frac{\rho_{\pm} \; v_{\pm}}{\Omega_\mathrm{Rot} \; B}\right]_\mathrm{O}     
\end{align*}

From Eq \ref{alpha for beam} we get
\begin{align*}
(\Omega_\mathrm{Rot} \; B)_\mathrm{O} \;=\; - \frac{2\pi c }{\Phi_\mathrm{b}} \; \rho_\mathrm{b,O}  \numberthis
\end{align*} 

Using this in the expression for $\Phi_{\pm}$ we get
\begin{align*}\label{alpha for electron-positron}
\Phi_{\pm} \;=\; \left[ \frac{\rho_{\pm}}{\rho_\mathrm{b}}\right]_\mathrm{O} \; \left[ \frac{v_{(\pm)}}{c} \right]_\mathrm{O} \; \Phi_\mathrm{b}  \numberthis
\end{align*}

From Eq \ref{GJ density} and \ref{charge density} we get for any arbitrary point `$\mathrm{A}$'
\begin{align*}\label{charge offset condition}
&\ 1 \;+\; \frac{\rho_{+}}{\rho_\mathrm{b} } \;+\; \frac{\rho_{-}}{\rho_\mathrm{b} } \;=\; - \frac{\vec{\Omega}_\mathrm{Rot}. \vec{B} f}{2\pi c \; \rho_\mathrm{b} }  \numberthis 
\end{align*}

Using Eq \ref{alpha for beam} and \ref{alternative current density for ith component} we obtain
\begin{align*}\label{beam charge density}
\rho_\mathrm{b} \;=\; - \left( \hat{\Omega}_\mathrm{Rot}.\hat{B} f \right)_\mathrm{O} \frac{\Omega_\mathrm{Rot} \; B}{2 \pi c} \numberthis 
\end{align*}

Using Eq  \ref{alternative current density for ith component}, \ref{alpha for electron-positron} , and \ref{beam charge density} in Eq \ref{charge offset condition}  we get
\begin{align*}
&\ \frac{\rho_{+}}{\rho_\mathrm{b}} \;+\; \frac{\rho_{-}}{\rho_\mathrm{b}} \;=\; \left[ \frac{\hat{\Omega}_\mathrm{Rot}.\hat{B} \; f}{(\hat{\Omega}_\mathrm{Rot}.\hat{B}\; f )_\mathrm{O}} \;-\; 1\right] \\\
&\ \Rightarrow \frac{1}{\rho_\mathrm{b} } \left[ \Phi_{+} \left( \frac{\Omega_\mathrm{Rot}\; B}{2\pi v_{+}}\right) + \Phi_{-} \left( \frac{\Omega_\mathrm{Rot}\; B}{2\pi v_{-}}\right) \right]\;=\;  \left[ \frac{\hat{\Omega}_\mathrm{Rot}.\hat{B} \; f }{(\hat{\Omega}_\mathrm{Rot}.\hat{B}\; f )_\mathrm{O}} \;-\; 1\right] \\\\
&\ \Rightarrow \frac{\Omega_\mathrm{Rot} B \; \Phi_\mathrm{b}}{2 \pi c \;\rho_\mathrm{b} } \left[ \left(\frac{\rho_{+}}{\rho_\mathrm{b}  } \right)_\mathrm{O} \left( \frac{v_\mathrm{+,O}}{v_{+}}\right) + \left(\frac{\rho_{-}}{\rho_\mathrm{b} } \right)_\mathrm{O} \left( \frac{v_\mathrm{-,O}}{v_{-}}\right)\right] \;=\; \left[ \frac{\hat{\Omega}_\mathrm{Rot}.\hat{B} \; f }{(\hat{\Omega}_\mathrm{Rot}.\hat{B}\; f )_\mathrm{O}} \;-\; 1\right] \\
&\ \Rightarrow -\left[ \left(\frac{\rho_{+}}{\rho_\mathrm{b} } \right)_\mathrm{O} \left( \frac{v_\mathrm{+,O}}{v_{+}}\right) + \left(\frac{\rho_{-}}{\rho_\mathrm{b} } \right)_\mathrm{O} \left( \frac{v_\mathrm{-,O}}{v_{-}}\right)\right] \;=\; \left[ \frac{\hat{\Omega}_\mathrm{Rot}.\hat{B} \; f}{(\hat{\Omega}_\mathrm{Rot}.\hat{B}\; f)_\mathrm{O}} \;-\; 1\right] 
\end{align*}

The pair plasma is neutral at point `$\mathrm{O}$' which translates to the condition   
\begin{align*}
\rho_\mathrm{-,O} \;=\; -\; \rho_\mathrm{+,O}  \numberthis
\end{align*}

This gives us the following expression, 
\begin{align*}
 \left( \frac{\rho_{+}}{\rho_\mathrm{b} } \right)_\mathrm{O} \left[ \frac{v_{+} v_\mathrm{-,O} \;-\; v_{-}v_\mathrm{+,O}}{v_{+} \;v_{-}}\right] \;=\;\left[ \frac{\hat{\Omega}_\mathrm{Rot}.\hat{B} \; f}{(\hat{\Omega}_\mathrm{Rot}.\hat{B}\; f)_\mathrm{O}} \;-\; 1\right]
\end{align*} 

Making use of the fact that the secondary plasma particles are highly relativistic we can reduce the above expression to the form
\begin{align*} \label{longdrift}
&\ \frac{v_{+} - v_{-}}{c} \;\approx\; \left(\frac{\rho_\mathrm{b} }{\rho_{+}} \right)_\mathrm{O} \;\left[ \frac{\hat{\Omega}_\mathrm{Rot}.\hat{B} \; f}{(\hat{\Omega}_\mathrm{Rot}.\hat{B}\; f)_\mathrm{O}} \;-\; 1\right] \\\\
&\ \Rightarrow |\Delta \beta| \;\approx\; \left| \left(\frac{\rho_\mathrm{b} }{\rho_+} \right)_\mathrm{O} \;\left[ \frac{\hat{\Omega}_\mathrm{Rot}.\hat{B} \; f}{(\hat{\Omega}_\mathrm{Rot}.\hat{B}\; f)_\mathrm{O}} \;-\; 1\right]\right| \numberthis
\end{align*}

We call the ratio $(\rho_\mathrm{b}/\rho_\mathrm{s}) = 1/\kappa_\mathrm{GJ}$ as the density term. We call the term in square brackets as the geomtrical term. The correction `$f_\mathrm{Rot} \approx 1 + \mathcal{O}({\Omega^2r^2}/{c^2})$' due to rotation can be taken to be 1, as the
higher order term $\mathcal{O} ({\Omega^2r^2}/{c^2}) \sim 0.01$ at
$r = 50 R_\mathrm{NS} $ for a pulsar with $P = $ 1 second. It can be
seen that the separation of the electron-positron distribution
function is a product of two terms viz., the density term
$({\rho_\mathrm{b} }/{\rho_\mathrm{s}})_\mathrm{O}$ and the geometrical term $[{(\hat{\Omega}_\mathrm{Rot}\cdot \hat{B}
    )_\mathrm{A}}/{(\hat{\Omega}_\mathrm{Rot}\cdot\hat{B} )_\mathrm{O}} - 1 ]$. The geometrical factor is zero only for very straight magnetic field lines.  Thus curved magnetic field line is
    a necessary requirement of longitudinal drift/ separation of
    $e^{\pm}$ distribution in a secondary plasma. \citealt{2020MNRAS.497.3953R} used the model by \citealt{2002A&A...388..235G} and found a wide parameter space for the separation of plasma distribution functions exists which depends on the arrangement of  non-dipolar surface magnetic field. 

\section{Combining Lorentz factor}

Let the moving frame of reference (MFR) and the plasma frame of reference (PFR) move with respect to the observer's drame of reference (OFR) with beta factor $\beta_\mathrm{MFR}$ and $\beta_\mathrm{s}$ in the outward direction along the open magnetic field lines. Let $\gamma_\mathrm{MFR}$ and $\gamma_\mathrm{s}$ be the corresponding Lorentz factor in OFR. Let $\beta_\mathrm{gr}$ be the beta factor of MFR with respect to PFR, then we have
\begin{equation}
    \beta_\mathrm{gr} = \frac{\beta_\mathrm{MFR} - \beta_\mathrm{s}}{1 - \beta_\mathrm{MFR} \beta_\mathrm{s} } \label{E1}
\end{equation}
The Lorentz factor $\gamma_\mathrm{gr}$ corresponding to $\beta_\mathrm{gr}$ is given by
\begin{equation}
    \gamma_\mathrm{gr} = \frac{1}{\sqrt{1 - \beta^2_\mathrm{gr}}} \label{E2}
\end{equation}

Substituting equation (\ref{E1}) into equation (\ref{E2}) we get,
\begin{equation}
    \gamma_\mathrm{gr} = \frac{\gamma_\mathrm{MFR}}{2 \gamma_\mathrm{s}}
\end{equation}
which on being re-arranged gives us
\begin{equation}
    \gamma_\mathrm{MFR} = 2 \gamma_\mathrm{s} \gamma_\mathrm{gr}
\end{equation}
In the the ultra-relativistic limit $\gamma_\mathrm{gr} = \sqrt{1 + p^2_\mathrm{gr}} \approx p_\mathrm{gr}$ and the expression reduces to the form 
\begin{equation}
    \gamma_\mathrm{MFR} \approx 2 \gamma_\mathrm{s} p_\mathrm{gr}
\end{equation}

\section{Notations and symbols used throughout the text.}
\begin{itemize}
   \item NLD: Non-linear Landau damping.
   \item GVD: Group-velocity dispersion.
   \item CNL: Cubic non-linearity. 
   \item $\kappa:$ Multiplicity of pair plasma. The ratio of the number density of the pair plasma to the co-rotational Goldreich-Julian number density. 
   \item  OFR: The observer's frame of reference. Can be identified with co-rotating frame of reference for normal period pulsars.
   \item PFR: Plasma frame of reference. The frame of reference where the mean velocity of the plasma particles is zero.
   %\item $\omega_\mathrm{p}:$ plasma frequency in PFR.
   \item $v_\mathrm{gr}:$ group velocity of linear Langmuir waves in PFR.
   \item MFR: Moving frame of reference that moves with velocity $v_\mathrm{gr}$ with respect to PFR. 
   \item $f^{(0)}_{\alpha}$: plasma particle distribution function (DF) of the $\alpha-$th species. Two-representative DF viz., a short-tailed Gaussian distribution DF and a long-tailed Lorentzian DF are used.
   \item $p$: Dimensionless momentum of plasma particles.
   \item ${p}_{\alpha}$: The mean momentum of the $\alpha-$ th species.
   \item $p_\mathrm{ph}$: Dimensionless momentum corresponding to wave phase velocity.
   \item $p_\mathrm{gr}$: Dimensionless momentum corresponding to wave group velocity.
   \item $\sigma:$ width (``temperature'') of the Gaussian distribution function.
   \item $\Delta p:$ width (``temperature'') of the Lorentzian distribution function.
   \item $E:$ Envelope electric field governed by the non-linear Schr\"{o}dinger equation along with non-linear Landau damping..
   \item $G:$ coefficient of the group velocity dispersion.
   \item $q:$ coefficient of cubic non-linearity.
   \item $s:$ coefficient of non-linear Landau damping term.
   \item $G_\mathrm{d}:$ dimensionless group velocity dispersion.
   \item $q_\mathrm{d}:$ dimensionless cubic non-linearity.
   \item $s_\mathrm{d}:$ dimensionless non-linear Landau damping term.
   \item $Q:$ equals $2 q_\mathrm{d}/ G_\mathrm{d}$. Decides the typical timescale associated with the emergence of solitons ($\mathcal{O} (1/Q)$)
   \item $\omega_\mathrm{p}:$ characteristic plasma frequency of Langmuir waves in PFR.
   \item $l:$ Characteristic length of the linear Langmuir waves in PFR.
   \item $\theta:$ Ratio of the spatial extent of the Langmuir envelope to the characteristic length $l$ of the linear Langmuir waves. 
   \item OFR: Observer's frame of reference. Identified with the co-rotating frame of reference.
   \item $\gamma_\mathrm{s}:$ Bulk Lorentz factor of the PFR with respect to OFR.
\end{itemize}



%%%%%%%%%%%%%%%%%%%%%%%%%%%%%%%%%%%%%%%%%%%%%%%%%%

% Don't change these lines
\bsp	% typesetting comment
\label{lastpage}
\end{document}